\documentclass[preprint,psfig]{aastex}

\newcommand{\be}{\begin{equation}}
\newcommand{\ee}{\end{equation}}
\newcommand{\nn}{\mbox{} \nonumber \\ \mbox{} }
\newcommand{\ba}{\begin{eqnarray}}
\newcommand{\ea}{\end{eqnarray}}

\newcommand{\Alfven}{ Alfv\'{e}n }

\newcommand{\E}{{\bf E}}
\newcommand{\B}{{\bf B}}

\newcommand\etal{et al.\ }
\newcommand\eg{e.g.\ }

\begin{document}

\title{Polarization and structure of relativistic parsec-scale AGN jets}
\author{M. Lyutikov$^{1,2,3,}$\altaffilmark{4}, V.I. Pariev$^{5,6}$,
D.C. Gabuzda$^{7}$}
\affil{$^1$ Physics Department, McGill University, 3600 rue University,
Montreal, QC,\\Canada H3A 2T8 \\
$^2$ Canadian Institute for Theoretical Astrophysics,\\ 60 St. George, Toronto, Ont,
M5S 3H8, Canada \\
$^3$ Kavli Institute for Particle Astrophysics and Cosmology, \\
2575 Sandhill Road, Menlo Park, CA 94305, USA}
\altaffiltext{4}{lyutikov@physics.mcgill.ca}
\affil{$^5$ Department of Physics and Astronomy,
University of Rochester, Rochester, NY 14627 \\
$^6$ Lebedev Physical Institute, Leninsky Prospect 53,
Moscow 119991, Russia}
\affil{$^{7}$ Department of Physics, University College Cork, Cork, Ireland}

\date{Received   / Accepted  }

\begin{abstract}
We consider the polarization properties of optically thin synchrotron
radiation emitted by relativistically moving electron--positron 
jets carrying large-scale helical magnetic fields. In our model, 
the jet is cylindrical, and the emitting plasma moves parallel to the 
jet axis with a characteristic Lorentz factor $\Gamma$. We draw 
attention to the strong influence that the bulk relativistic 
motion of the emitting relativistic particles has on the observed
polarization. 

Our computations
predict and explain the following behavior. 
(i) For jets unresolved in the direction
perpendicular to their direction of propagation, the position angle of
the electric vector of the linear polarization has a bimodal
distribution, being  oriented either parallel or perpendicular to the
jet.  (ii) If an ultra-relativistic jet with $\Gamma\gg 1$
whose axis makes a small angle to the line of sight,
$\theta \sim 1/\Gamma$, 
experiences a relatively small change in the direction of propagation, 
velocity or pitch angle of the magnetic fields, the polarization is 
likely to remain parallel or perpendicular; on the other hand, in some cases,
the degree of polarization can exhibit
large variations and the polarization position angle can experience
abrupt $90^{\circ}$ changes. This  change  is more likely to occur in jets
with flatter spectra.
(iii) In order for the jet polarization to be oriented along the jet axis,
the intrinsic toroidal magnetic field (in the frame of the jet) should
be of the order of or stronger than the 
intrinsic poloidal field; in this case, the
highly relativistic motion of the jet implies  that, in the observer's frame, 
the jet is strongly dominated by the toroidal 
magnetic field $B_\phi / B_z \geq \Gamma$. 
(iv) The emission-weighted average pitch angle of the 
intrinsic helical field in the jet
must not be too small to produce polarization along the jet axis. 
In force-free jets with a smooth distribution 
of emissivities, the emission should be generated  in a limited range of
radii not too close to the jet core. 
(v) For mildly relativistic jets, when a counter jet can be seen, the 
polarization of the  counter jet is preferentially orthogonal to the axis, 
unless the jet is strongly dominated by the toroidal magnetic field in 
its rest frame.
(v) For resolved jets, the
 polarization pattern  is not symmetric with respect to jet axis. 
Under certain conditions, this can be used to deduce the direction 
of the spin of the central object
(black hole or disk), whether it is aligned or anti-aligned with the jet axis.  
(vi) In  resolved ``cylindrical shell'' type  jets,  the central parts 
of the jet
are polarized along the axis, while the outer parts are polarized
orthogonal to it, in accordance with observations.

We conclude that large-scale magnetic fields can explain the
salient polarization  properties of parsec-scale AGN jets.  Since
the typical degrees of polarization are $\leq 15\%$, the
emitting parts of the  jets  must have
comparable rest-frame toroidal and poloidal fields. In this case, 
most relativistic jets are  strongly dominated
by the toroidal magnetic field component in the observer's frame,
$B_\phi/B_z \sim \Gamma$.  We also discuss the possibility that
relativistic AGN jets may be electromagnetically (Poynting flux) dominated.
In this case, dissipation of the toroidal magnetic field (and not fluid
shocks) may be responsible for particle acceleration.
\end{abstract}

\section{Introduction}

\begin{figure}[ht]
\includegraphics[width=0.9\linewidth,angle=270]{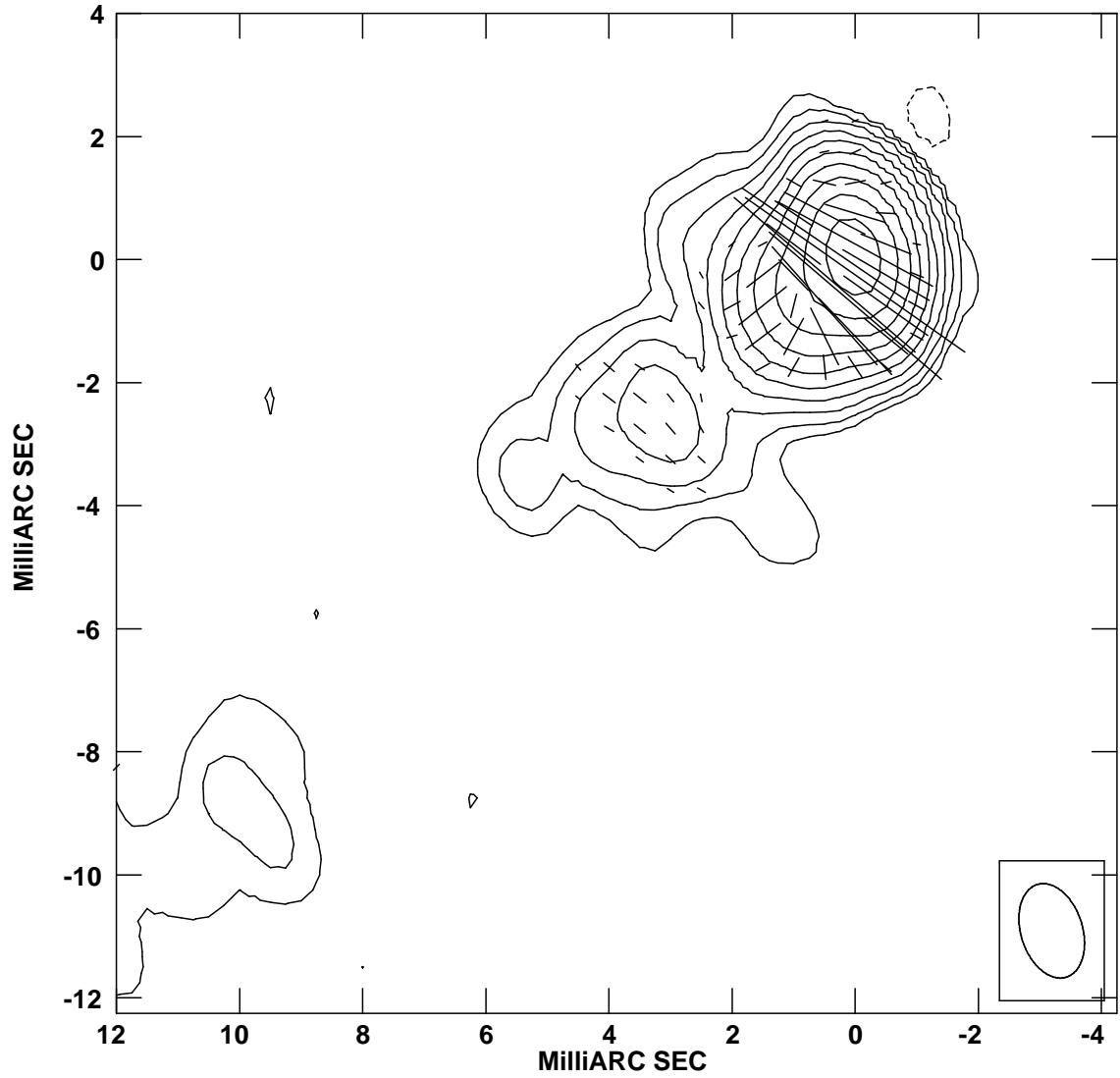}
\caption{Observed 6~cm total-intensity contours and superposed
linear-polarization vectors for the BL~Lac object 1418+546: the orientation
of the polarization relative to the jet displays jumps between being
aligned with and orthogonal to the jet (reproduced from
\protect\citealt{gab03a} and \protect\citealt{push04}). 
}
\label{alternate_1418}
\end{figure}

\begin{figure}[ht]
\includegraphics[width=0.9\linewidth,angle=270]{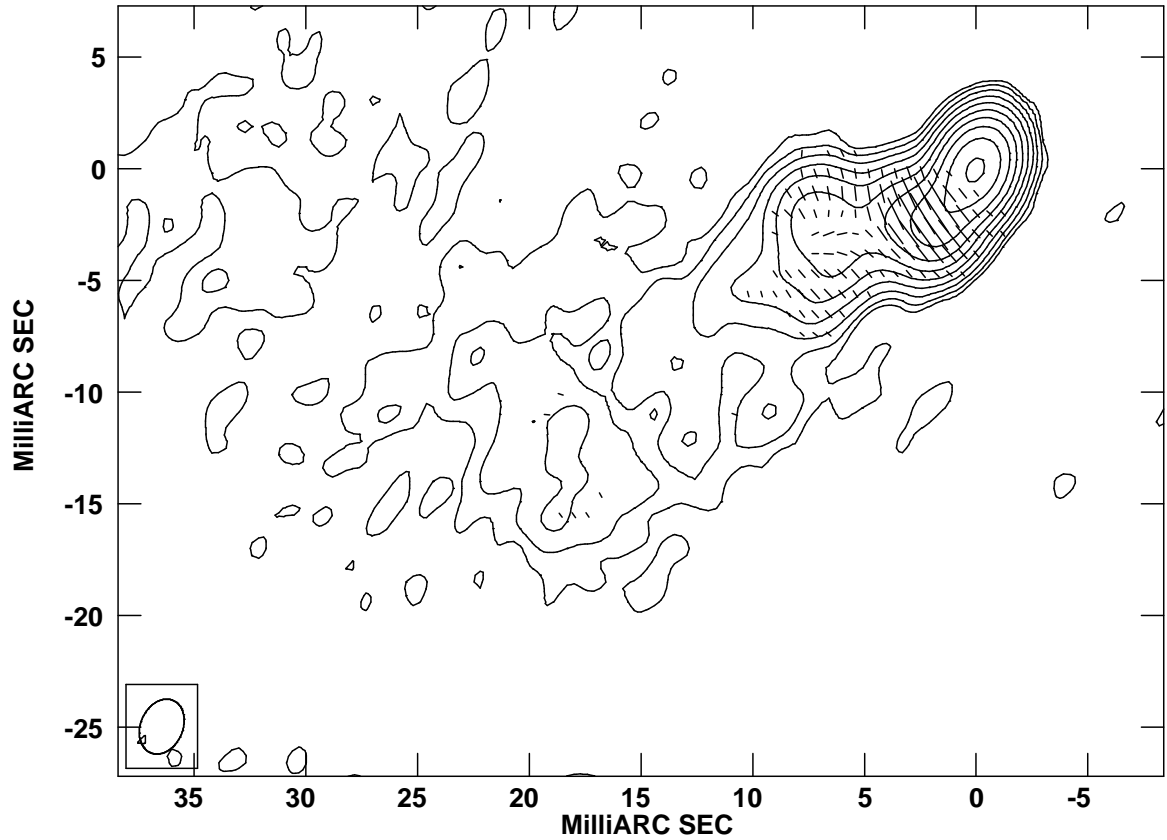}
\caption{Observed 4~cm total-intensity contours and superposed
linear-polarization vectors for the resolved jet of Mrk501:  the
EVPAs are aligned with the jet in the central region of the jet and
orthogonal to the jet at its edges (reproduced from \protect\citealt{gab03a}
and \protect\citealt{push04}).
}
\label{resolved_1652}
\end{figure}

\begin{figure}
\includegraphics[width=0.7\linewidth,angle=270]{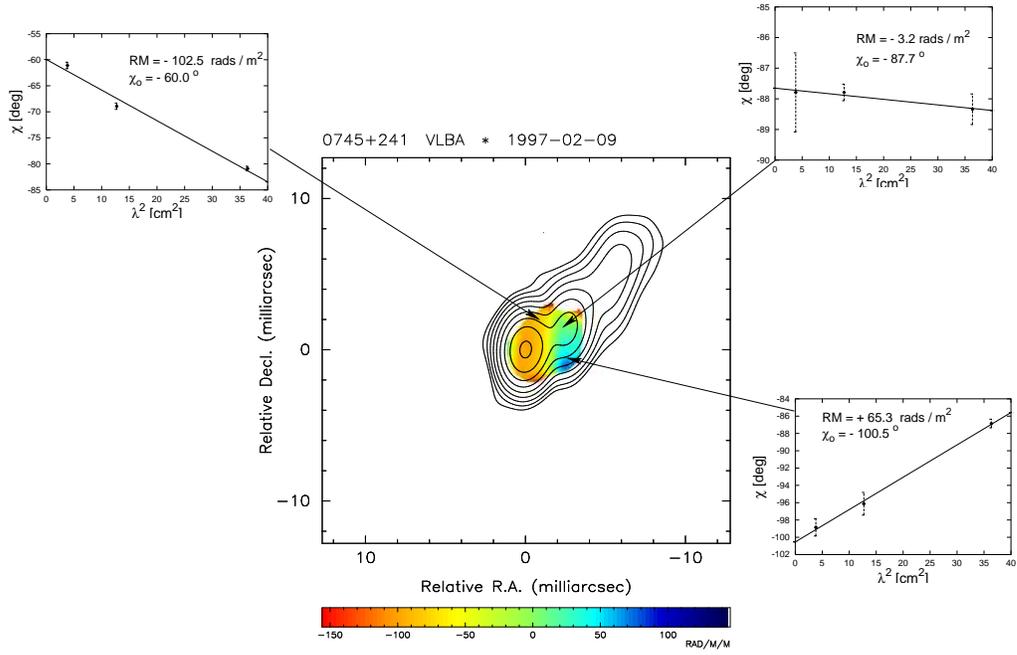}
\caption{Observed 6~cm total-intensity contours and superposed
distribution of the parsec-scale rotation measure for the BL~Lac
object 0745+241. A gradient of the rotation measure across the jet
is clearly visible (reproduced from \protect\citealt{gab04}).
}
\label{rmgrad_0745}
\end{figure}

VLBI linear polarization studies of compact parsec-scale jets in radio-loud
AGNs have revealed a number of general tendencies
\citep[\eg][and references therein]{caw93,gab00,gab03a}.  These tendencies
include (i) a roughly bimodal distribution of the electric vector position
angles (EVPAs) for the jet linear polarization with respect to jet direction,
with quasars tending to have polarization orthogonal to the jet and BL~Lac
objects tending to have polarization along the 
jet \citep{mjmw02}; (ii) the EVPAs sometimes
``follow'' the jet as it bends, keeping its relative orientation with
respect to the jet direction (Fig.~\ref{follow_1749}, 
also \citet{owen89} for the kiloparsec jet in M87); 
(iii) the jet EVPAs
sometimes experience orthogonal jumps along the jet, changing from aligned
to orthogonal or vice versa (\citealt{gab01a}; 
Fig.~\ref{alternate_1418});
(iv) in some resolved jets, the EVPA in the central part of the jet lies along
the jet, while the edges are polarized orthogonally to the jet direction
(\citealt{att99,moellenbrock00};  
Fig.~\ref{resolved_1652}); (v) Faraday-rotation
gradients are frequently observed across the jets of BL~Lac objects
(Fig.~\ref{rmgrad_0745}).  All  of these  properties
are statistical, each having a  number of known exceptions. 
%(The images
%shown here were all derived from VLBA observations obtained in February 1997,
%reduced as is described by \citealt{gpg01}).

The large-scale magnetic fields are almost universally accepted to play
an important role in the production \citep[\eg][]{bla77}, acceleration
\citep[\eg][]{bp82,cl94,cam95} and, especially, collimation of the jets
\citep[\eg][]{hn89,sauty02,hn03}; see also the review by \cite{begelman84}. 
 Yet their importance for jet propagation and emission 
generation 
is underestimated.  
Most commonly, polarization structure is attributed to transverse or oblique
relativistic shocks propagating in the jet and compressing the upstream
turbulent magnetic field, so that the compressed downstream magnetic
field becomes anisotropically distributed
\citep[\eg][]{laing80,haa89,haa89b}.
This mechanism produces fields compressed in the plane of the shock, and is
expected to produce polarization along the jet for the case of a transverse
shock.  Numerous observations of polarization orthogonal to the jet are
attributed to a sheared component of magnetic field in a sheath surrounding
the jet.  This overall scenario is open to criticism. First, since shocks are
intrinsically transient events, it is difficult to see how the jet could retain
its polarization orientation over extensive lengths, even in the
presence of appreciable bending; it is also hard to see how weaker bridges
with the same characteristic polarization connecting bright knots could be
produced: this would require a quasi-steady sequence of weak shocks.
Secondly, since internal shocks can be oblique, with a range of orientations,
it is not natural to have a bimodal distribution of the relative EVPAs
(the EVPAs in BL~Lac seem to be in disagreement with the shock model
\cite{dmm00}).
Thirdly, some features showing longitudinal polarization are
far from being compact (\eg, the resolved VLBI jets of 1219+285 \cite{gab94}
and 1803+784 \cite{gab03b}).

An alternative interpretation of the jet polarization, which we favor, is
that the flow carries large-scale helical magnetic fields.
In this paper, we
calculate the  polarization  properties of a relativistic jet
carrying helical magnetic fields and compare them with observations.
The results are encouraging. We can both reproduce the average  properties
of the jet polarization, such as the bimodal distribution of the observed
EVPAs, as well as allow for possible exceptions, such as orthogonal jumps
in the EVPA along the jet.

There have been two approaches to studies of the synchrotron polarization
of AGN jets. On the one hand, \cite{laing80,laing81} developed detailed
models for optically thin synchrotron emission from 
{\it non-relativistic jets}. Recently, such non-relativistic models were 
generalized to include plasma effects on the propagation of the 
radiation inside the jet, Faraday rotation and the Cotton--Mouton effect
\citep{zheleznyakov02, beckert02, beckert03, ruszkowski03}. In these 
works, although plasma is relativistic, the radiation and polarization
transport are considered in a static medium. In actual jets, the 
plasma is in a state of bulk motion with relativistic speeds.
On the other hand, the
polarization properties of the synchrotron emission from relativistic AGN jets
have been discussed by \citet{blandford79,bjornsson82,kc85}, but
not in the context of large-scale spiral magnetic fields. \citet{pariev03}
has investigated the polarization properties of a specific configuration 
of a spiral magnetic field with a uniform axial field, zero total 
poloidal current in the jet, and with sheared relativistic spiral
motion of the plasma perpendicular to the magnetic field lines. 
In this paper, we  present the results of
fully relativistic calculations for various jet 
internal structures and the resulting polarization structures.  We take
a step back and try to answer the question of whether we can reproduce the
most salient polarization features of AGN jets under generic assumptions
about the internal helical magnetic field in the jet (basically generalizing
the work of \cite{laing81} to relativistic jets).  In order to answer this
question, one must know the internal structure of the jet (distribution
of poloidal and toroidal magnetic fields) and the distribution of synchrotron
emissivities. These are highly uncertain in terms of theoretical principles,
and are also subject to averaging over the whole jet (for unresolved jets)
and along the line of sight (for resolved jets).

Conventionally (and erroneously for a relativistically moving plasma!), the
direction of the observed polarization for optically thin regions and the
associated magnetic fields are assumed to be in one-to-one correspondence,
being orthogonal to each other, so that some observers chose to plot the
direction of the electric vector of the wave, while others plot vectors
orthogonal to the electric vectors and call them the
direction of the magnetic field. {\it This  is  correct only for
non-relativistically moving sources, and thus cannot be applied to AGN jets.}
First, relativistic boosting changes the relative strength of the magnetic 
field components along and orthogonal to the line of sight, which transform
differently under the Lorentz boost. Thus, if we could see the magnetic fields
directly, the strength of the jet poloidal and toroidal fields measured
in the laboratory frame would be different from those measured in the jet
frame.  Second, since the emission is boosted by the relativistic motion
of the jet material, the polarization EVPA rotates parallel to the plane 
containing the velocity vector of the emitting volume ${\bf v}$ and the unit
vector in the direction to the observer ${\bf n}$, so that 
{\it the observed electric field of the wave is not, in
general, orthogonal to the observed magnetic field} 
\citep{blandford79,lyu03}.
Thus, to reconstruct the internal structure of relativistic jets from
polarization observations, one needs to know both the polarization
{\it and } the velocity field of the jet.  Overall, plotting the direction
of the electric field (EVPA) should be considered the only acceptable way
to represent polarization data for sources where relativistic motion may
be involved.

Because of the finite resolution of real polarimetric observations,
the observed intensities of the polarized and unpolarized components
are averages over the effective beam size. When the resolution is 
relatively low, the transverse distribution of the intensity is unresolved.
In this case, only the intensity and polarization averaged over the
jet size is observed. If the jet is circular in cross section
and the distributions of the magnetic fields, emitting particles, and
velocity fields are axisymmetric across the jet, then the whole
jet can be decomposed into a collection of infinitesimally thin
shells. These shells have constant components of the spiral
magnetic field and constant emissivity along their perimeter, and
are moving uniformly with a constant velocity. The
summation of the polarization from the front and back crossing
points of a line of sight with such a shell results in the observed EVPA
of each shell either being parallel or perpendicular 
to the jet axis. Consequently,
the integrated polarization from a cylindrical axisymmetric jet
is either parallel or perpendicular to the jet axis. This is true
regardless of relativistic effects and is also true in the presence
of rotation of the jet \citep{pariev03}. Real jets are not cylindrical,
and diverge with distance from the core, and they may have an elliptical
or otherwise nonaxisymmetric cross section. We qualitatively expect
that these effects will lead to deviations of the EVPA from a strictly bimodal
distribution, and even to the appearance of $\pi/4$ EVPAs with respect
to the jet axis.
 
Polarization along or orthogonal to the jet is often interpreted in terms
of toroidally or poloidally dominated jets. Since {\it the ratio of the
toroidal to poloidal fields  depends on the reference frame}, one
should be careful in defining what is meant by, for example, a toroidally
dominated jet. A strongly 
relativistic  jet with comparable toroidal and poloidal fields
in its rest frame (defined, for example, as a frame in which the particle
distribution is isotropic) will be strongly toroidally dominated
in the observer frame.

The choice of reference  frame in which the structure of the jet is considered
 depends on the questions to be asked.  For example,
the intrinsic stability of the jet (\eg  with respect  to kink modes)
depends mostly on the properties of the jet in its rest frame.
It is expected that strongly toroidally dominated jets (in their rest frame)
should become unstable.  Though there are stable laboratory field
configurations with dominant toroidal fields (reverse field pinches),
they require a careful arrangement of currents which is not likely to
occur under astrophysical conditions. In addition, 
relativistic force-free,
differentially rotating pinches \footnote{Here, by a 
relativistic force-free pinch, we mean that the electric
fields are of the order of the magnetic fields, and the charge densities
are of the order of $j/n ec$, see below.} are likely to be more
stable than the non-relativistic analogue due to the stabilizing effects
of differential rotation \citep{istpar94,istpar96}. 
These two arguments show that the
field configuration in the rest frame may be somewhat dominated by the
toroidal field, but is unlikely to be strongly toroidally dominated.
The observational appearance of jets, such as their intensity and linear
polarization, also depends mostly on the internal structure of the jet and
the emissivity distribution, but a proper transformation to the observer
frame is needed.  On the other hand, interactions of a jet with the
surrounding medium and the corresponding instabilities (\eg, 
Kelvin--Helmholtz)
depend mostly on the jet properties in the observer frame.

Our polarization calculations indicate that, in the jet rest frame,
the toroidal and poloidal fields are generally comparable, so that {\it
in the  observer frame, the relativistic jets are dominated by
the toroidal magnetic field}: the more relativistic the jet, the
larger the observed toroidal magnetic field.  
Note that this conclusion  is also consistent
with the dynamical evolution of the large-scale poloidal and toroidal
fields: it is expected that the ratio of the poloidal to the toroidal 
magnetic field 
in the observer frame decreases linearly with the
cylindrical radius of the jet as it expands from the central disk 
(sizes of the order of hundreds of A.U.)  
to a thickness of $\sim 0.1$ pc; (see Section~\ref{discussion}).

We also assume that the emission is optically thin and neglect possible
plasma propagation effects, such as Faraday and Cotton--Mouton effects.
This approximation is usually well justified \citep{pariev03}. 
Faraday rotation is absent in a pair plasma, as well as intrinsic 
depolarization inside the jet. VLBI polarization measurements
are done at frequencies in the range $1.6\,\mbox{GHz}$ to $43\,\mbox{GHz}$.  
Synchrotron self-absorption
is very small at GHz and higher frequencies. 
For $B\sim 10^{-2}\,\mbox{G}$, 
$n\sim 0.1\,\mbox{cm}^{-3}$, and $p\approx 2.5$, an approximate 
estimate for the synchrotron self-absorption coefficient 
is $\displaystyle \kappa_{s}=
(5\cdot 10^3\,\mbox{pc})^{-1}\left(\nu/1\,\mbox{GHz}\right)^{-2-p/2}$
(e.g., \citealt{zhelezniakov96}). The corresponding optical depth through the 
$1\,\mbox{pc}$ jet is only $2\cdot 10^{-4}$. 
The typical energy of particles emitting in this frequency range in a
$10^{-2}\,\mbox{G}$ magnetic field is $100\,\mbox{MeV}$ to $1\,\mbox{GeV}$.
The conversion of linear polarized into circular polarized radio waves
(the Cotton--Mouton effect) is also small. For both
thermal and relativistic power-law particle distributions with 
$p\approx 2.5$ and a minimum Lorentz factor $\sim 10$, the estimate for 
the conversion coefficient is $\displaystyle\kappa_{c}=(500\,\mbox{pc})^{-1}
\left(\nu/1\,\mbox{GHz}\right)^{-3}$ \citep{sazonov69}. This is too small to 
influence the transfer of linear polarization inside the jet, but could be 
the source for a small circular polarization, at the level of a fraction 
of a per cent. We focus on linear polarization in this work.

As a first step, we consider the synchrotron  emission of  an unresolved,
thin, circular  cylindrical shell populated by relativistic electrons
with a power-law distribution and moving uniformly with constant velocity
$\beta=v/c$.  The properties of the synchrotron emission
are then determined by {\it three parameters}:
the internal pitch angle of the magnetic field $\psi'$,
the Lorenz factor of the shell in the laboratory frame  $\Gamma$
and the viewing angle, $\theta$, that the line of sight to the observer makes
with the jet axis in the observer's reference frame.
Note that, in the case of unresolved jets, this is equivalent to
calculating the polarization from a jet with a pitch angle 
properly averaged by the emissivity.

By performing the correct Lorentz transformations of the jet and radiation
electromagnetic fields, we construct polarization maps for various
observation angles $\theta$ and
jet parameters. We find that, in order to
produce polarization along the jet axis, the pitch angle of the helical
magnetic field {\it in the jet frame} must be of the order of unity or greater,
$B'_\phi/B_z' \sim 1$.  By virtue of the Lorentz transformation, this
implies that the relativistic jets are observed to be dominated by the
toroidal magnetic field $B_\phi/B_z \geq \Gamma$.  In addition, we find that
a change of the EVPA in the jet can occur only for a narrow range of
rest-frame pitch angles, $\pi/4 \leq \psi' \leq \pi/3$,  which allows one to
measure (and not only derive a lower limit for) the ratio of the toroidal
and poloidal magnetic fields $B_\phi /B_z  \sim  \Gamma \gg 1$.

Further, we consider the emission from 
''filled'' jets, i.e., when most of the
jet volume contributes to the emissivity.  To  find the emission properties
of such  jets, one also needs to integrate the jet emissivity over the
jet volume. This requires knowledge of the internal structure of the jet and
the distribution of  synchrotron emissivity throughout the jet, which
introduces large uncertainties into the model.  To test the sensitivity
of the results to a particular  jet model, we  investigate  three  possible
force-free models for the jet structure (diffuse pinch,  pinch with zero
total poloidal magnetic flux and high-order reverse field pinch) and two
prescriptions for the number density of relativistic particles 
(proportional to the second
power of the current density and to the magnetic energy density).  We
conclude that the emission weighted average pitch angle in the jet
must not be too  small if we wish to produce longitudinal polarization.  This
implies that a quasi-homogeneous emissivity distribution 
 {\it cannot reproduce the
variety of position angle behavior observed} -- the average polarization
remains primarily orthogonal to the jet, dominated by the central parts
of the jet where 
the magnetic field in the comoving frame 
is primarily poloidal (since the toroidal
field must vanish on the axis).  Since this is not generally the case,
the synchrotron emission in such jets must be generated in a  narrow
range of radii, where the toroidal and poloidal magnetic fields are of
the same order of magnitude in the jet frame. Though \cite{laing81} argued
against this possibility on the grounds that it should produce strongly
''edge-brightened'' profiles in resolved jets, subsequent observations
of the resolved jets of Cen~A \citep{cbf86} and  M87  \citep{boh83} did,
in fact, show such an intensity distribution.

\section{Synchrotron emission from steady
relativistic flows}
\label{sec_2}

\begin{figure}[ht]
\includegraphics[width=0.9\linewidth]{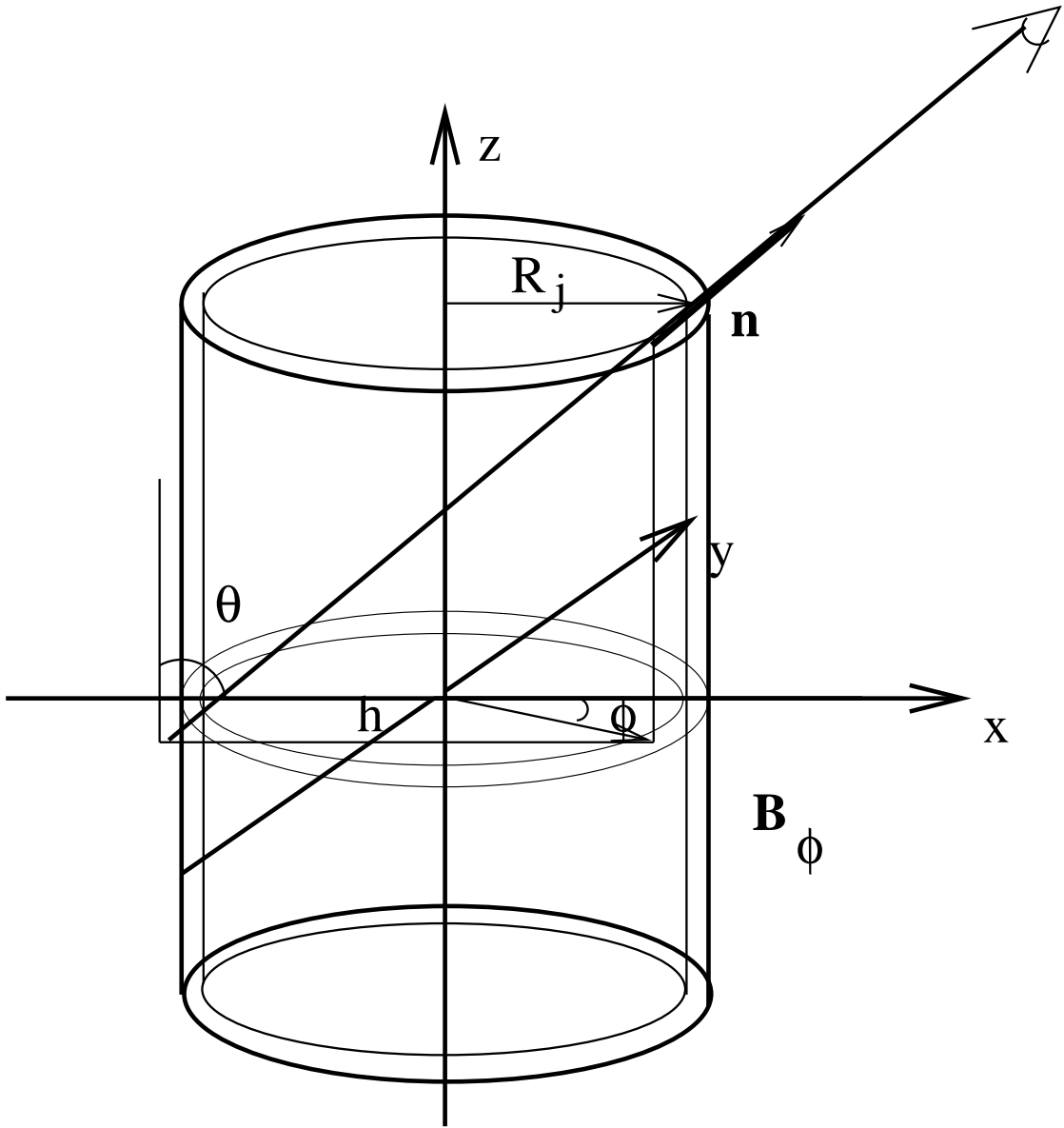}
\caption{Geometry of the model.}
\label{polariz-geomAGN}
\end{figure}

Consider a cylindrical jet  (Fig.~\ref{polariz-geomAGN})
viewed by an observer.
Below we denote all quantities measured in the local frame comoving with
an elementary emitting volume with primes,
while unprimed quantities are those measured in the
observer's  frame.
Let $r$, $\phi$, and $z$ be cylindrical coordinates  centered
on the jet axis and $x$, $y$, and $z$ be the  corresponding
rectangular coordinates. The magnetic field is in the $\phi-z$ direction.
The components of all vectors written below are components with
respect to the rectangular coordinates $x$, $y$, and $z$.
The velocity $\beta=v/c$ of the bulk motion is directed along the 
$z$ direction (the corresponding Lorentz factor is 
$\Gamma= 1/\sqrt{1-\beta^2}$). We do not consider bulk rotation 
of the jet in this work. An observer is located 
in the $x-z$ plane, so that
the unit vector along the direction of the emitted photons can be written
${\bf n} = \{\sin \theta, 0, \cos \theta \}$.
We assume that the distribution function for the emitting particles
in the frame comoving with an element of the jet is isotropic
in momentum and is a power law in energy:
\be
dn=K_{\rm e} \epsilon^{-p}d\epsilon dVd\Omega_{\bf p}\mbox{.}
\ee
Here, $dn$ is the number of particles in the energy interval
$\epsilon,\epsilon+d\epsilon$, $dV$ is the elementary volume,
$d\Omega_{\bf p}$ is the elementary solid angle in the direction of the
particle momentum ${\bf p}$, $K_{\rm e}=K_{\rm e}(r)$ and $p={\rm constant}$.

Since the emitting particles are ultra-relativistic and we neglect the
generation  of circular polarization, we do not include circular
polarization in our model (Stokes $V=0$).  We also neglect a possible
tangled component of the magnetic field in the emission region.  Under
these assumptions, our estimates provide an upper limit on the possible
polarization.

The Stokes parameters per unit jet length 
for a stationary flow are then given by (see
also \cite{lyu03,pariev03})
\ba &&
{ I}=\frac{p+7/3}{p+1}\frac{\kappa(\nu)}{D^2 (1+z)^{2+(p-1)/2}}
 \int { dS \over \sin \theta} \, K_{\rm e}\,
\,{\cal D}^{2+(p-1)/2} |B'\sin \chi'|^{(p+1)/2} \mbox{,}
\nn &&
{ Q}=\frac{\kappa(\nu)}{D^2 (1+z)^{2+(p-1)/2}}
 \int  { dS \over \sin \theta} \, K_{\rm e}  \,
\,{\cal D}^{2+(p-1)/2} |B'\sin \chi'|^{(p+1)/2}
\cos 2{\tilde\chi} \mbox{,}
\nn &&
 {U}=\frac{\kappa(\nu)}{D^2 (1+z)^{2+(p-1)/2}}
 \, \int  { dS \over \sin \theta} \, K_{\rm e} \,
\,{\cal D}^{2+(p-1)/2} |B'\sin \chi'|^{(p+1)/2}
\sin 2{\tilde\chi} \mbox{,}
\nn &&
{V}=0\mbox{.}
\label{Stokes}
\ea
The function $\kappa(\nu)$ is
\be
\kappa(\nu) =
\frac{\sqrt3}4\Gamma_E\left(\frac{3p-1}{12}\right)
 \Gamma_E\left(\frac{3p+7}{12}\right)\frac{e^3}{m_{\rm e} c^2}\left[
   \frac{3e}{2\pi m_{\rm e}^3c^5}\right]^{(p-1)/2}\nu^{-(p-1)/2}
\mbox{,}
\ee
where $e$ and $m_{\rm e}$ are the charge and mass of an electron,
$c$ is the speed of light,
$\Gamma_E$ is the Euler gamma-function,  $z$ is the cosmological 
redshift and $D$  is the luminosity distance to the jet.  The integration
in (\ref{Stokes}) is over $dS= r dr d \phi $, $0<\phi < 2 \pi$, $0<r <R_j$
for unresolved jets ($R_j$ is the jet radius), while, for resolved jets
$dS =  h dh d\phi / \sin^2 \phi$, $\arcsin h/R <\phi 
< \pi\mbox{sgn}\, h - \arcsin h/R$,
where $ h$ is the projected distance of the line of sight from the
jet axis \citep{pariev03}.  In Eq.~(\ref{Stokes}),
${\cal D}$ is the Doppler boosting factor
\be
{\cal D} = {1 \over \Gamma (1- \beta \cos \theta ) },
\ee
$\chi'$ is the angle between the line of sight and the magnetic field in
the rest frame of a plasma element,
 and
$\tilde{\chi}$ is the observed EVPA in the plane of the sky seen by 
the observer, measured clockwise from some reference direction.

We next introduce a unit vector ${\bf l}$ normal to the plane containing
${\bf n}$ and the reference direction in the plane of the sky. 
We choose the direction of the projection of the jet axis to the 
plane of the sky as a reference direction, so that ${\bf l}=\{ 0,1,0 \}$. Then,
\be
\cos \tilde{\chi} = {\hat{\bf e}}  \cdot ({\bf n}\times {\bf l}), \quad
\sin \tilde{\chi} = {\hat{\bf e}}  \cdot {\bf l} \mbox{,}
\ee
where ${\hat{\bf e}}$ is a unit vector in the direction of the 
electric field of the wave.

Calculation of the Stokes parameters using Eqs.~(\ref{Stokes}) is not
as straightforward as in the case of a plasma at rest.  Several key
ingredients need to be taken into account in the case of relativistically
moving synchrotron sources (\citealt{cocke72}, \citealt{blandford79},
\citealt{bjornsson82}, \citealt{ginzburg89}).  First, the synchrotron
emissivity depends on the angle between the direction of an emitted photon
and the magnetic field in the plasma rest frame. Second, since  emission
is boosted by the bulk relativistic motion, the position angle
of the linear polarization rotates parallel to the ${\bf n} - {\bf v}$ plane.
The fractional polarization emitted by each element remains the same, but
the directions of the polarization vector of the radiation emitted by
different elements are rotated by different amounts.
This leads to the effective depolarization of the total emission. The
theoretical maximum fractional polarization for a homogeneous field
can be achieved only for a uniform plane-parallel velocity field.  Third,
integration along the line of sight and over the image of the jet in
the plane of the sky for unresolved jets is best carried out in the
laboratory frame (the frame where the source is at rest as a whole).

Let ${\bf n}'$ be a unit vector in the direction of the wave vector
in the plasma rest frame, and ${\hat {\bf B}}'$ be a unit vector along
the magnetic field in the plasma rest frame.  The electric field of a
linearly polarized electromagnetic wave ${\bf e}'$ is directed along the unit
vector ${\hat{\bf e}}' ={\bf n}' \times {\hat{\bf B}}'$  and the magnetic
field of the wave ${\bf b}'$ is directed along the unit vector
${\hat{\bf b}}' = {\bf n}' \times  {\hat{\bf e}}'$.
Making the  Lorentz boost to the observer frame, we find \citep{lyu03}
\ba &&
{\bf n}' =
\frac{ {\bf n} + \Gamma {\bf v} \left( {\Gamma\over \Gamma+1} ({\bf n} \cdot {\bf v})
-1\right)}{ \Gamma \left(1- ({\bf n} \cdot {\bf v})\right)}\mbox{,}
\nn &&
{\hat {\bf e}} ={  {\bf n} \times {\bf q}' \over
\sqrt{ q^{\prime 2} - ( {\bf n} \cdot {\bf q}')^2} } \mbox{,}
\nn &&
{\bf q}' = {\hat{\bf B}}' +
 {\bf n}  \times (  {\bf v} \times  {\hat{\bf B}}')
-{ \Gamma \over 1+\Gamma} ( {\hat{\bf B}}' \cdot {\bf v} ) {\bf v} \mbox{,}
\label{ee0}
\ea
where here and in all further expressions we set $c=1$.  
We can express the rest-frame  unit vector ${\hat{\bf B}}'$ in terms of
the unit vector ${\hat{\bf B}}$ along the magnetic field in the 
laboratory frame. Assuming ideal MHD,
there is no electric field in the plasma rest frame, ${\bf E}'=0$. We then
obtain
\ba &&
{\hat{\bf B}} =
{1 \over \sqrt{ 1 - ({\hat{\bf B}}' \cdot {\bf v})^2} }
\left( {\hat{\bf B}}' - { \Gamma \over
 1+  \Gamma} ({\hat{\bf B}}' \cdot {\bf v}) {\bf v} \right)  \mbox{,}
\nn &&
{\hat{\bf B}}' = { (1+\Gamma) {\hat{\bf B}} +
\Gamma^2 ({\hat{\bf B}} \cdot {\bf v}) {\bf v} \over
(1+\Gamma) \sqrt{ 1+  \Gamma^2 ({\hat{\bf B}} \cdot  {\bf v})^2 }}  \mbox{,}
\ea
 to get
\ba &&
{\hat {\bf e}} ={  {\bf n} \times {\bf q} \over
\sqrt{ q^2 - ( {\bf n} \cdot {\bf q})^2} } \mbox{,}
\nn &&
{\bf q} = {\hat{\bf B}} +
 {\bf n}  \times (  {\bf v} \times  {\hat{\bf B}})  \mbox{.}
\label{eee}
\ea
This is a general expression giving the polarization vector in terms
of the observed quantities ${\hat{\bf B}}$, ${\bf n}$ and ${\bf v}$.

Relativistic aberration makes the observed electric field 
in the wave non-orthogonal
to the observed B field and to its projection on the sky. Due to the
conservation of ${\bf  e}'\cdot( {\bf b} ' + {\bf B}') $, we find that (see
also \cite{blandford79})
\be
{\bf e} \cdot {\bf B} =({\bf v}\times{\bf B})\cdot({\bf n}\times{\bf e})
\mbox{.}
\ee
For example, the angle between ${\bf \hat{e}}$ (which lies in the plane
of the sky) and ${\bf \hat{B}}$ is
\be
\cos \zeta=
{\bf \hat{e}} \cdot {\bf \hat{B}} =
{ ({\bf \hat{B}} \cdot {\bf n}) ( {\bf \hat{B}} 
\cdot ( {\bf n}  \times {\bf v})) \over
\sqrt{ q^2 - ( {\bf n} \cdot {\bf q})^2} }
\label{zetaexpr}
\ee
which is, in general, not equal to zero.  The observed electric field is
orthogonal to the observed B field if either ${\bf \hat{B}}$ lies in the
${\bf n}  - {\bf v}$ plane or $ ({\bf \hat{B}} \cdot {\bf n})  =0$.

Next, we apply the above general relations to the case of synchrotron
emission by a thin cylindrical shell carrying toroidal magnetic field
and moving uniformly with velocity $\beta$ parallel to its axis
in the $z$ direction. The shell is viewed at an angle $\theta$ with
respect to its axis, so that
\ba &&
{\bf v} = \beta \{0,0,1 \}
\nn  &&
{\bf n}= \{\sin \theta, 0, \cos \theta \}
\ea

Let the magnetic field $\B'$ in the emitting shell in the jet frame be helical:
\ba &&
\B' = B_\phi' \{ -\sin \phi, \cos\phi, 0\}
+ B_z' \{0,0,1\} = B' \sin \psi'  \{ -\sin \phi, \cos\phi, 0\} +
B'  \cos  \psi'  \{0,0,1\}
\nn &&
\B' = B' {\hat{\bf B}}', \,\,\,
{\hat{\bf B}}' = \sin \psi'  \{ -\sin \phi, \cos\phi, 0\} 
+ \cos  \psi'  \{0,0,1\} \label{helical_b}
\ea
where $\psi'$ is the magnetic field pitch angle in the shell rest frame,
${\hat{\bf B}}'$ is a unit vector along the magnetic 
field in the shell frame, and
$B'$ is the magnitude of the field.
At this point, we are interested in the emission of a single shell and
do not specify, for the moment, the radial dependence of the 
magnetic fields and emissivities.
This is considered in Section~\ref{filled}.

For this particular choice of geometry, we find
\ba &&
{\bf n}'= {\cal D}  \{ \sin \theta, 0, 
\Gamma ( \cos \theta - \beta )
 \}\mbox{,}\label{eq_nprime} \\
  &&
{\hat{\bf B}} = { 1\over \sqrt{ 1- \beta^2 \cos^2 \psi'}}
\{ -\sin \psi' \sin \phi,  \cos\phi \sin \psi', 
{ \cos  \psi'  \over \Gamma} \} \mbox{,}\label{eq_bhat} \\
&&
\cos \chi' = ({\bf n}' \cdot {\hat{\bf B}'})=
 {\cal D} \left( \Gamma \cos\psi' 
( \cos \theta - \beta) - \sin \theta
\sin \phi \sin \psi' \right) \mbox{,}\label{eq_coschipr}
\ea
where ${\hat{\bf B}}$ is the unit vector along the magnetic 
field in the laboratory
(observer) frame. Note that
\ba &&
{\bf \hat{B}}  \cdot  {\bf n}= 
{ \cos \theta \cos \psi' - \Gamma \sin  \theta \sin \phi 
\sin \psi' \over
\Gamma \sqrt{1- \beta^2 \cos^2 \psi'} }, \nonumber \\
&& {\bf \hat{B}}  \cdot  {\bf v} = { \beta \cos \psi' \over 
\Gamma \sqrt{1- \beta^2 \cos^2 \psi'} }, \\
&& {\bf \hat{B}}  \cdot  {\bf n} \times  {\bf v} =
-{ \beta  \sin  \theta \sin \psi' \cos\phi \over 
\sqrt{1- \beta^2 \cos^2 \psi'} } \nonumber
\ea

The  pitch angle in the laboratory frame, $\psi'$ 
is larger than in the
rest frame:
\be
\tan \psi = \Gamma \tan \psi'\mbox{.}\label{eq_psipr}
\ee

\begin{figure}[ht]
\includegraphics[width=0.9\linewidth]{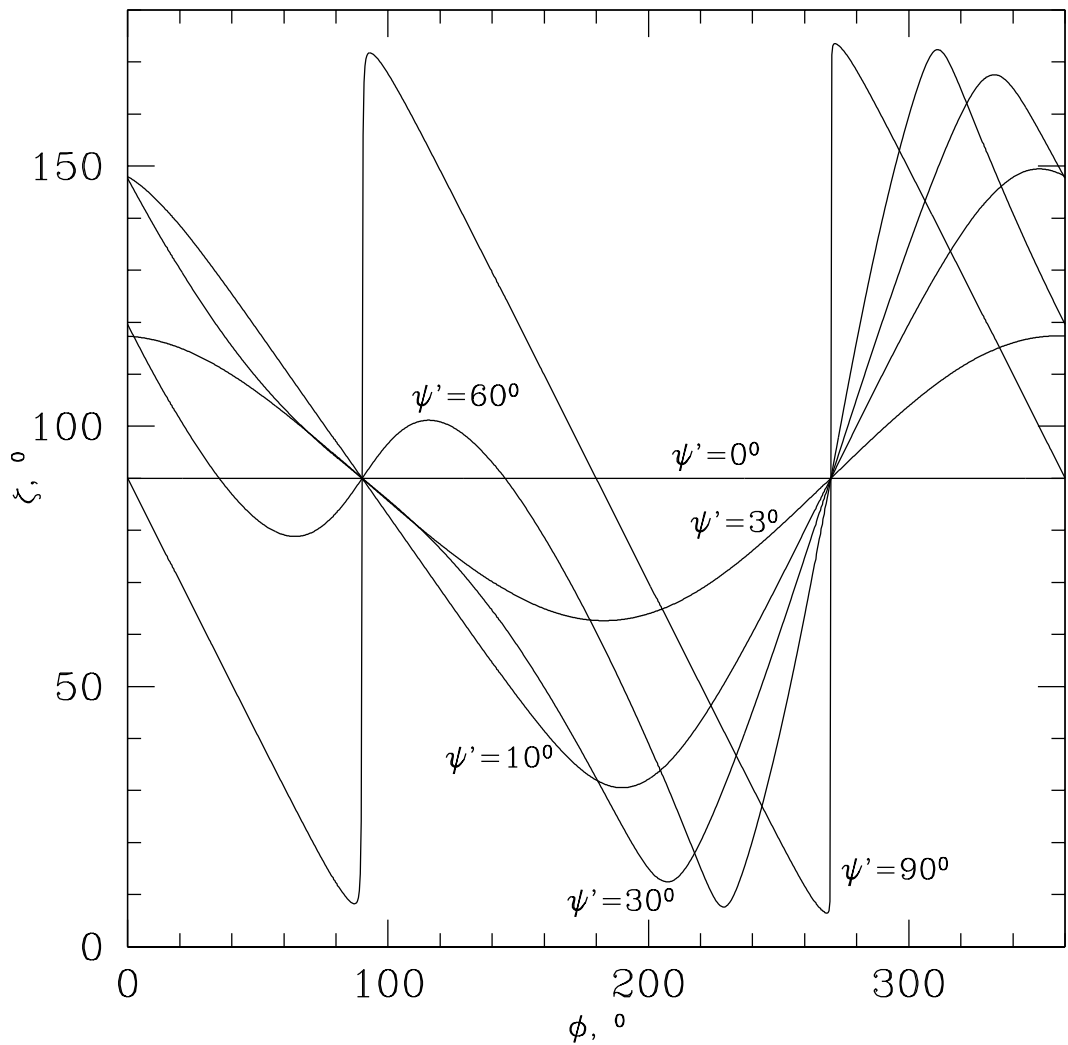}
\caption{Angle $\zeta$ between the observed direction
of the magnetic field ${\hat{\bf B}}$
and the polarization vector in the wave as a function of $\phi$ for
a cylindrical shell with
$\Gamma =10$, $\theta =0.1$ and six values of $\psi'$. In accordance
with formula~(\protect\ref{zetaexpr}) $\zeta=90^0$ when either
the magnetic field is pure axial ($\psi'=0$) or at the visible edges
of the jet ($\phi=90^0$ and $\phi=270^0$).
}
\label{zeta0}
\end{figure}

\begin{figure}[ht]
\plotone{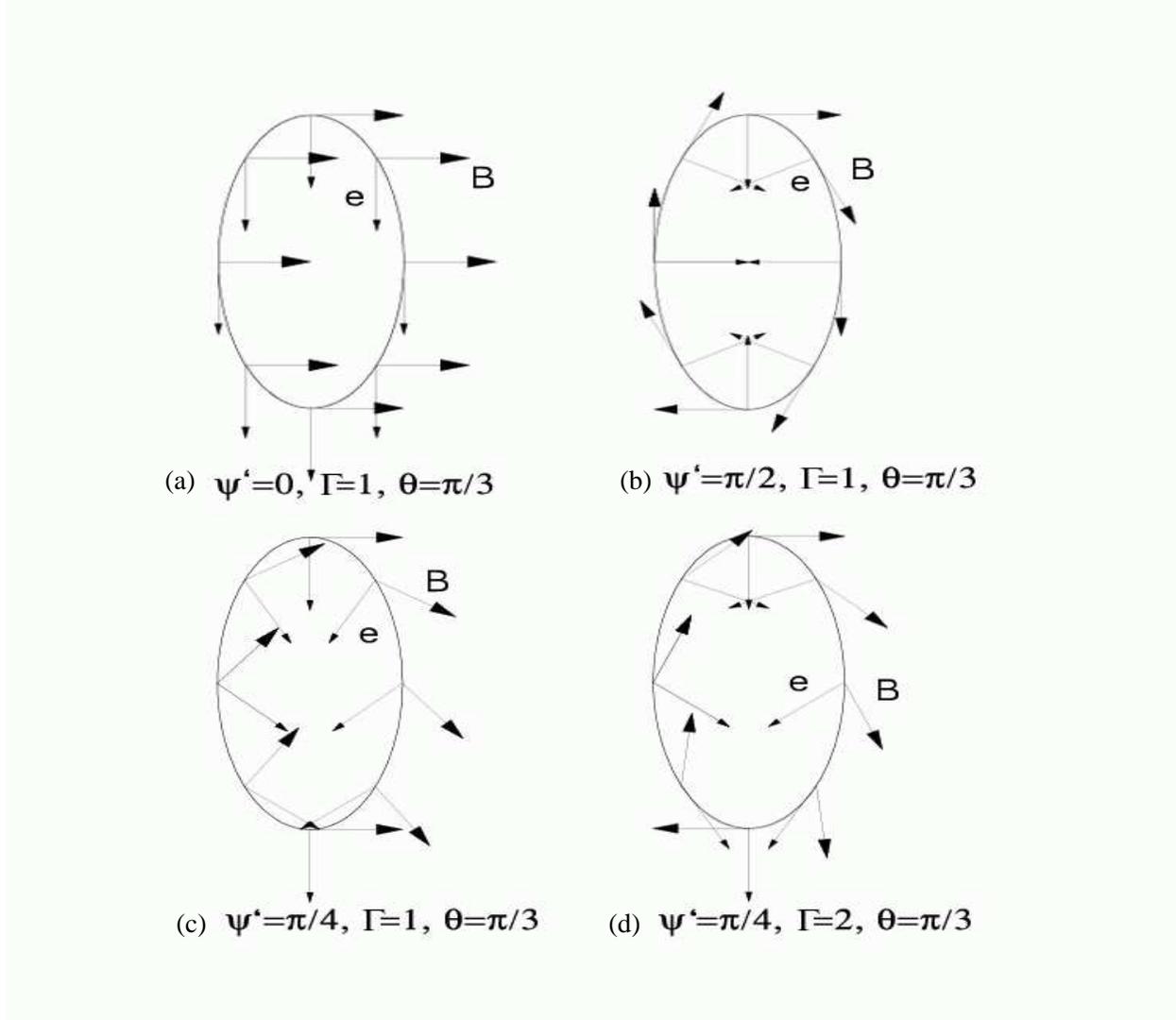}
\caption{Examples of observed  electric field in the wave and  
magnetic field for $\theta =\pi/3$.
Projections of the
magnetic field on the plane of the sky are denoted by arrows with large head,
electric fields are denoted by arrows with small heads.  (a) Purely
poloidal magnetic field $\psi' =0$, 
(b) toroidal  magnetic field   $\psi' = \pi/2$,
(c) helical  field with  $\psi' = \pi/4$ viewed in the rest frame,
(d)  helical  field with  $\psi' = \pi/4$  moving with $\Gamma=2$.
For stationary jets ((a)-(c))
electric vector of the wave is always orthogonal to the magnetic field 
in the jet in the observer frame, 
for moving jets (case (d))
this is not true anymore.
}
\label{rotation}
\end{figure}

As we have already pointed out, one of the main effects of the relativistic 
Lorentz transformation is that
the observed magnetic field and polarization vectors are not, in general,
orthogonal \citep{blandford79}.  For cylindrical jets, the expression for
the angle $\zeta$ is too lengthy to be reproduced here (see also 
Appendix~\ref{BK}). The dependence of $\zeta$ on $\phi$ 
is plotted in Fig.~\ref{zeta0}.
In Fig.~\ref{rotation}, we plot the observed relative orientations of
the electric field in the wave and the magnetic field 
projected onto the plane of the sky.
Note that, for a stationary jet, the electric field is always perpendicular
to the magnetic field, while this is no longer true for a relativistically
moving jet. In fact, {\it  the electric field of the wave may become almost
parallel to the magnetic field for some viewing angles}.

The degree of linear polarization of the observed radiation is
$\Pi=\sqrt{{ Q}^2+{ U}^2}/{ I}$.  The resultant EVPA measured by the observer,
$\tilde{\chi}_{\rm res}$, is obtained from
\be
 \cos 2\tilde{\chi}_{\rm res}=\frac{{ Q}}
{\sqrt{{ Q}^2+{ U}^2}}\mbox{,} \,\,\,
 \sin 2\tilde{\chi}_{\rm res}=\frac{{ U}}
{\sqrt{{ Q}^2+{ U}^2}}\mbox{,}
\quad 0\leq \tilde{\chi}_{\rm res} < \pi\mbox{.}
\ee
It can be verified that, for a cylindrically symmetric flow and 
magnetic field, the change of $\phi$ to $\pi-\phi$
in the integrals~(\ref{Stokes}) does not change the value of ${ Q}$ 
and the sign of ${ U}$ is reversed. Therefore, the Stokes parameter
${ U}$ integrates to zero. Consequently, if ${ Q}>0$ then
$\tilde{\chi}_{\rm res}=0$, and if ${ Q}<0$ then
$\tilde{\chi}_{\rm res}=\pi/2$. Thus, {\it  for circular jets, the observed
EVPA can be either parallel or perpendicular to the projection of the
jet axis onto the plane of the sky}. This is true for differentially 
rotating jets as well \citep{pariev03}. This naturally explains the 
observed bimodal distribution of the jet EVPAs.

Since $U =0$ for a circular jet, we define the fractional polarization
as $ \Pi =  Q/I$,  retaining the sign of $ Q$, so that $\Pi$ can be smaller
than zero. For $\Pi >0$, the polarization is along the jet, while the
polarization is orthogonal to the jet for $\Pi < 0$.

Given that the angle between the observed EVPA and observed
magnetic field for individual radiating plasma elements are
not, in general, orthogonal, depending on the viewing conditions,
we can ask why it is the case that reasonably clear patterns
have appeared in the relationship between the observed EVPAs
and the local jet direction. This can be understood, in part,
as a consequence of the fact that the resolution provided by
VLBI observations is often insufficient to fully resolve the
jets in the transverse direction, so that the EVPA integrated
over the jet cross section is observed. This is essentially
equivalent to the statement above that the observed EVPA
{\em integrated over a cylindrical jet} is expected to
be either parallel or perpendicular to the projection of the
jet axis onto the plane of the sky. The observed EVPA will essentially 
be determined by the ratio $B'_\phi/B'_z$; if the jets are approximately 
cylindrical, the observed EVPA
can be used to draw conclusions about the ratio of the toroidal
and poloidal magnetic field components in the rest frame, but it is not
always possible to translate these into unambiguous conclusions
about this ratio in the observer's frame.

\section{Hollow cylindrical jets}
\label{sec_hollow}

\subsection{Unresolved hollow jets (cylindrical shell)}
\label{subsec_hollow_unres}

We first consider cylindrical 
jets with their emissivity confined to a
narrow cylindrical shell. In order to interpret observations that do not
resolve the transverse dependence of the polarization 
across the jet, we integrate 
the emissivity of the shell over the azimuthal angle in the observer's
frame.
For a constant velocity field of the shell, all relativistic effects 
can be accounted for by the Lorentz transformation: 
the degree of polarization can be calculated in 
the jet rest frame for the viewing angle in the jet rest frame 
$\theta'$ obtained from expression~(\ref{eq_nprime})
\be
\sin\theta' =\frac{\sin\theta}{\Gamma\left(1-
\beta\cos\theta \right)}, \quad 
\cos\theta' =\frac{\cos\theta -\beta}{1-\beta\cos\theta}
\label{eq_thetaobpr}\mbox{,}
\ee
where $0<\theta <\pi$ and $0<\theta' <\pi$.
In addition, the Lorentz transformation of the jet magnetic field
is $B_z=B'_z$, $B_{\phi}=\Gamma B'_{\phi}$ and results in the 
change of the pitch angle described by expression~(\ref{eq_psipr}). 
Though corresponding non-relativistic calculations
have been done by \citet{laing81}, the interpretation of these 
calculations for a relativistically moving shell is different.

\begin{figure}[ht]
\includegraphics[width=0.3\linewidth]{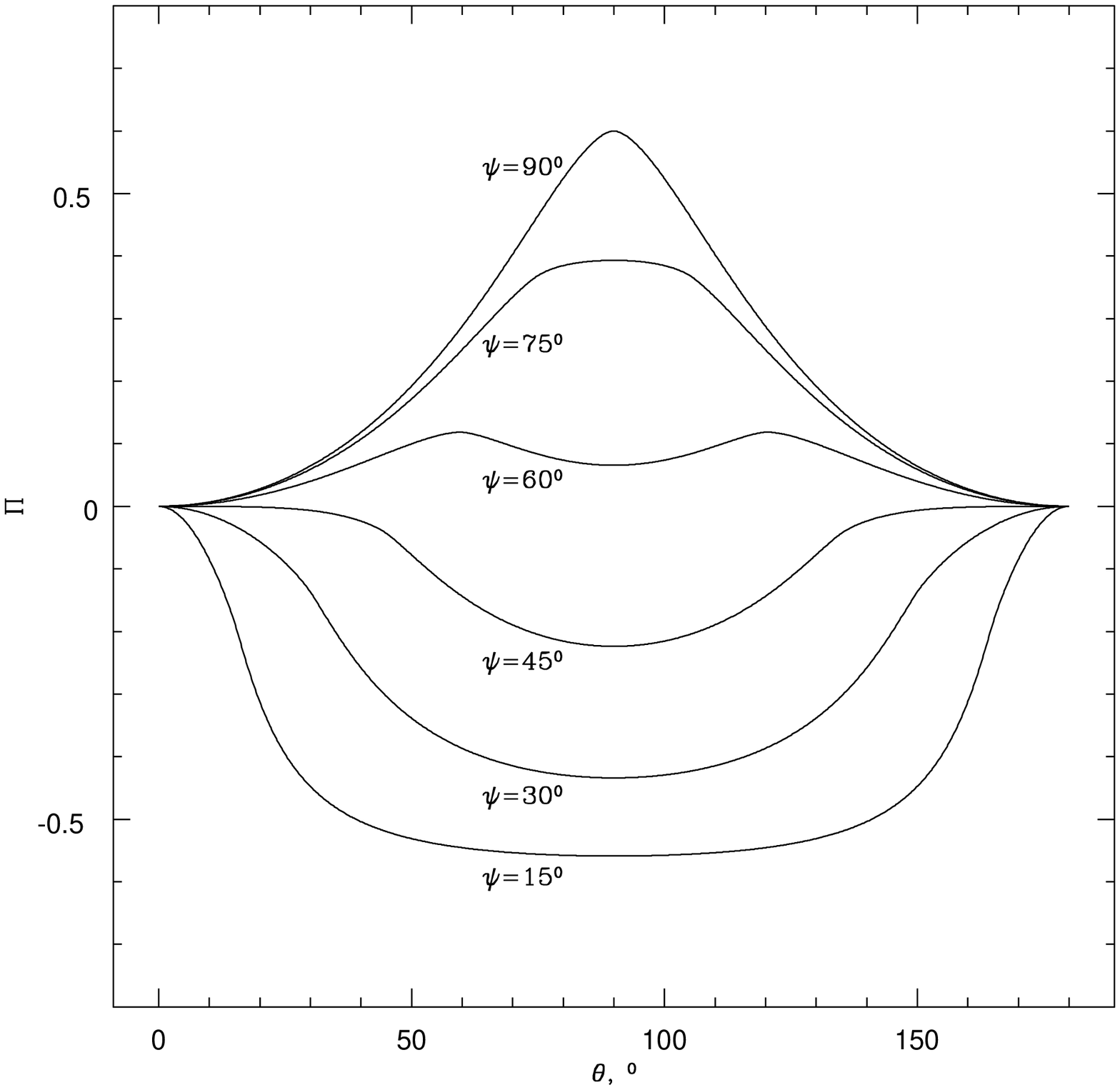}
\includegraphics[width=0.3\linewidth]{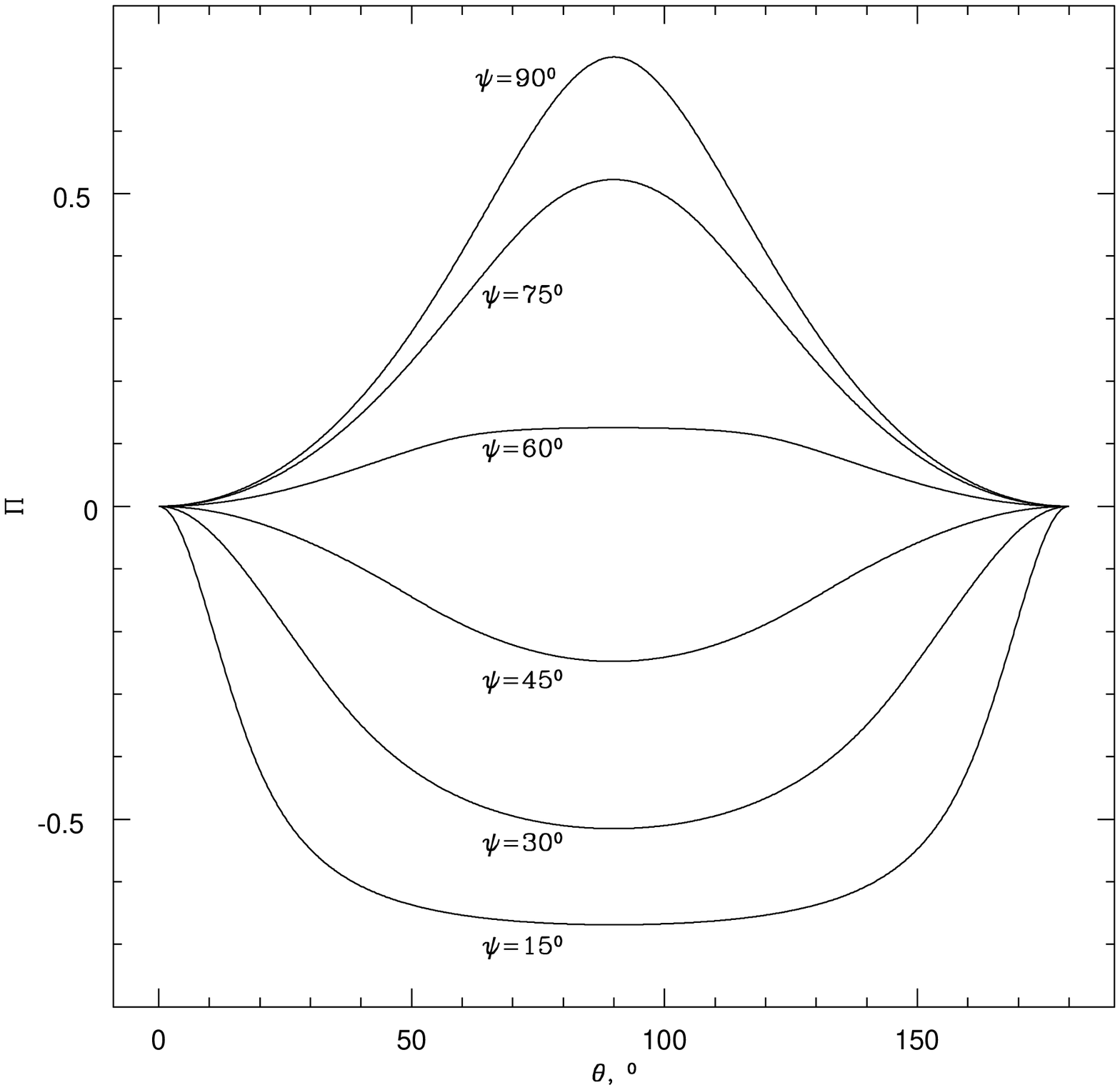}
\includegraphics[width=0.3\linewidth]{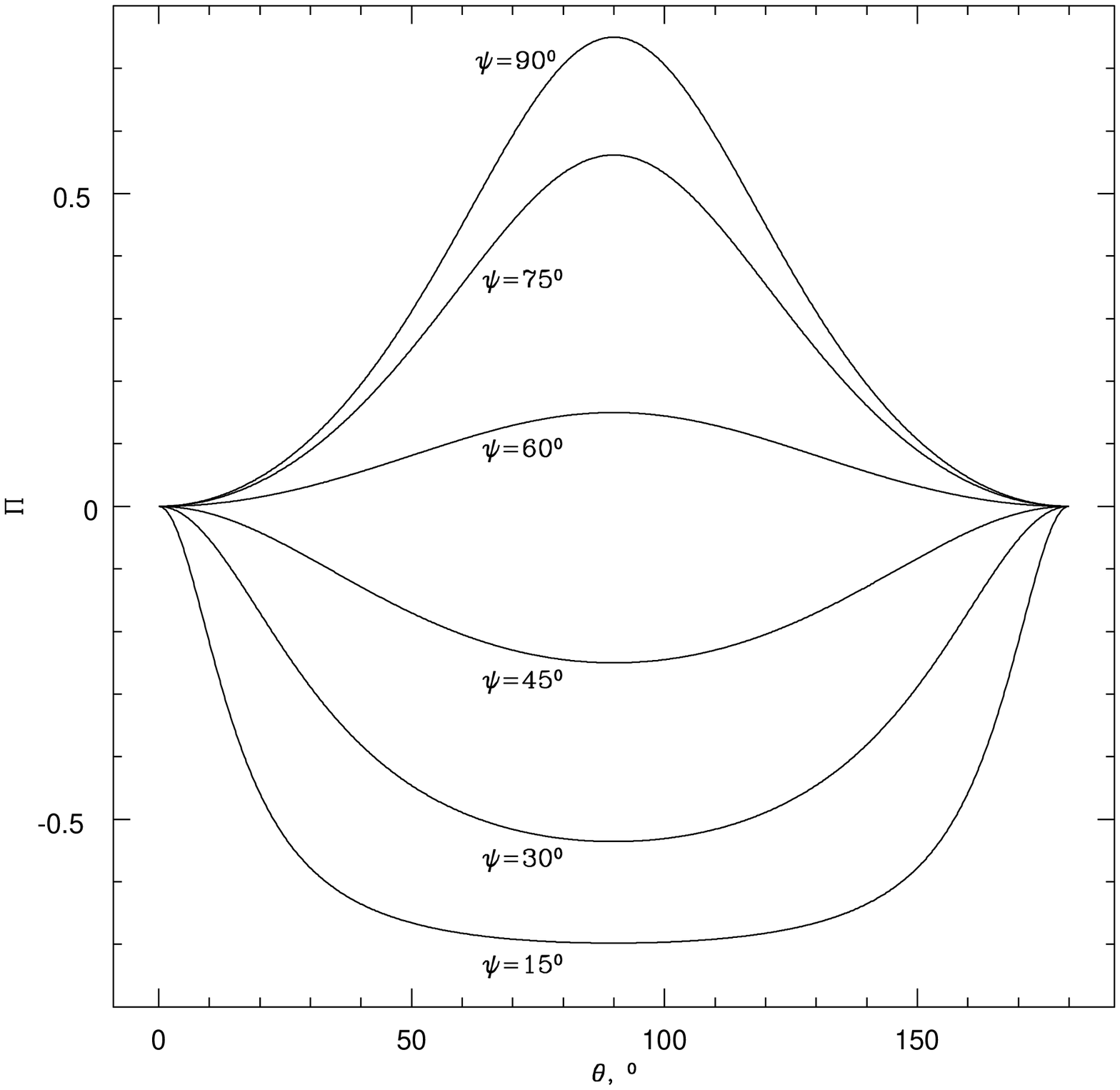}\\
\vskip .01 truein
\hskip 1 truein (a) \hskip 1.8 truein (b) \hskip 1.8 truein (c) \\
\vskip 0.2 truein
\includegraphics[width=0.3\linewidth]{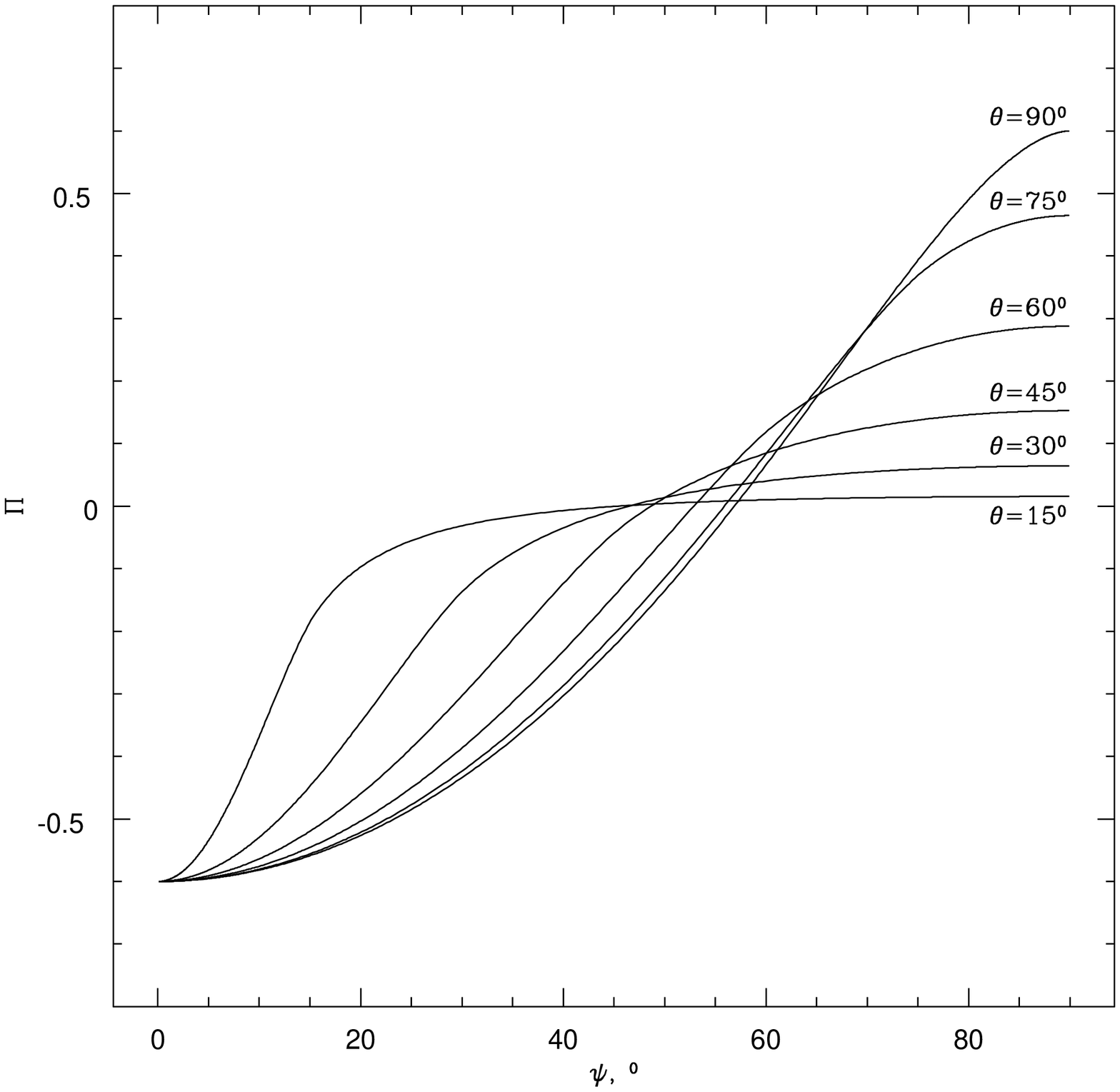}
\includegraphics[width=0.3\linewidth]{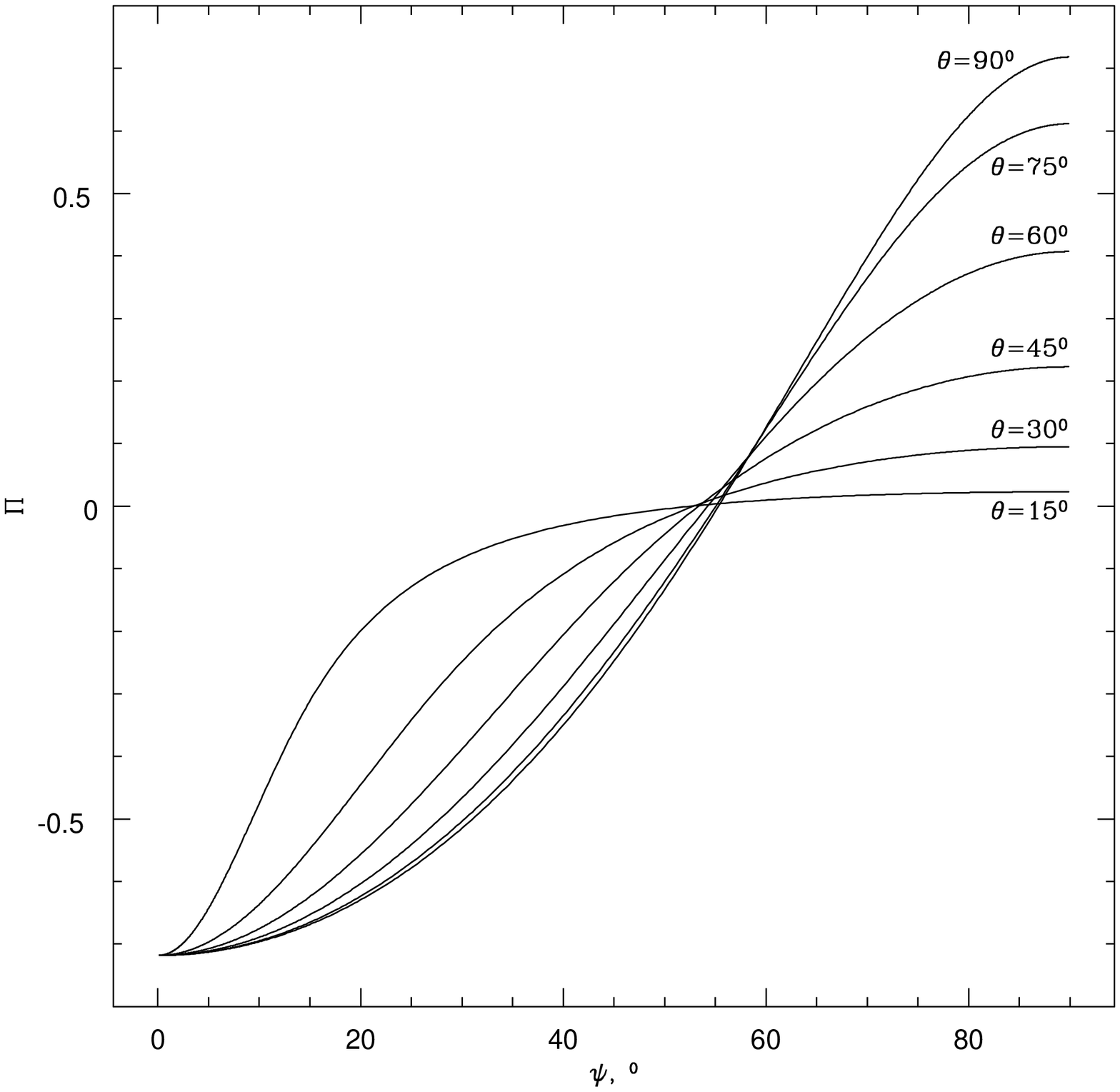}
\includegraphics[width=0.3\linewidth]{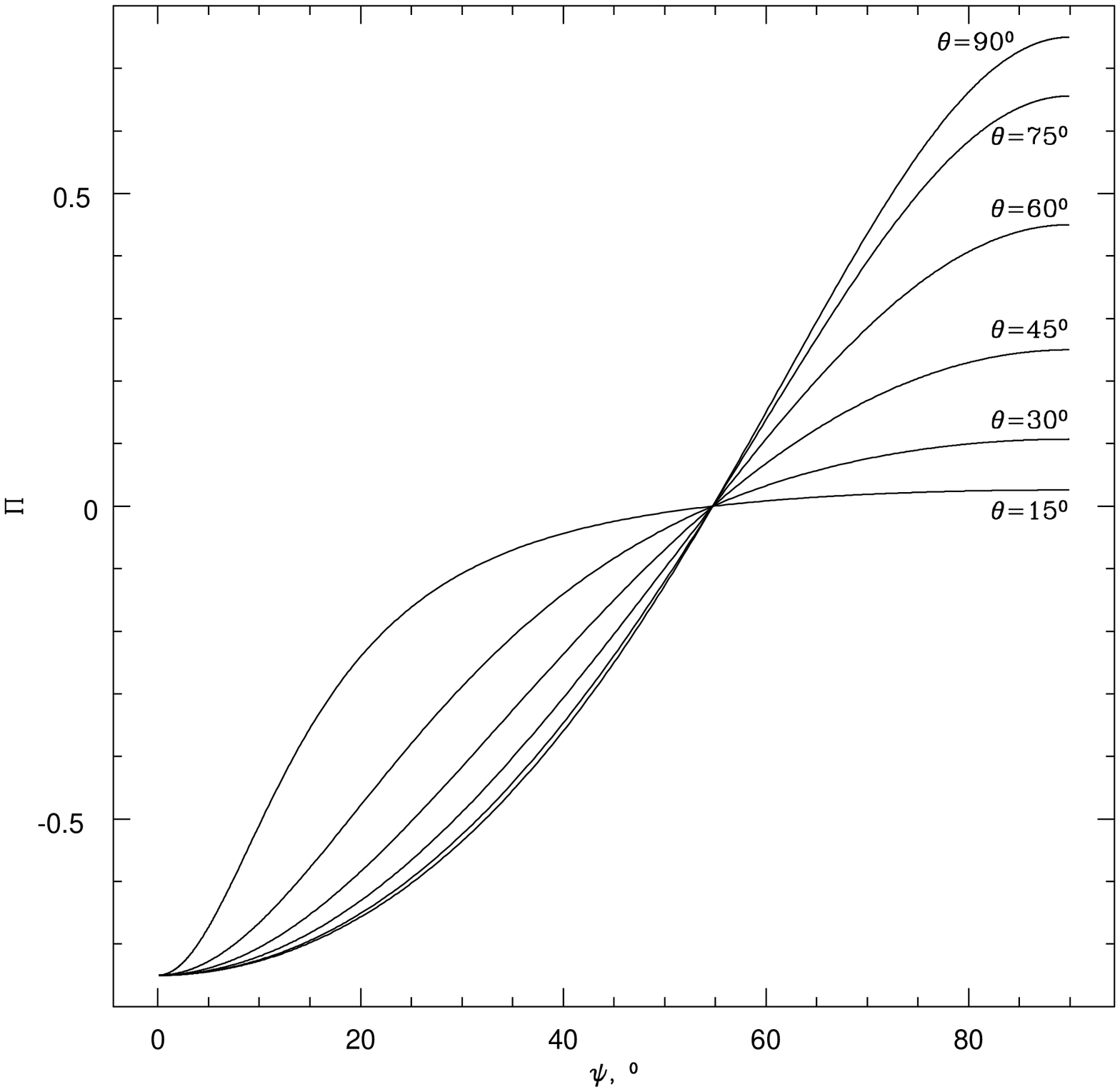}
\vskip .01 truein
\hskip 1 truein (d) \hskip 1.8 truein (e) \hskip 1.8 truein (f)  \\
\caption{Polarization fraction $\Pi= Q/I$  for non-relativistic
cylindrical shells. First raw of plots are $\Pi$ 
as a function of the observer angle $\theta$ for $p=1$ (a), $2.4$
(b), $3$ (c); second raw of plots are $\Pi$ as a function of the pitch  
angle $\psi'$  for $p=1$ (d), $2.4$ (e), $3$ (f).
Positive values indicate polarization along the jet, negative ---
polarization orthogonally to the jet.
}

\label{jetnonrelat}
\end{figure}

The general expressions for polarization given in Section~\ref{sec_2}  
can be simplified considerably for 
non-relativistic jets. 
Setting
$\beta = 0$, we find
\ba &&
\sin ^2 \chi' = \cos^2 \psi' \sin ^2  \theta +
 { 1 \over 2} \sin 2 \theta \sin 2 \psi' \sin \phi +
(  \cos^2 \theta + \cos^2   \phi \sin^2 \theta )  \sin^2  \psi'
\nn &&
\cos 2 \tilde{\chi} =
{ \cos ^2 \phi \sin ^2 \psi' 
-( \cos \psi' \sin \theta + \cos \theta \sin \phi \sin \psi')^2
\over 1- ( \cos \theta \cos \psi' -  
\sin \theta \sin \phi \sin \psi')^2 }.
\ea
The degree of polarization $\Pi$ from a cylindrical shell 
populated by particles with various power-law indices $p=1$, $2.4$, $3$ 
are plotted in Fig.~\ref{jetnonrelat} as a function of the viewing angle
and pitch angle \citep[see also][]{laing81}. 
(A typical power-law index for pc-scale jets is $p=2.4$, which
corresponds to a spectral index of $\alpha = -0.7, S_{\nu}\propto 
\nu^{\alpha}$; see, for example, \cite{gab01a,gab01b}).
\footnote{
A particularly simple case is $p=3$, when the integrations in 
Eqs.~(\ref{Stokes}) can be done
exactly (see also \citet{pariev03} for another exactly integrable case).  
We find for the average polarization of an
unresolved shell
\be
\Pi=- { 3 (1 + 3 \cos 2  \psi') \sin^2 \theta  \over
2(5 - \cos 2 \theta -\cos 2\psi' - 3\cos 2\theta \cos 2\psi') }
\ee
}
As can be seen from  Fig. \ref{jetnonrelat} 
(positive values indicate polarization along the jet, and negative values
polarization orthogonal to the jet), 
there is a weak dependence of the
polarization on the spectral index. 
In particular, flatter spectra (smaller $p$)
produce more complicated polarization structure, 
so that {\it sources with flatter spectra
are more likely to produce a change of polarization from parallel 
to perpendicular}.

\begin{figure}[ht]
\includegraphics[width=0.45\linewidth]{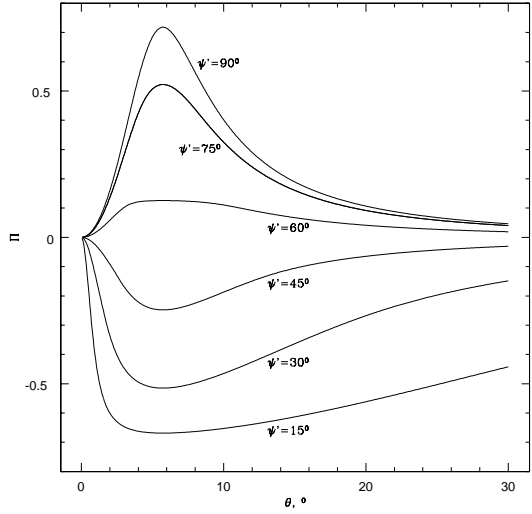}
\includegraphics[width=0.45\linewidth]{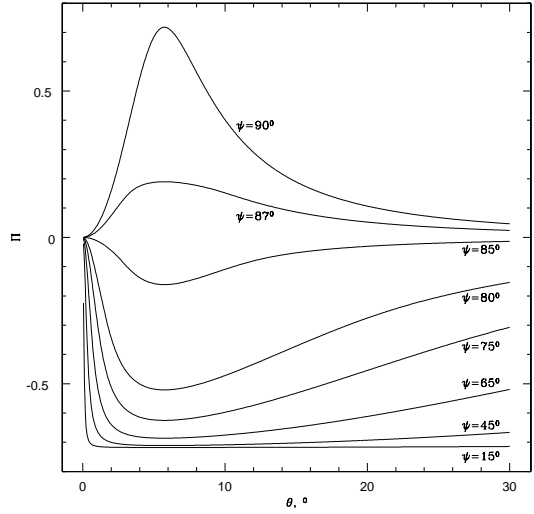}\\
\hskip 3 truein (a) \hskip 3 truein (b) \\
\includegraphics[width=0.45\linewidth]{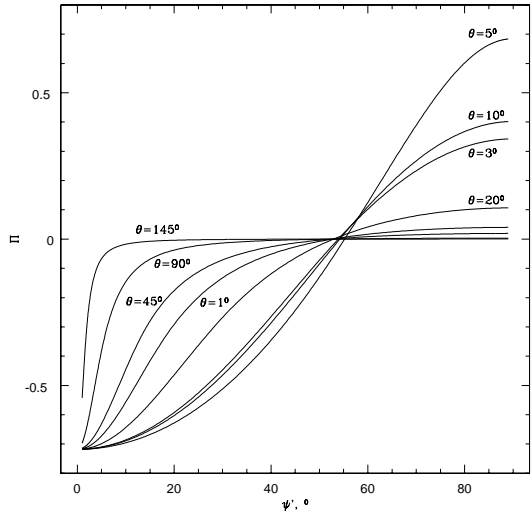}
\includegraphics[width=0.45\linewidth]{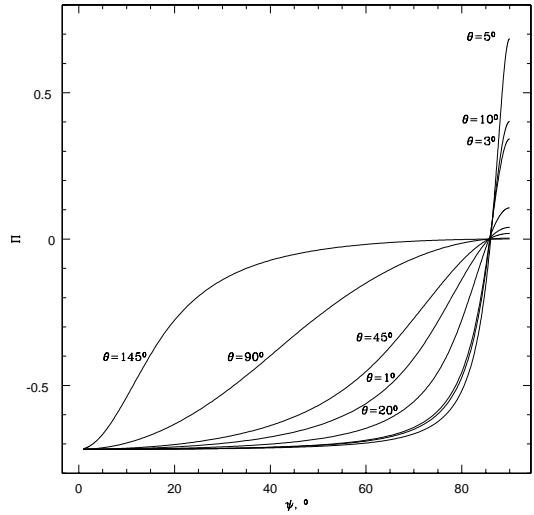}\\
\hskip 3 truein (c) \hskip 3 truein (d) \\
\caption{ Polarization fraction  for cylindrical shells with $\Gamma =10$ and
particle index $p=2.4$
as a function of the observer angle $\theta$ for different values
of rest frame pitch angle $\psi'$ (a) and  for different values
of laboratory frame  pitch angle $\psi$ (b); 
  for different values of the observer angle  $\theta$
 as a function of the  rest frame  pitch angle   $\psi'$ (c) and
laboratory frame  pitch angle $\psi$ (d).
}
\label{polariztot}
\end{figure}

\begin{figure}[ht]
\includegraphics[width=0.45\linewidth]{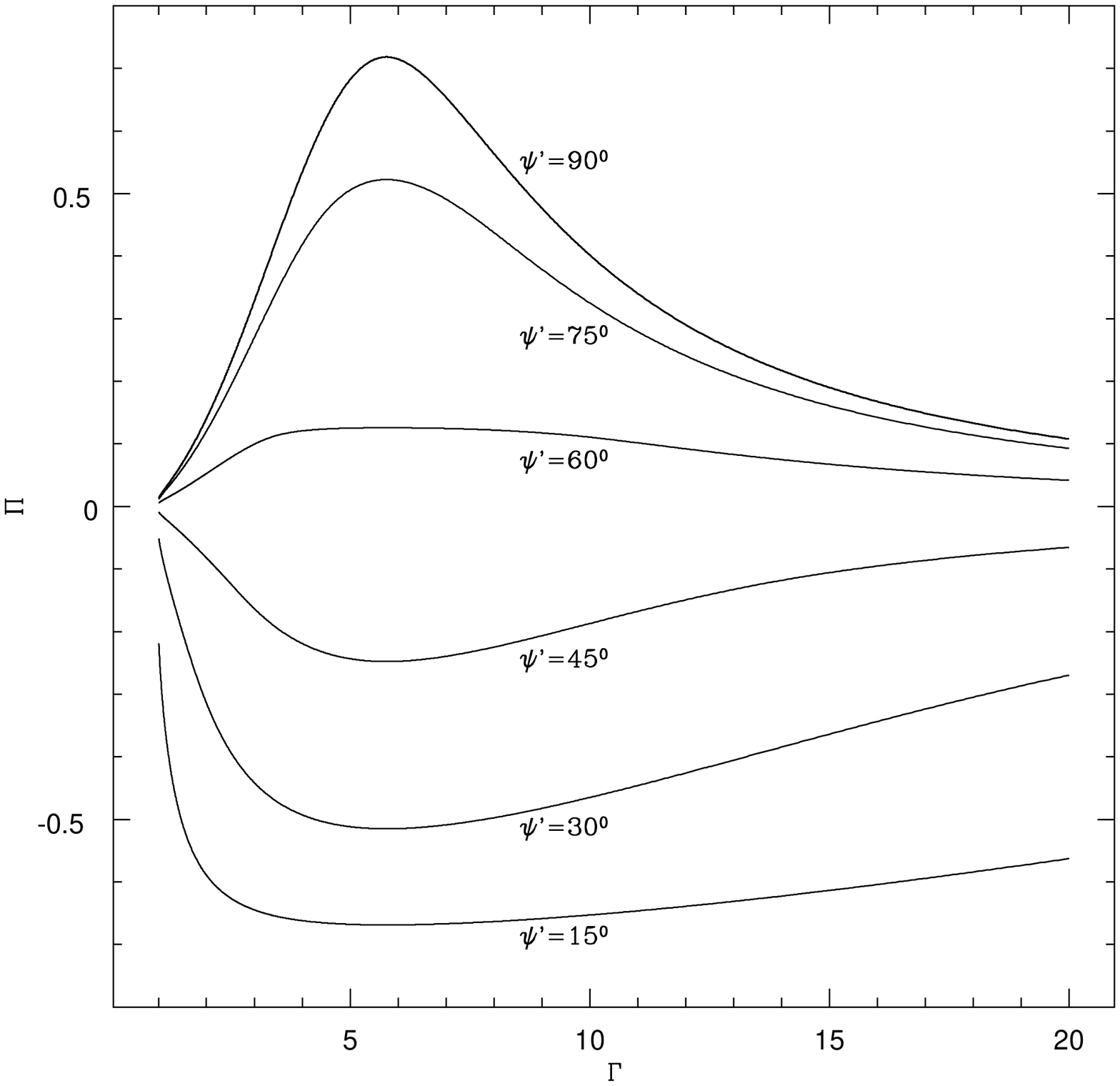}
\includegraphics[width=0.45\linewidth]{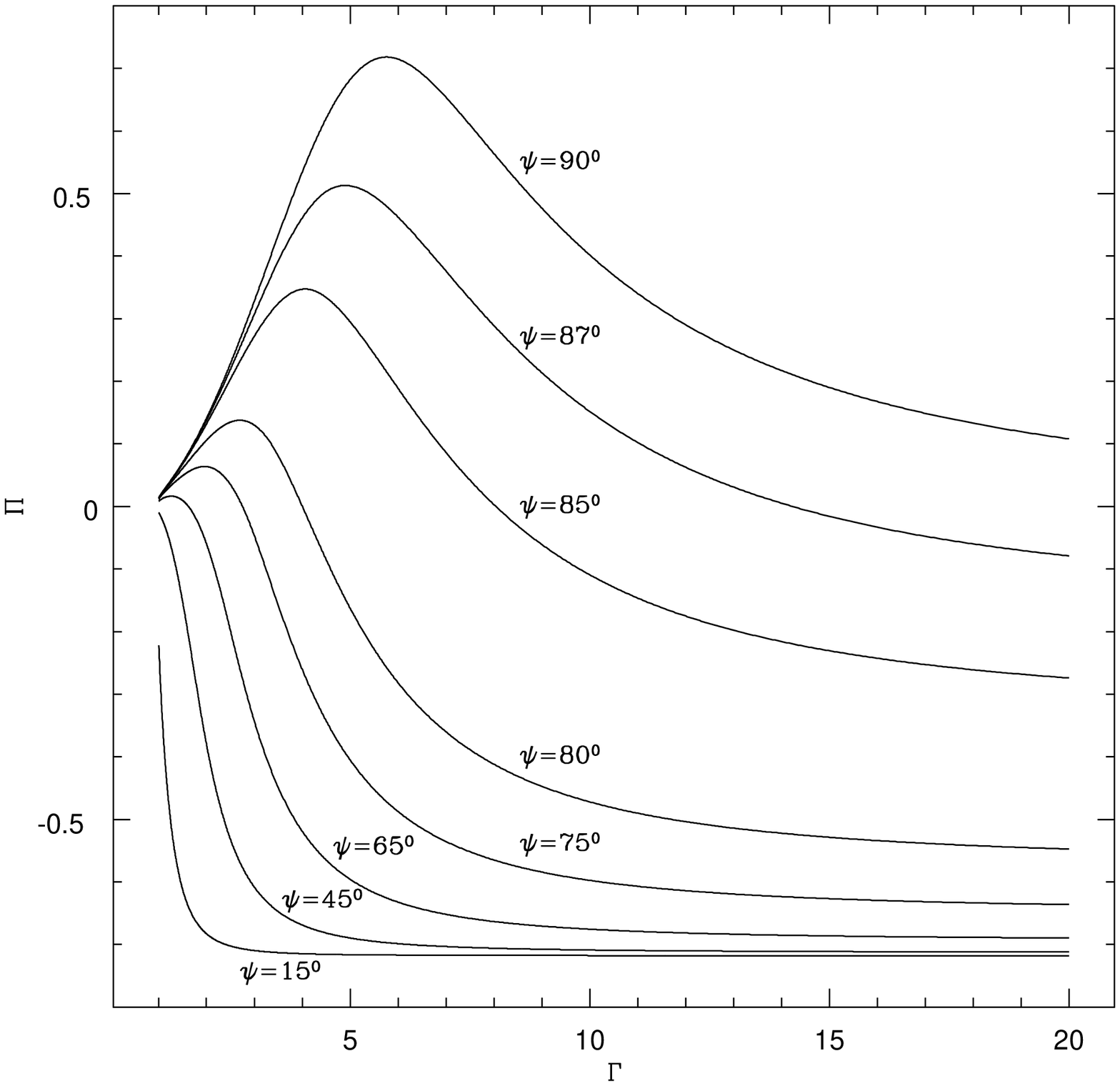}
\caption{Polarization fraction  for a cylindrical shell as a function
of Lorentz factor $\Gamma$ for different rest frame
and  laboratory frame pitch angles $\psi'$ and  $\psi$.
Viewing angle $\theta=10^0$ and $p=2.4$.
}
\label{polariztotg}
\end{figure}

For relativistic jets, we adopt $p=2.4$ as a fiducial number and
plot the fractional polarization for various viewing angles and pitch angles 
in the rest frame and the laboratory frame (Fig.~\ref{polariztot}). 
In addition, we plot in Fig.~\ref{polariztotg} the fractional polarization 
as a function of the bulk Lorentz factor $\Gamma$ for $\theta = 10^\circ$.

Several conclusions can be drawn upon analyzing these plots.  Consider
first the polarization as a function of the viewing angle for various  
rest-frame pitch angles
(Fig.~\ref{polariztot}a). For extreme values of the pitch angle
in the rest frame
(purely toroidal
magnetic field, $\psi'=\pi/2$,  and purely poloidal magnetic field $\psi'=0$),
the corresponding polarization is along (${ \Pi }>0$) and orthogonal to 
($  \Pi  <0$) the jet.  For large pitch angles, 
$ \pi/3 \leq  \psi'  \leq \pi/2$, the polarization
always remains along the axis,  peaking at $\theta \sim 1/\Gamma$
and decreasing sharply for $\theta \Gamma \gg 1$.  For intermediate
pitch angles,  $ \psi' \leq \pi/3$, the polarization can change from
parallel to orthogonal depending on the viewing angle, with such a change 
being more likely to occur for lower values of $p$. For  smaller
pitch angles, $\psi' \leq \pi/4$, the polarization is high, weakly
dependent on the viewing angle, and oriented orthogonal to the jet.

Thus, the average polarization is a very sensitive function of the jet
parameters. For some jet parameters, the polarization properties are
fairly constant, while, in some cases, they can change appreciably due to
small variations in the jet parameters.

Most importantly, the observation
of polarization along the jet requires that the toroidal magnetic field
{\it in the jet frame} be at least of the order of the poloidal field.
For relativistic jets, this means that such jets are {\it strongly
dominated by the toroidal field in the observer's frame},
$B_\phi/ B_z \sim \Gamma  B_\phi'/ B_z' \gg 1 $.
In cases when a change of sign is seen, we can place strong constraints
on the internal pitch angle, since such transitions can be seen only
for a limited range of pitch angles, $\pi/3 \leq \psi' \leq \pi/4$.
A $90^{\circ}$ change of the EVPA can be initiated by a small deviation
of the jet direction (if $\theta \sim 1/\Gamma$, in order to change
the observed EVPA by an angle of the order of unity, the real direction
 may  change by $\Delta\theta \ll  1/\Gamma$), acceleration
of the jet or injection of a toroidal magnetic  flux. 
On the other hand,
a relativistic jet viewed at a large angle, $\theta\gg 1/\Gamma$, 
will be either weakly polarized
along the jet (for $B'_\phi \geq B_z'$) or strongly polarized orthogonally
to the jet (for $B'_\phi \leq B_z'$); see Fig \ref{polariztot}a.

\subsection{Resolved hollow jets (cylindrical shells)}
\label{subsec_hollow_res}

\begin{figure}[ht]
\includegraphics[width=0.45\linewidth]{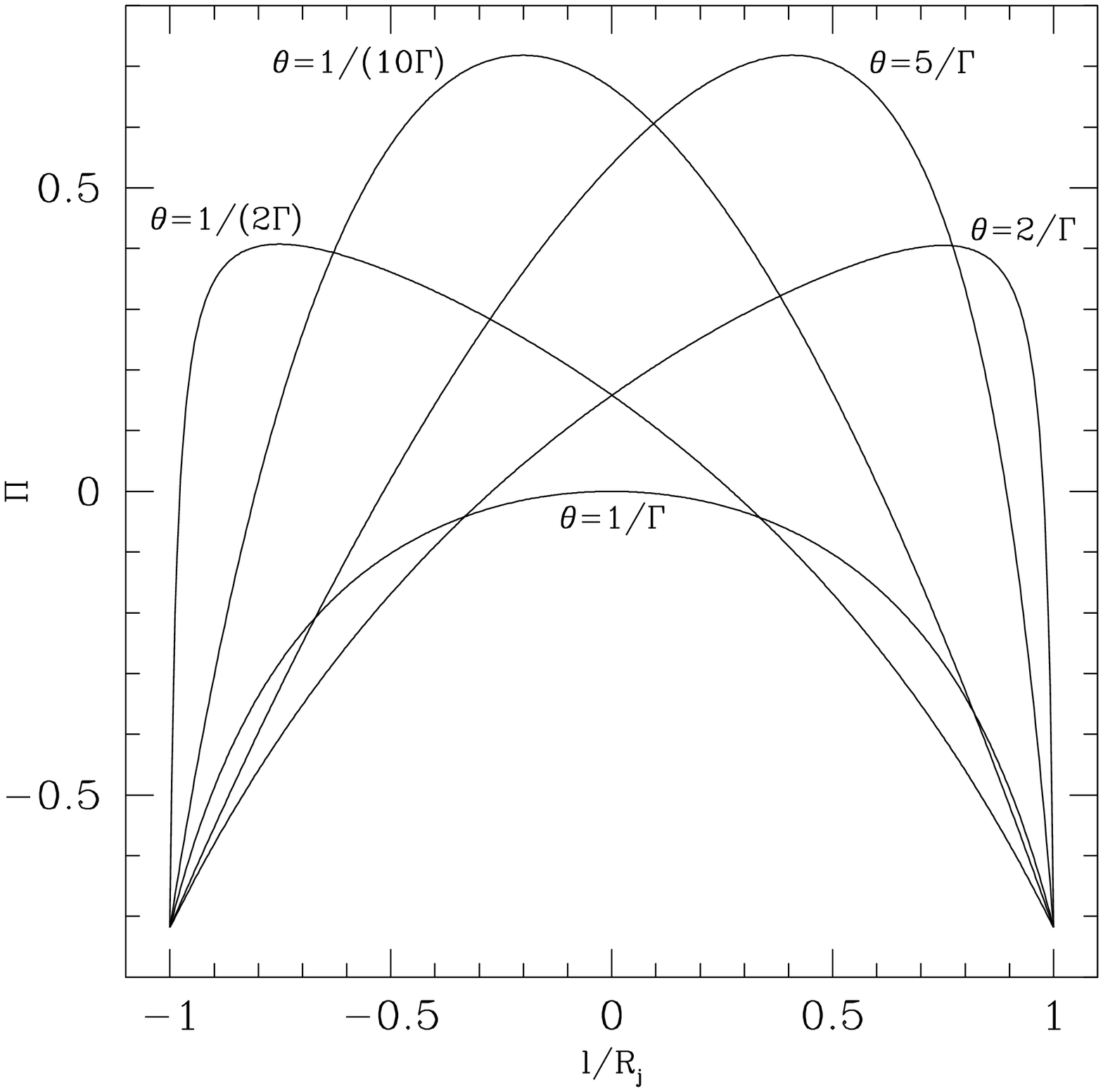}
\includegraphics[width=0.45\linewidth]{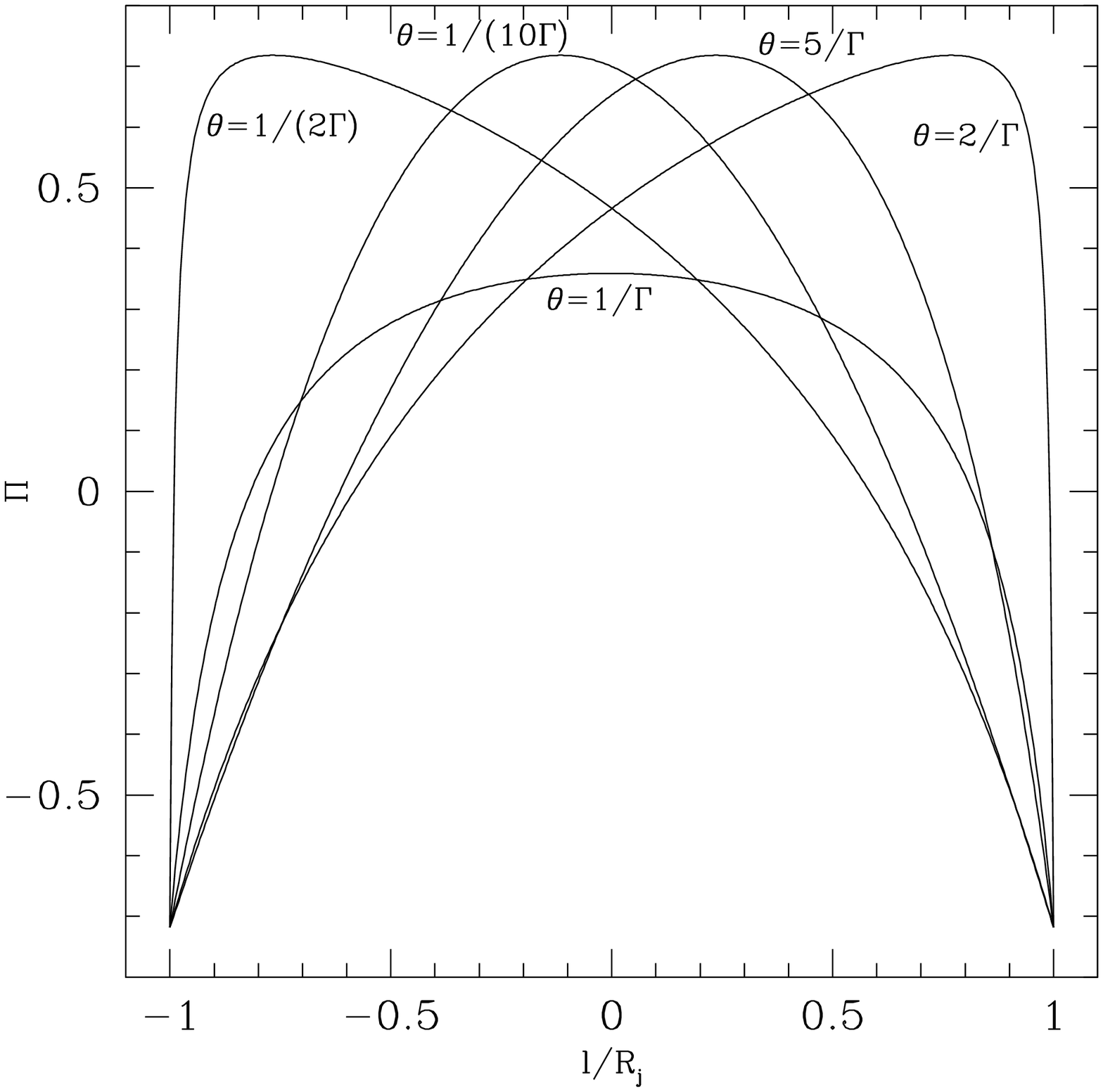}\\
\hskip 3 truein (a) \hskip 3 truein (b) \\
\includegraphics[width=0.45\linewidth]{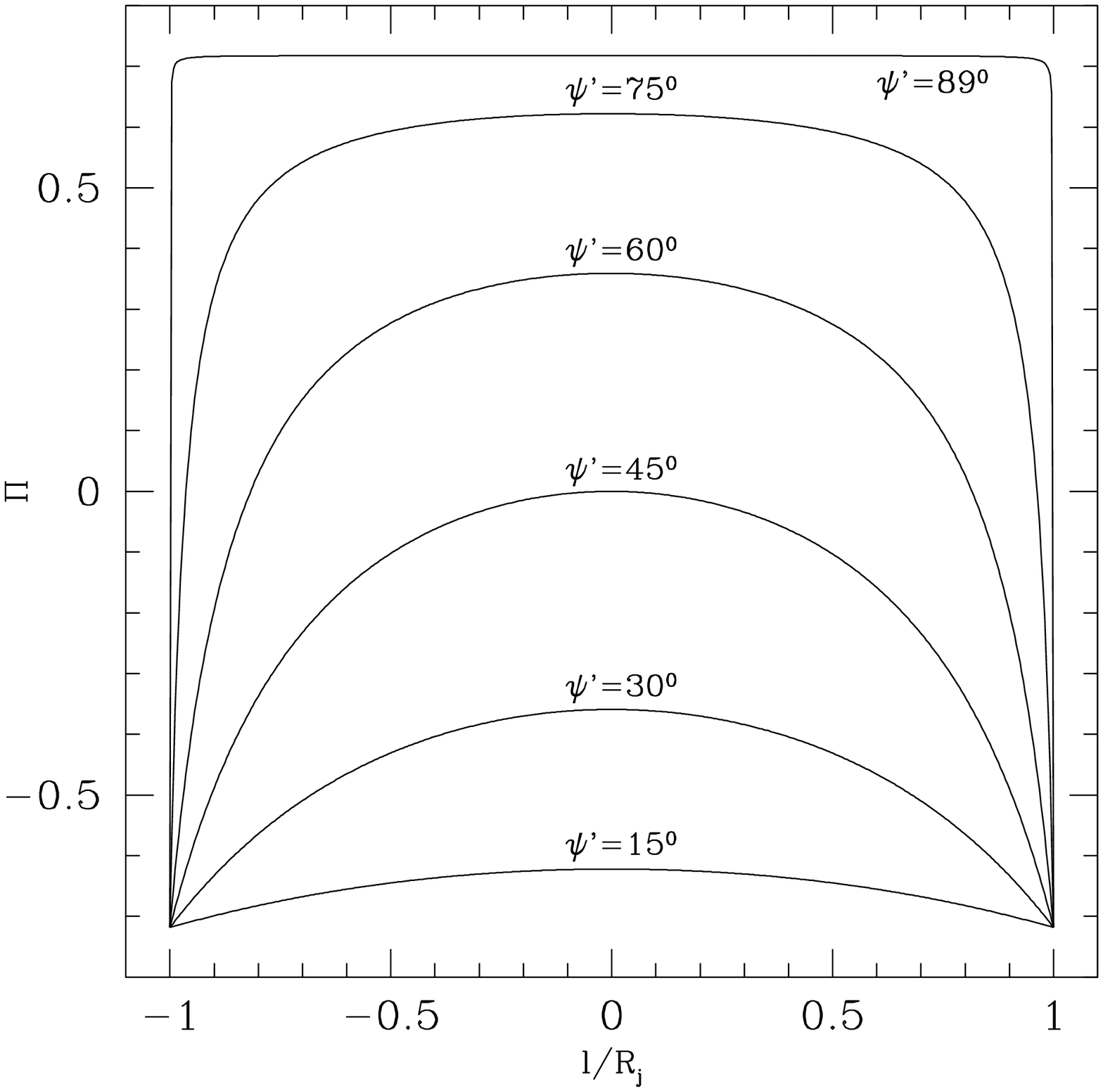}
\includegraphics[width=0.45\linewidth]{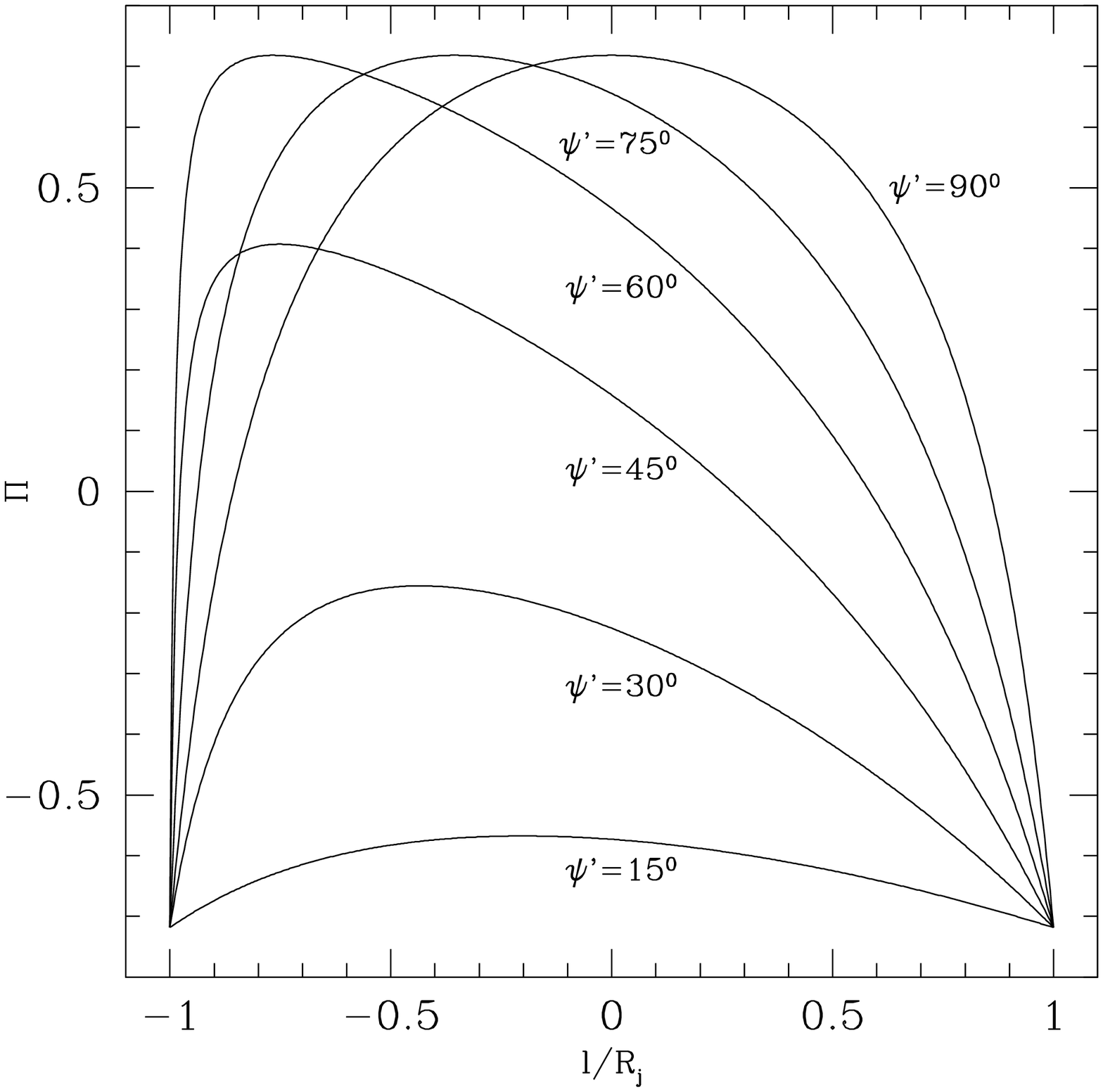}\\
\hskip 3 truein (c) \hskip 3 truein (d) 
\caption{
Profiles of the polarization degree $\Pi$ for 
resolved cylindrical shells 
as a function of the distance across the jet
projected on the sky, $l$. The radius of the shell, $R_j$, 
is arbitrary.
$\Gamma = 10$ and $p=2.4$. The plots (a) are for $\psi' = 45^0$, 
and plots (b) are for $\psi'=60^0$ while for the different values
of the viewing angle $\theta$. Plots (c) are for $\theta=1/\Gamma$,
and plots (d) are for $\theta=1/(2\Gamma)$ while for the different 
values of the pitch angle $\psi'$. For small and large $\theta$, the 
polarization in the central part of the image is along the axis of 
the jet regardless of the value of $\psi'$. For $\theta \sim 1/\Gamma$
the polarization can be longitudinal only for $\psi'>45^0$. 
At the edge of the jet the polarization is orthogonal to the jet 
axis for any values of $\theta$ and $\psi'$. 
}
\label{jetresol}
\end{figure}

We next turn to resolved jets.
If emission is confined to a narrow cylindrical 
shell, $K_{\rm e} \propto \delta(r-R_j)$
then integration over $dS $ gives a multiplier $h/\arcsin h/R_j$, so that
polarization becomes
\be
\Pi= { p+1 \over p+7/3} \cos 2 \tilde{\chi}
\label{2chinr}
\ee
In  Fig.~\ref{jetresol} we plot
the fractional polarization for shells moving with $\Gamma=10$, 
for various pitch angles $\psi'$ and viewing angles $\theta$.

First, note that the polarization close to the edge of the observed jet is 
always orthogonal to the jet. This can easily be understood from 
simple geometrical
considerations: at the visible  edge of the cylindrical shell, the projection
of the toroidal magnetic field onto the plane of the sky becomes zero,
so that only the poloidal magnetic field contributes to the observed
synchrotron emission, and this part of the jet is observed to be polarized
orthogonal to the jet axis. For $\theta\ll 1/\Gamma$ and 
$\theta \gg 1/\Gamma$, the polarization
in the central part of the jet is along the symmetry axis. For intermediate 
values of $\theta\sim 1/\Gamma$, the polarization in the central parts 
of the jet is longitudinal if the pitch angle $\psi'> 45^0$ and orthogonal
if $\psi'< 45^0$. This can explain in a natural way the fact that resolved 
jets sometimes have polarization aligned with the jet in the central region 
and orthogonal to the jet at the edge.  

Second, {\it polarization observations of resolved jets can be used to 
infer the relative orientation of the  spin of the central object that 
launched the jet (black hole or disk): whether it is aligned or 
counter-aligned with the jet axis.} This possibility comes form the fact 
that the {\it left and right-handed  helices
produce different polarization signatures, } see Figs.~\ref{jetresol}.
In order to make a distinction between the two choices, it is necessary 
to determine
independently the angle at which the jet is viewed in its rest frame. 
For relativistic jets, 
this amounts to determining 
the product $\theta \Gamma$: if $\theta \Gamma < 1$ then the jet is 
viewed ``head-on,'' while the jet is viewed ``tail-on'' if  
$\theta \Gamma > 1$. 
A right-handed magnetic helix viewed head-on produces the same polarization
signature as a left-handed helix viewed tail-on.

\begin{figure}[ht]
\includegraphics[width=0.9\linewidth]{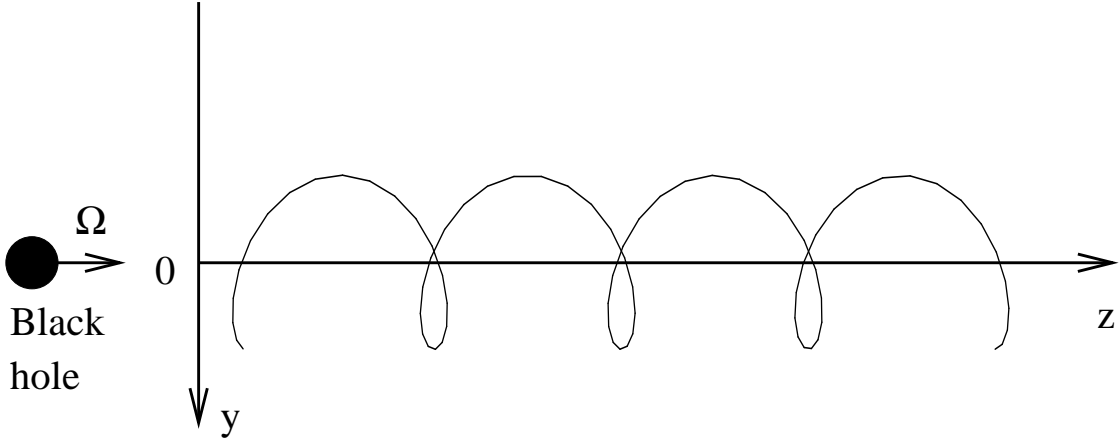}
\caption{A ''head-on'' ($\theta \Gamma < 1$)
view of a left-handed helical magnetic field in the reference frame comoving 
with the jet in the positive $z$-direction. 
Polarization $\Pi$ of the lower, $y>0$, part
is larger, i.e. it is more likely to be along the jet. The location of the 
observer is the same as on Fig.~\protect\ref{polariz-geomAGN}.
}
\label{helix}
\end{figure}

We expect that the magnetic field lines in the jet are retrograde with 
respect to the black hole spin. Thus, if the spin of the central 
object is  oriented along the $z$ direction,
the magnetic field lines will form a left-handed helix (see Fig.~\ref{helix}).
Suppose next that we view the jet at angles $\theta \Gamma < 1$,
so that, in the frame of the jet, a circular left-handed
helix is moving toward an observer along the $z$ axis. 
Let us denote by $l$ the coordinate along the unit vector ${\bf l}=\{ 0,1,0 \}$
in the plane of the sky parallel to the $y$ axis.
In this case, the parts of the jet located at positive $l$ will be more 
preferentially 
polarized along the jet (that is, $\Pi$ will be larger), than the parts 
of the jet located at negative $l$.
This conclusion is invariant with respect to sign changes of both the
$B_z$ and $B_\phi$ components of the magnetic field. A spiral magnetic 
field given by Eqs.~(\ref{helical_b}) forms a right-handed spiral, 
which corresponds to the spin of the central object being counter-aligned with 
the direction of the jet, i.e. the black hole or accretion disk rotates 
clockwise when viewed by an observer looking down the jet. Then, parts 
of the jet located at positive $l$ (the sign of $l$ is the same as the sign 
of $y$) will have smaller 
$\Pi$ than parts of the jet located at negative $l$. The latter behavior 
is seen for the curves $\theta=1/(2\Gamma)$ and $\theta=1/(10\Gamma)$ in 
Fig.~\ref{jetresol}a,b. 

In the absence of rotation, the jet axis can be defined by the symmetry of 
the intensity
profile, so that the direction of the  black hole spin can be determined from
the asymmetry of the polarization profile with respect to this axis (see 
Fig.~\ref{jetresol}a, b, and d). 
If an observer positions himself along the projection of the jet onto the 
plane of the sky ($z$-axis on Fig.~\ref{helix}) with his head toward the 
end point of the jet and his feet toward the core of the AGN, then 
the side of the jet with negative $l$ will be on his left side, and 
the side of the jet with positive $l$ will be on his right side.
Then, a head-on viewed jet 
($\theta < 1/\Gamma$) emanating from a clockwise rotating central engine 
has $\Pi$ larger on the left side of its image on the sky, 
while a head-on viewed jet emanating from a counter-clockwise rotating central 
engine has $\Pi$ larger on the right side of its image on the sky.
The opposite holds for a tail-on viewed jet ($\theta > 1/\Gamma$):
jets emanating from a clockwise rotating central engine 
have $\Pi$ larger on the right side of their image on the sky, 
while jets emanating from a counter clockwise rotating central 
engine have $\Pi$ larger on the left side of its image on the sky.
When $\theta \approx 1/\Gamma$, i.e. when the jet is viewed orthogonal 
to its axis in the frame comoving with the jet, the $\Pi(l)$ curve becomes 
symmetric with respect to the jet axis (Fig.~\ref{jetresol}c). 
The caveat here is how one can 
know if a jet is viewed head-on or tail-on in its rest frame.
 Because the parsec-scale counterjet is almost always not observed, 
we do not readily see how to overcome this caveat.

\section{Emission from ''filled'' jets}
\label{filled}

To find the total emission from a jet, we must know the distribution of
the emissivity and the internal structure of the jet, both of which are
highly uncertain functions. It is probable that 
jets are electromagnetically dominated on parsec scales
(\citealt{lyndenbell96, lyndenbell03, ustyugova00, li01, lovelace02,
lovelace03}, see also Section \ref{EMjet}), so that
their structure is determined by relativistic force-free electrodynamics.
In this work, we assume that the jets are strongly magnetized and are 
described by relativistic force-free electrodynamics. 
In Appendix~\ref{FF}, we present equations of relativistic 
force-free electrodynamics and suggest a simple method that can be used to find
a broad class of solutions for cylindrical, relativistic, force-free pinches
given the corresponding non-relativistic pinches.

There are many ways in which  the poloidal magnetic field, electric charge
and poloidal current can be distributed inside a jet.  This should depend
primarily on the jet launching conditions (as boundary conditions) and
the evolution of the jet as it propagates. At the moment,
there is no general agreement about the structure of jets. Furthermore,
it is not even clear if jets carry a total poloidal current and total poloidal
magnetic flux.  It is likely that jets do carry total current, as
models for the launching and collimation of jets show
\citep{bp82,cl94, ustyugova00, lovelace02}. 
Whether there is a total poloidal magnetic flux
is less  clear.  Most models for jet launching assume that there is a
large poloidal magnetic flux at the base of the jet, but physically it is
hard to see how this large-scale field survives in a turbulent disk.
An alternative possibility is that magnetorotational instability 
operating in the disk produces
complicated field structures with negligible total poloidal flux. One
advantage of these models is that the energy associated with current
reversals can be used to power the jet \citep[\eg][]{lnr97}.

There are a multitude of possible jet profiles, each giving a different
polarization structure (also with a number of different prescriptions for
the jet emissivity).  Given these  uncertainties, we consider below three
types of jet structure: (i) a diffuse pinch with large poloidal and
toroidal field fluxes, (ii) a particular example of an infinite (single)
reverse-field pinch with vanishing total poloidal magnetic flux and
(iii) a high-order reverse-pinch configuration.
We believe that these three sample jet structures are generic
examples and somewhat
at the extreme ends of expected cases in terms of their total poloidal
field flux, which is logarithmically divergent for (i) and zero for (ii) and 
(iii). The current density is concentrated toward the center of the jet in 
all cases considered, forming the core of the jet. The total energy of 
the magnetic field is also finite, and the spatial distribution of the 
magnetic-field energy is concentrated inside a core of the same typical
radius as the current-carrying core of the jet. Below we denote the 
radius of the jet core as $R_j$ and its inverse as $a=1/R_j$.

In addition, the distribution of emissivity in the jet must also be specified.
We consider here the consequences of the hypothesis that dissipation
and particle acceleration occur throughout the jet volume, as opposed to 
the dissipation being localized in current reconnection layers (Lyutikov \&
Uzdensky 2003) or resonant layers of electromagnetic perturbations
(Beresnyak, Istomin \& Pariev 2003). Although the exact dissipation
rate may depend on micro-physical and kinetic processes, it is reasonable
to assume that the dissipation rate is given by the Ohmic heating in the
rest frame of a plasma element, $j'^2/\sigma' $, where $j'$ is the 
absolute value
of the current density and $\sigma' $ is the conductivity
in a frame comoving with the plasma element. Some
fraction of this dissipated energy is then converted into heat and another
fraction to accelerated particles. 
We assume that the power-law
exponent for the accelerated particles is the same throughout the jet
volume, but that the number density of relativistic particles scales as
$K_{\rm e} \propto j'^2$.

However, we cannot be certain that the dissipation rate of the magnetic
field is proportional to $j'^2$. 
Therefore, we consider another prescription for the
number density of emitting particles in the comoving frame, $K_{\rm e}
\propto B'^2$. This corresponds to the condition of minimizing the
sum of the energy densities of the magnetic field and relativistic
particles, frequently used to obtain estimates of physical conditions
in radio sources (e.g., Burbidge 1956; 
Pacholczyk 1970; Kronberg et al. 2001). In our case of strongly 
magnetized jets, the energy of the relativistic particles should be much 
smaller than the energy of the magnetic fields, but it may still be 
reasonable to assume that the particle energy density is proportional
to $B'^2$.

\subsection{Diffuse pinch}
\label{diffuse}

The non-relativistic force-free diffuse pinch is  given by:
\be
B_\phi^{nr}= {a r   \over  1+( a r)^2} B_0
,\,\,\,
B_z ^{nr}= {1  \over  1+( a r)^2} B_0
\label{diffus}
\ee
To calculate the polarization properties of such a jet, we can 
perform the integration of the Stokes parameters~(\ref{Stokes}) and calculate
the degree of polarization $\Pi$ for the whole jet in a frame 
comoving with the jet, then use Eqs.~(\ref{eq_thetaobpr}) to express 
$\Pi$ as a function of the viewing angle in the laboratory 
reference frame. Alternatively, we can find the jet
structure in the observer's frame, use relations~(\ref{eee}) to 
calculate the polarization vector in the observer's frame and perform 
the integration of the Stokes parameters~(\ref{Stokes}) in the observer's 
frame. The equivalence of the results of both calculations provides
a consistency check.

The diffuse-pinch configuration can be generalized to the relativistic
case using the procedure described in Appendix~\ref{FF}. The Lorentz
transformation along the $z$ axis will lead to a radial electric field
and distributed charge densities (if a non-relativistic 
pinch had a line
current, a line charge would also appear in the relativistic  case).
First, we consider jets with $\Gamma=\mbox{constant}$. The 
fields in the laboratory frame are then:
\be
B_z = { B_0 \over 1+( a r)^2}\mbox{,} \hskip .3 truein
B_\phi =  { a  r  \Gamma B_0 \over 1+( a r)^2}\mbox{,}  \hskip .3 truein
E_r = \sqrt{\Gamma^2 -1  }
{a  r B_0 \over 1+( a r)^2}\mbox{.}
\label{diff1}
\ee
The corresponding current density in the frame 
comoving with the plasma element is
\be
j'_z = { a B_0  \over 2 \pi  (1+ a^2 r^2)^2}\mbox{,} \hskip .3 truein
j'_\phi = {  a^2 B_0 r \over  2 \pi (1+ a^2 r^2)^2}\mbox{,} \hskip .3 truein
j'^2=\frac{a^2 B_0^2}{4\pi^2 (1+a^2 r^2)^3} 
\mbox{.} 
\label{diff2}
\ee
The diffuse pinch has a logarithmically divergent total poloidal flux
$\Phi_z  =  \ln(1 + ( a r)^2) {B_0 \pi /a^2} $, a pitch angle that
increases with radius, $B_{\phi}/B_z = \Gamma ar$, 
and a total poloidal current
$I_0 = B_0 \Gamma/ (2 a)$.

\begin{figure}[ht]
\includegraphics[width=0.3\linewidth]{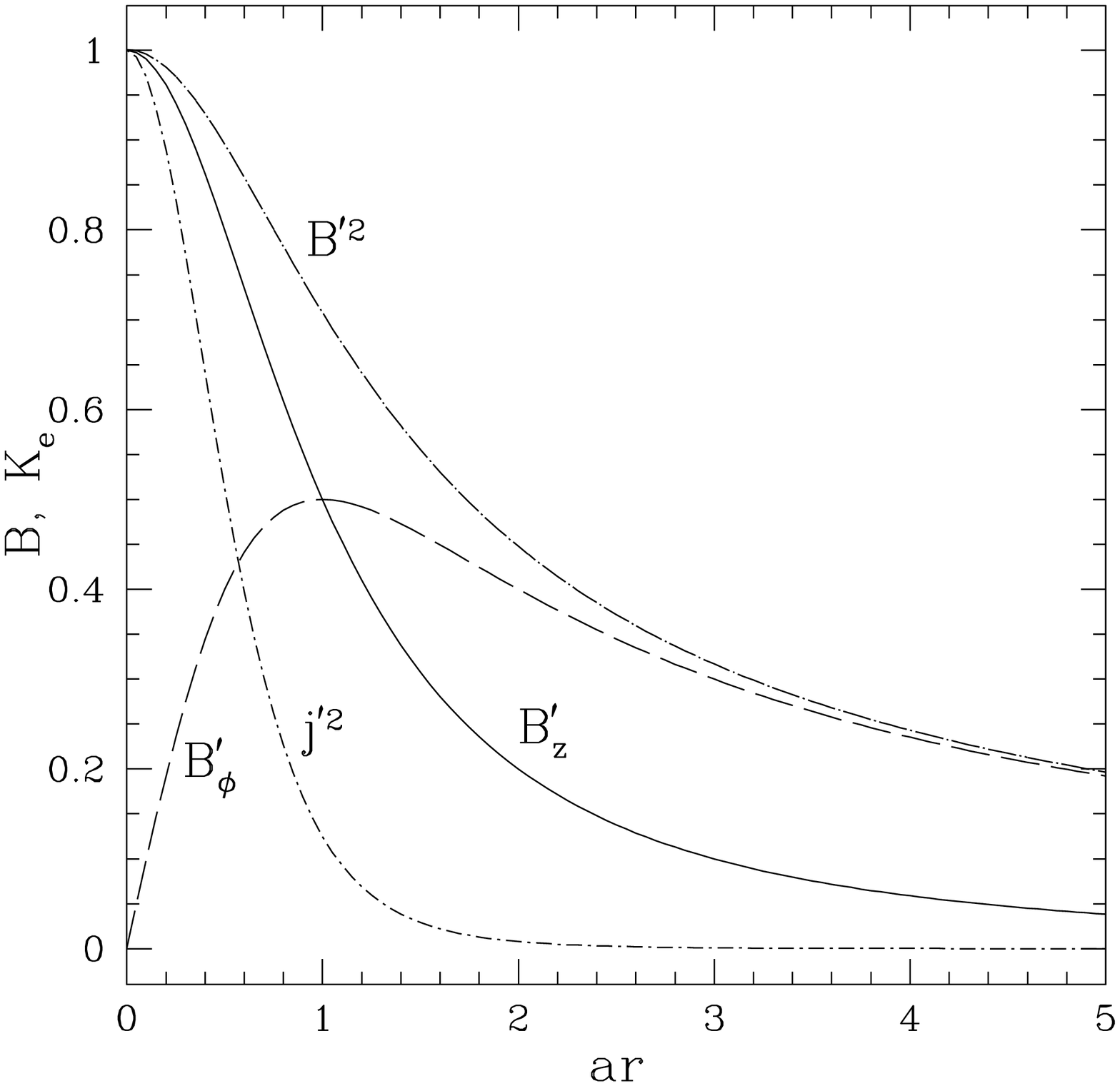}
\includegraphics[width=0.3\linewidth]{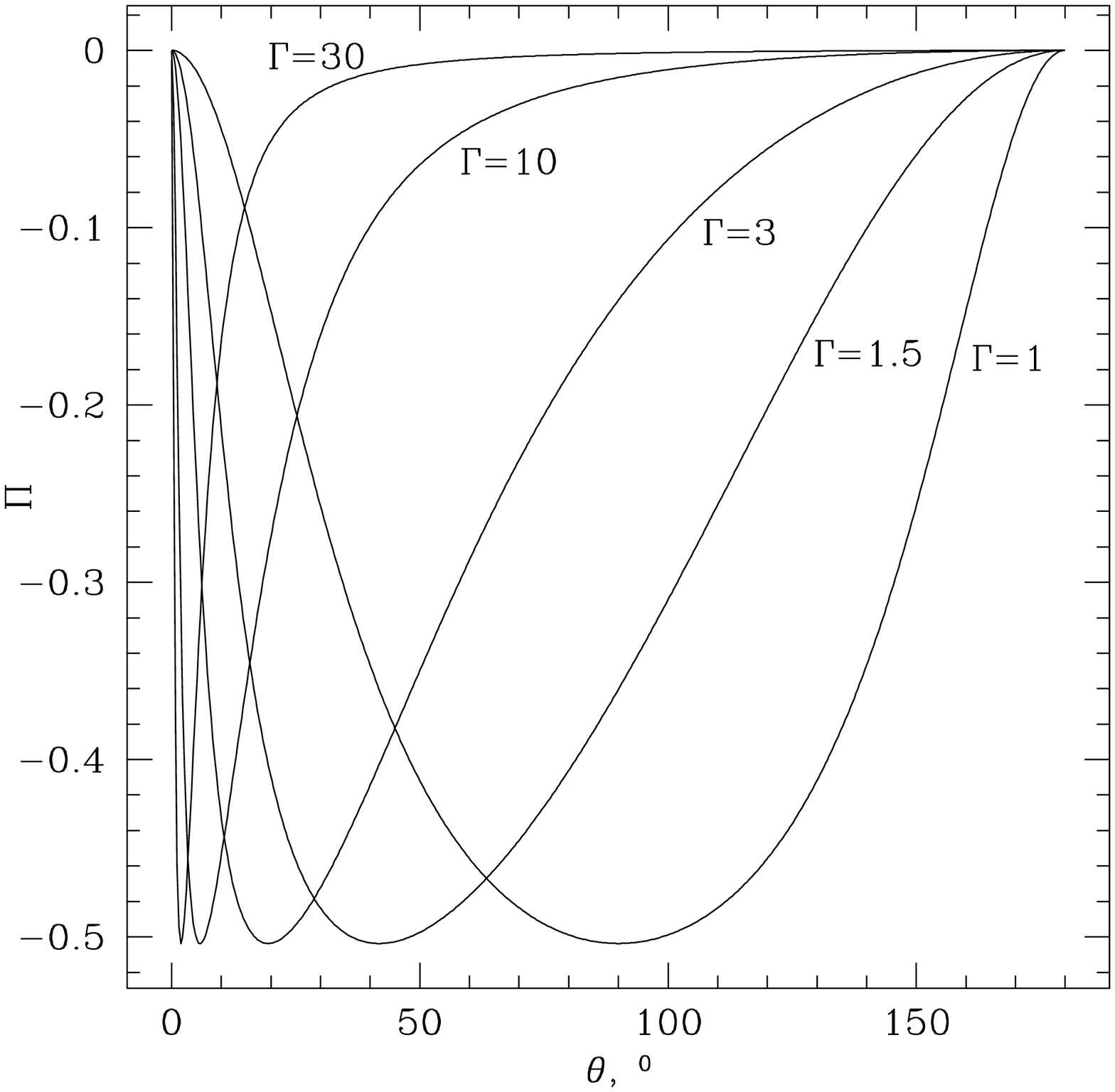}
\includegraphics[width=0.3\linewidth]{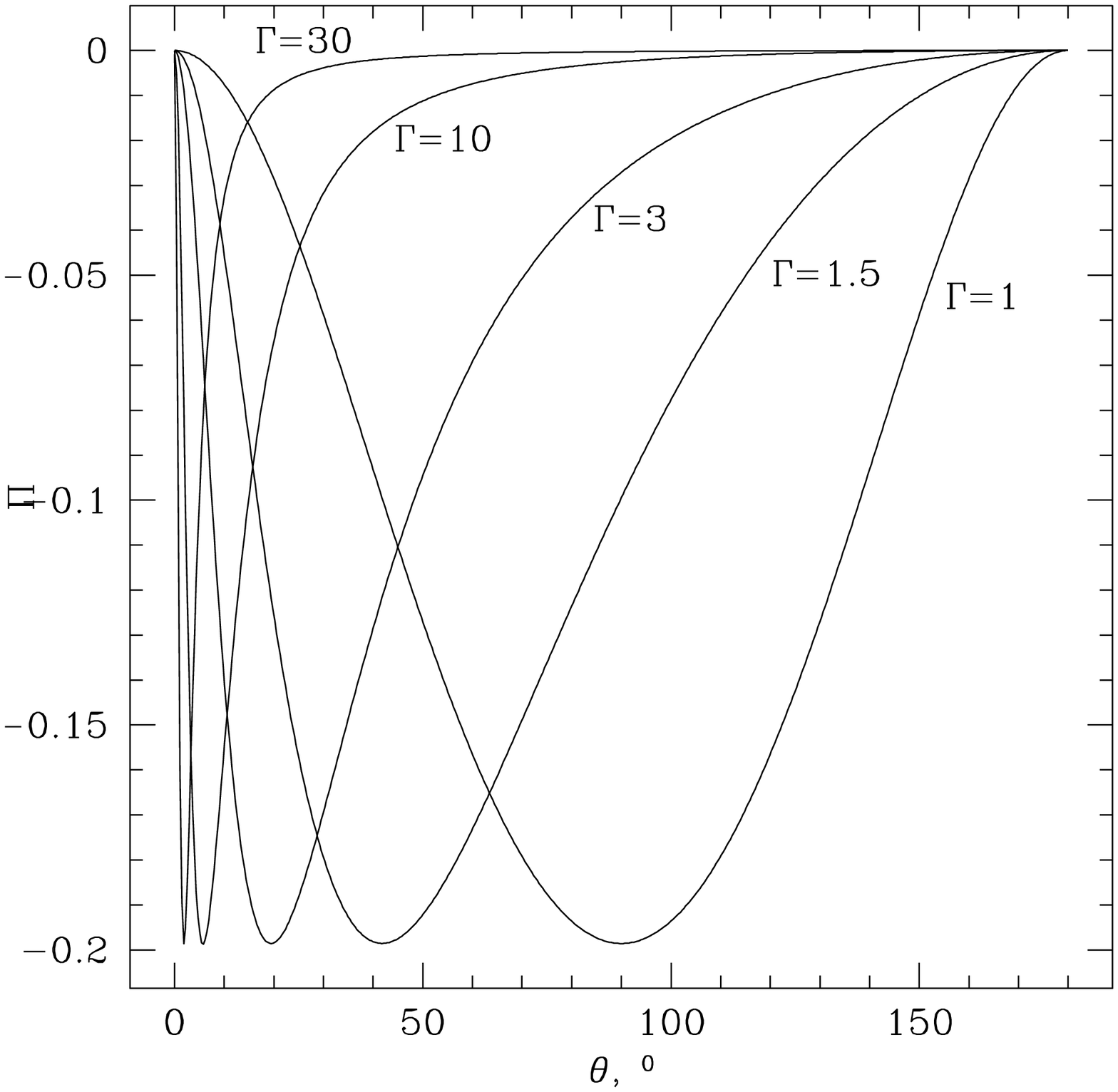}
\vskip .01 truein
\hskip 1 truein (a) \hskip 1.8 truein (b) \hskip 1.8 truein (c) \\
\caption{Polarization for a diffuse pinch, Section \ref{diffuse}.
(a) Radial profiles of $B_z'$, $B_\phi'$, $B^{\prime 2}$, and $j^{\prime 2}$
(Eqs.~(\ref{diff1}-\ref{diff2})). Units are arbitrary:
the profile of $B_z$ is normalized to unity, the units of $B_z$ and 
$B_{\phi}$ are the same, the profiles of the densities of emitting 
particles, $B^{\prime 2}$ and $j^{\prime 2}$, are also normalized to unity.
(b) Degree of polarization for $p=2.4$ for the density of relativistic 
particles $K_{\rm e} \propto j^{\prime 2}$, as a function of the observer angle
$\theta$ for different bulk Lorentz factors $\Gamma$.
(c) Same as (b), but for $K_{\rm e} \propto B^{\prime 2}$.
}
\label{myfig1}
\end{figure}

We integrated this expression from $r=0$ to large radii, $r\gg 1/a$,
for two cases: when the emissivity is  $\kappa \propto j'^2$ and  
$\kappa \propto B'^2$.  The results
are presented in Fig.~\ref{myfig1}. The polarization vector is always
perpendicular to the jet axis.

\subsection{Jets with reversed field pinch and zero poloidal flux}

\begin{figure}[ht]
\includegraphics[width=0.3\linewidth]{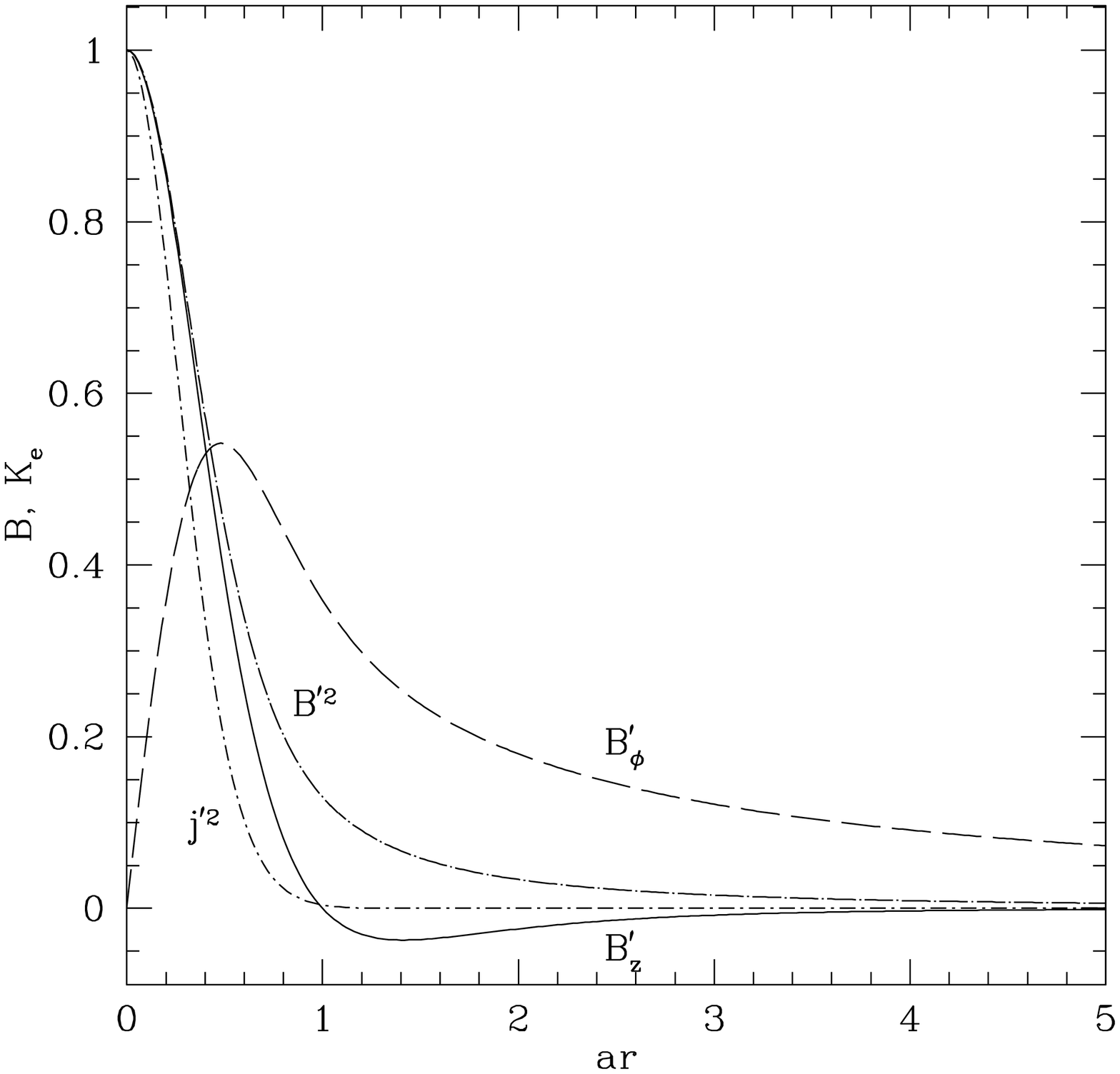}
\includegraphics[width=0.3\linewidth]{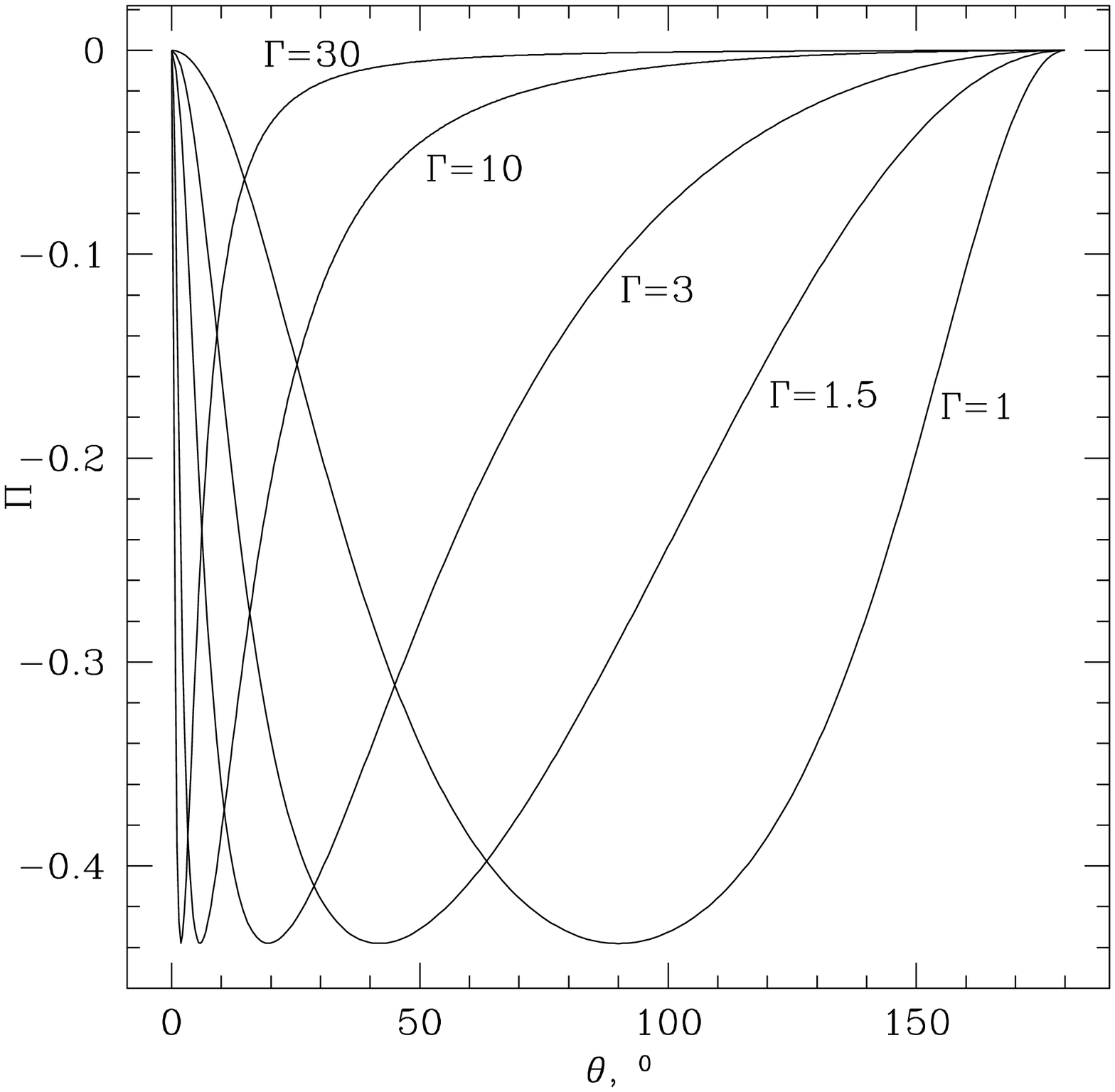}
\includegraphics[width=0.3\linewidth]{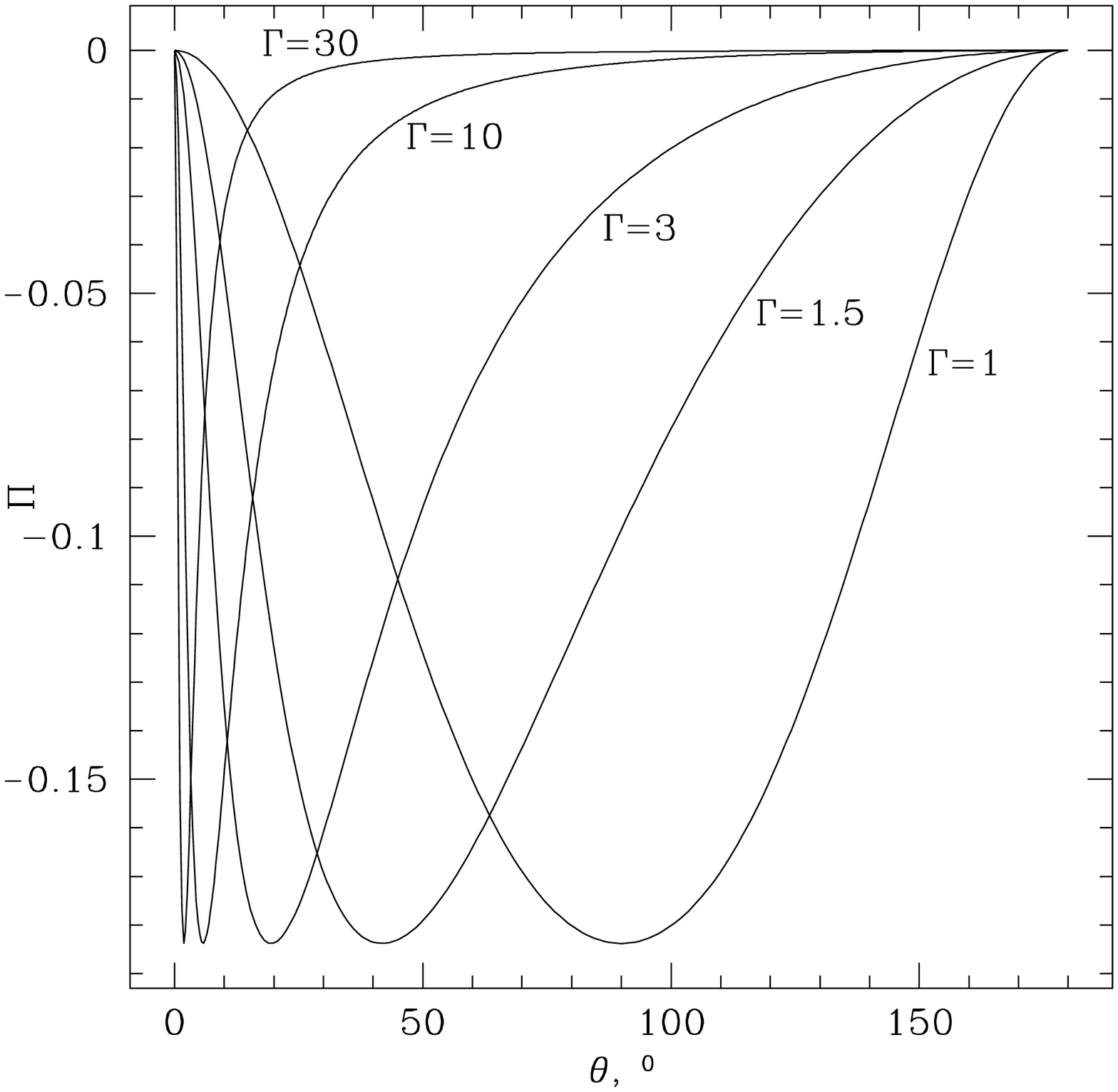}
\vskip .01 truein
\hskip 1 truein (a) \hskip 1.8 truein (b) \hskip 1.8 truein (c) \\
\caption{
Polarization for a reversed field pinch
with vanishing total poloidal flux for the fields given by
equations~(\ref{myeqn2}). Notations are the same as in Fig.~\ref{myfig1}.
(a) Radial profiles. (b) Degree of polarization for $p=2.4$ for the 
density of relativistic particles $K_{\rm e} \propto j^{\prime 2}$.
(c) Degree of polarization for $p=2.4$ for the 
density of relativistic particles $K_{\rm e} \propto B^{\prime 2}$.
}
\label{myfig4}
\end{figure}

Next we calculate the polarization from a jet with zero total poloidal flux,
$\displaystyle \int_0^\infty B_z rdr =0$.  This case corresponds to a
large-scale magnetic field at the base of the jet generated by the disk
itself, as opposed to the case of non-vanishing poloidal flux of an
advected magnetic field. The following choice of force-free magnetic
field (in the laboratory frame) is representative of the case with
vanishing total flux:
\begin{equation}
B_z^{nr}=B_0\frac{1-a^2 r^2}{(1+a^2 r^2)^3}, \quad
B_\phi^{nr} = B_0 \Gamma \frac{ar \sqrt{2/15}}{(1+a^2 r^2)^3}
\sqrt{30+10 a^2 r^2 + 15 a^4 r^4 + 6 a^6 r^6 + a^8 r^8}
\mbox{.}\label{myeqn2}
\end{equation}
(see Fig.~\ref{myfig4}).
We have the
following expressions for the components of the current:
\begin{eqnarray}
&& j'_\phi=\frac{cB_0}{4\pi}4a^2 r \frac{2-a^2 r^2}{(1+a^2 r^2)^4},
\label{myeqn3} \\
&& j'_z=\frac{cB_0}{4\pi}\frac{2\sqrt{30}a(a^2 r^2 -1)(a^2 r^2 -2)}
{(1+a^2 r^2)^4 \sqrt{30+10 a^2 r^2 + 15 a^4 r^4 + 6 a^6 r^6 + a^8 r^8}}
\mbox{.}
\label{myeqn5}
\end{eqnarray}
The results of the polarization calculations for the number density 
of particles 
$K_{\rm e} \propto j'^2$ and $K_{\rm e} \propto B'^2$
 are presented in Fig.~\ref{myfig4}.  Similar to the case
of a diffuse pinch, the electric vector of the wave is orthogonal to the
jet axis.

\subsection{Jets with multiple reversals of the axial magnetic
field}

\begin{figure}[ht]
\includegraphics[width=0.3\linewidth]{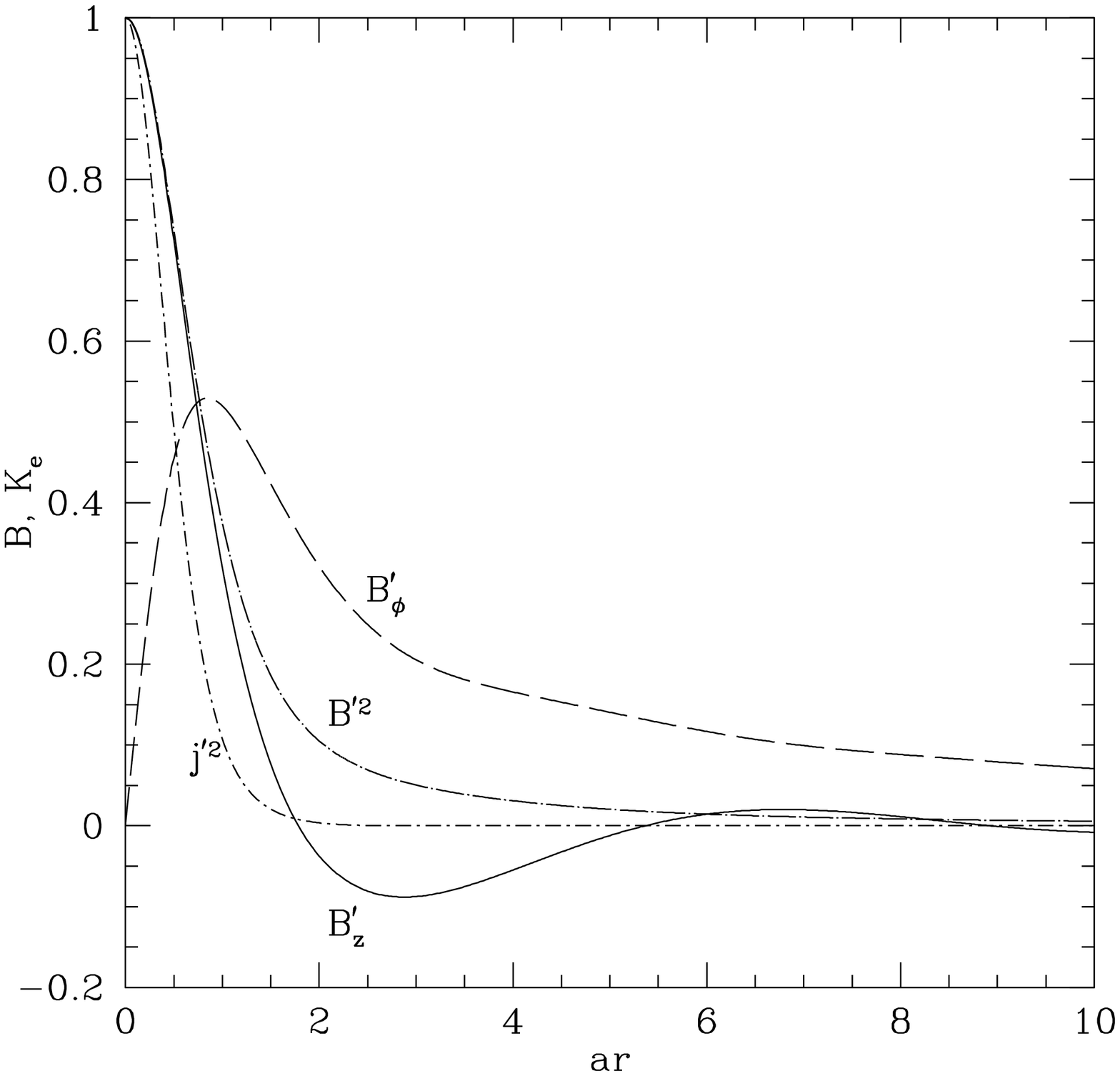}
\includegraphics[width=0.3\linewidth]{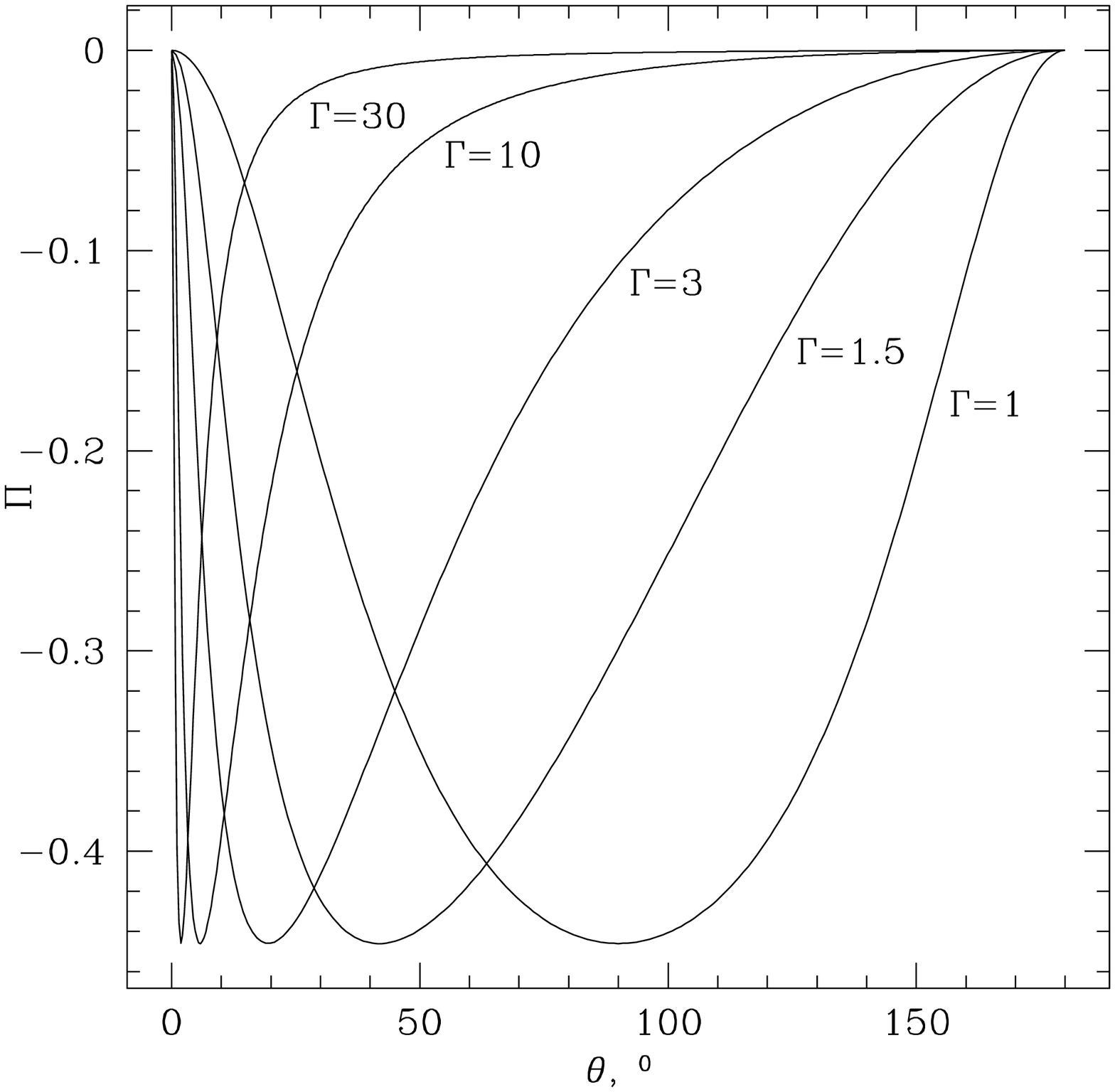} 
\includegraphics[width=0.3\linewidth]{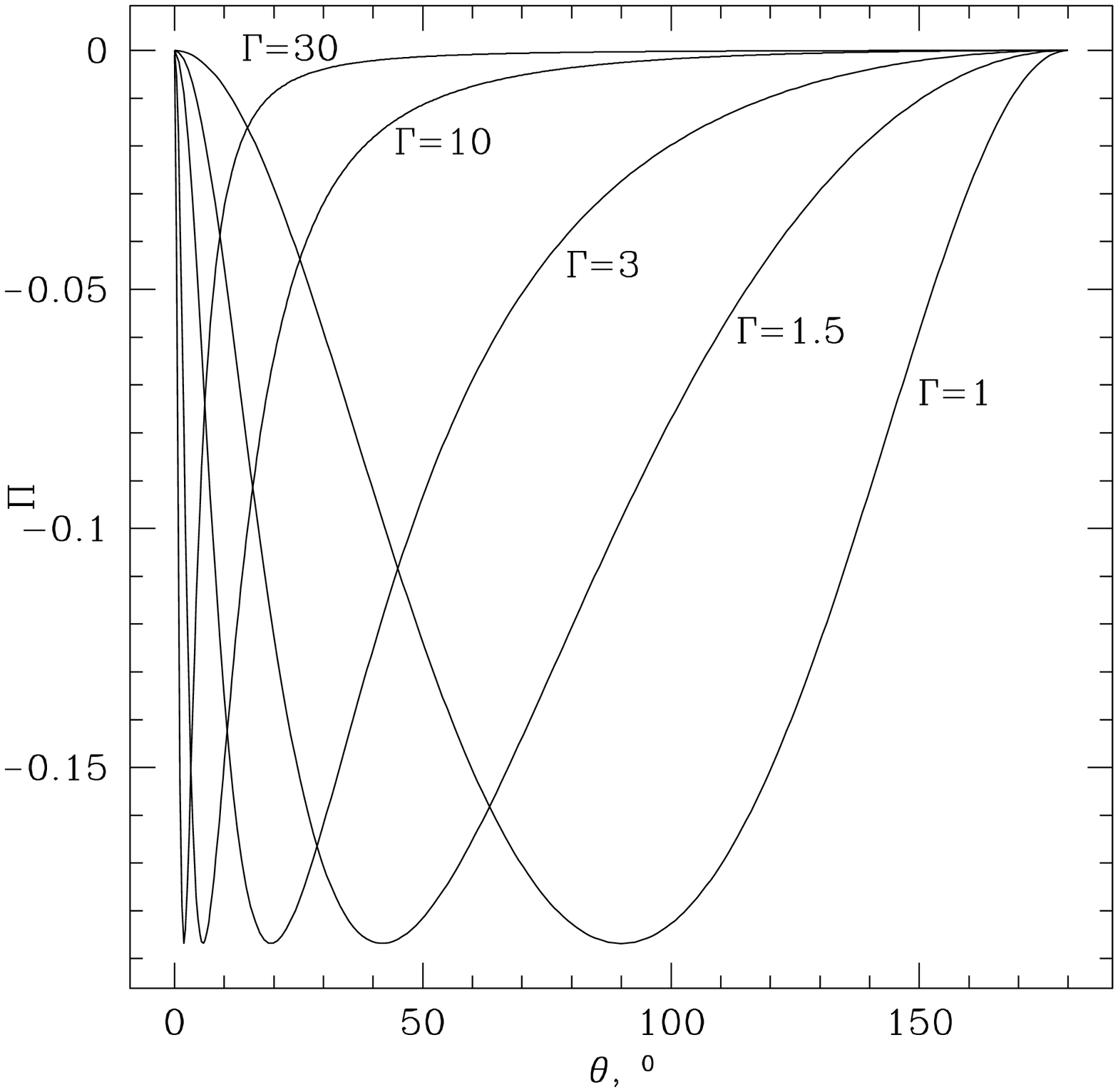}
\vskip .01 truein
\hskip 1 truein (a) \hskip 1.8 truein (b) \hskip 1.8 truein (c) \\
\caption{Polarization for a pinch with multiple axial
field reversals for the fields given by 
Eqns.~(\ref{eqn_4.3.1}-\ref{eqn_4.3.2}).  
Notations are the same as in Fig. \ref{myfig1}.
(a) Radial profiles. (b) Degree of polarization for $p=2.4$ for the 
density of relativistic particles $K_{\rm e} \propto j^{\prime 2}$.
(c) Degree of polarization for $p=2.4$ for the 
density of relativistic particles $K_{\rm e} \propto B^{\prime 2}$.
}
\label{oscill}
\end{figure}

As a further illustration, we consider the possibility that the axial
magnetic field in the jet has multiple reversals. We also impose
the condition that the total flux of the axial magnetic field
vanishes, $\int_0^{\infty} B_z rdr = 0$. This situation may correspond
to a magnetic field produced as a result of amplification by the
magnetorotational instability \citep{balbus91,balbus98} or a magnetic dynamo
\citep{pariev04} operating in the accretion disk. The loops
of magnetic field can rise above the accretion disk, reconnect, and form
multiply reversed magnetic field structure carrying no total poloidal
flux. We choose the following form of the axial magnetic field in
the frame comoving with the jet:
\begin{equation}
B_z^{nr}=B_0 \frac{\cos\lambda a r}{(1+a^2 r^2)}\label{eqn_4.3.1}\mbox{,}
\end{equation}
where $\lambda=0.879$ is chosen to satisfy the condition
$\int_0^{\infty} B_z rdr = 0$. $B_\phi^{nr}$ is then found by integrating
numerically the nonrelativistic force-free equilibrium equation.
$B_{\phi}^{nr} \propto 1/r$ for $r\gg a$. The toroidal magnetic field and
electric field in the laboratory frame can be calculated from
Eq.~(\ref{b}).The current density in the frame comoving with the jet
is calculated as
\begin{equation}
j^{\prime}_{\phi}=\frac{2a^2 r \cos(\lambda ar)}{(1+a^2 r^2)^2}
+\frac{\lambda a \sin(\lambda a r)}{1+a^2 r^2}\mbox{,} \quad
j^{\prime 2}=j^{\prime 2}_{\phi} \left(1+\frac{B_z^{nr 2}}
{B_{\phi}^{nr 2}}\right)
\label{eqn_4.3.2}\mbox{.}
\end{equation}
The radial dependencies of the fields in the frame comoving with the jet,
$B_z^{\prime}(r)$, $B_{\phi}^{\prime}(r)$, and the two prescriptions
for the number density of emitting particles, $K_{\rm e} 
\propto j^{\prime 2}$
and $K_{\rm e}\propto B^{\prime 2}$, are shown in Fig.~\ref{oscill}a.

We perform the integration of the Stokes parameters for a constant emissivity
from $r=0$ to large radii, $r\gg 1/a$.
The results are presented  in Fig.~\ref{oscill}. The polarization vector is
perpendicular to the jet axis. The profile of $j^{\prime 2}(r)$ is more
concentrated toward the inner core of the jet than the profile of 
$B^{\prime 2}$. Therefore, the degree of polarization in the case 
$K_{\rm e}\propto j^{\prime 2}$ is lower (more orthogonal to the jet axis) 
than the degree of polarization in the case $K_{\rm e}\propto B^{\prime 2}$.
This is true for all three cases of force-free fields in the ``filled'' 
jets considered in this section.

\subsection{Emission from ''filled'' jets without shear: conclusion}

The three  examples of jet profiles and two prescriptions for  emissivity  
we have considered illustrate that the polarization remains  perpendicular 
to the jet axis in all cases.  Parallel polarization cannot be 
produced in models in which the energy density of the relativistic 
particles is $\propto j^{'2}$.
In the case when the energy density of the emitting particles is
$ \propto B^{'2}$, the degree of polarization is small. The parallel 
polarization observed in some (many) sources
cannot be explained using models in which the relativistic particles
are distributed across the jet. 
The reason is that the current distribution and magnetic fields are 
highest on the axis,  where the magnetic field in the jet rest frame
is dominated by the poloidal field, so that the average emission is
dominated by regions with large poloidal field.  
Thus, in order to produce 
parallel polarization {\it the emitting particles must be concentrated
in the regions of stronger toroidal field}, closer to the edges of the jet.
\footnote{More precisely, the emission-weighted average pitch angle in the jet
rest frame must not be small. In the case of force-free jets, this can occur only not
too close to the jet axis.} 

The results of this section  indicate the possibility that {\it  particle
acceleration occurs selectively at the periphery of the jet}.
Another possibility are strongly sheared jets, in which the emission 
from the central parts (which are dominated by 
poloidal magnetic field
in the comoving frame) is beamed away from us. 
We investigate this possibility next.

\section{Emission from sheared jets}
\label{sheared}

The jets considered in the previous section were all non-sheared, so that
the degree of polarization can be derived in their rest frame and then 
the viewing angle transformed using the light-aberration formulas. 
In this section, we consider sheared
jets,  when the axial bulk velocity 
 is a function of radius in the jet.   
In this case, there is no reference frame in which the entire jet is 
at rest. The fully relativistic
approach for calculating the Stokes parameters for an arbitrarily moving
medium described in Section~\ref{sec_2} is needed. 
As a model problem, we assume that the jet
internal structure corresponds to a diffuse pinch 
(Section~\ref{diffuse}). In addition, 
we chose the 
following scaling of the Lorentz factor with radius:
\be 
\Gamma = 1+ { 1 \over 1+( a r)^2} (\Gamma_0-1)\mbox{,}
\label{eqn_gammar}
\ee
so that the axis of the jet moves with Lorenz factor $\Gamma_0$. 
Then, we follow the procedure for constructing the equilibrium of
a relativistic force-free jet given in Appendix~\ref{FF}.

\begin{figure}[ht]
\plottwo{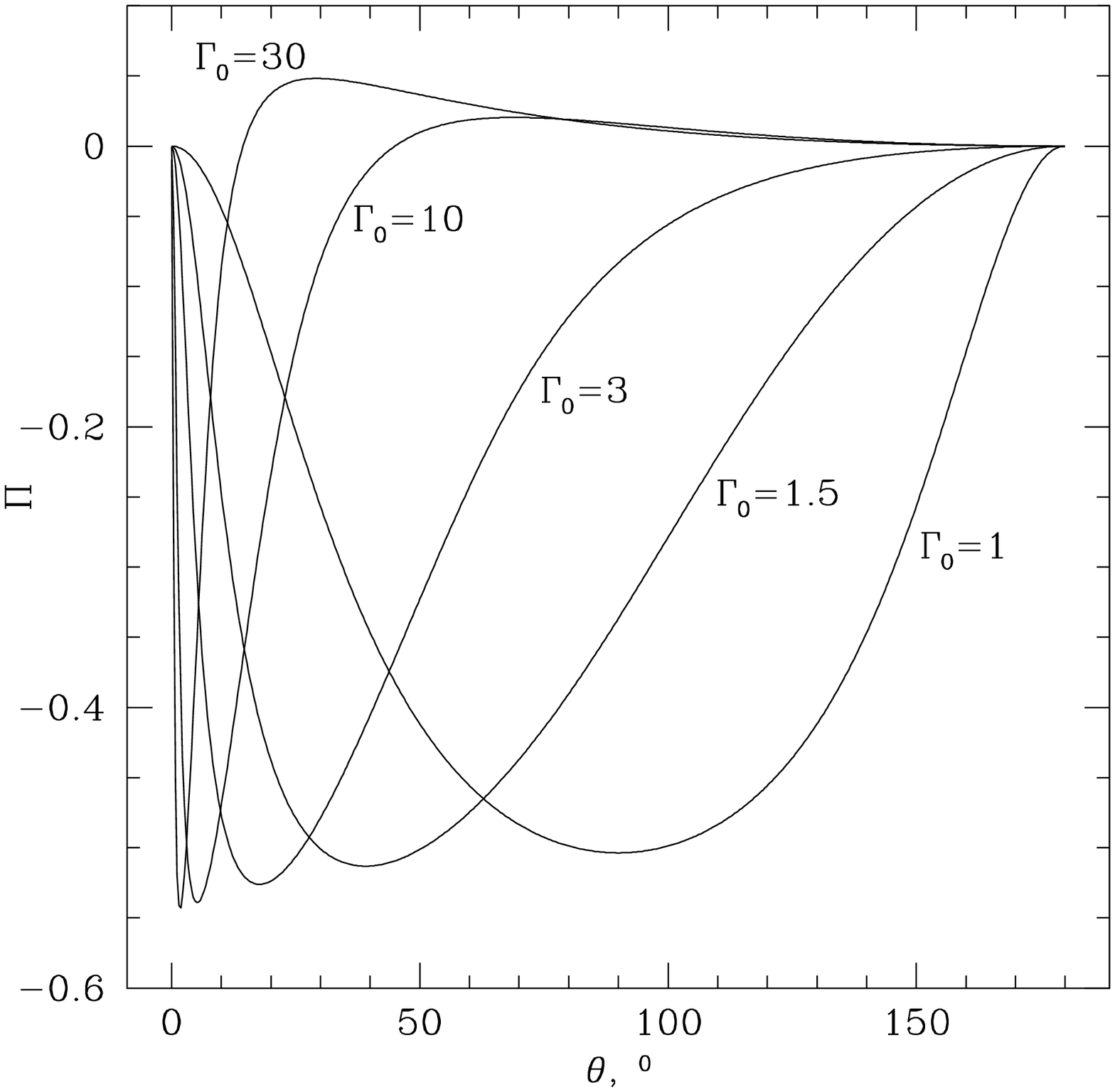}{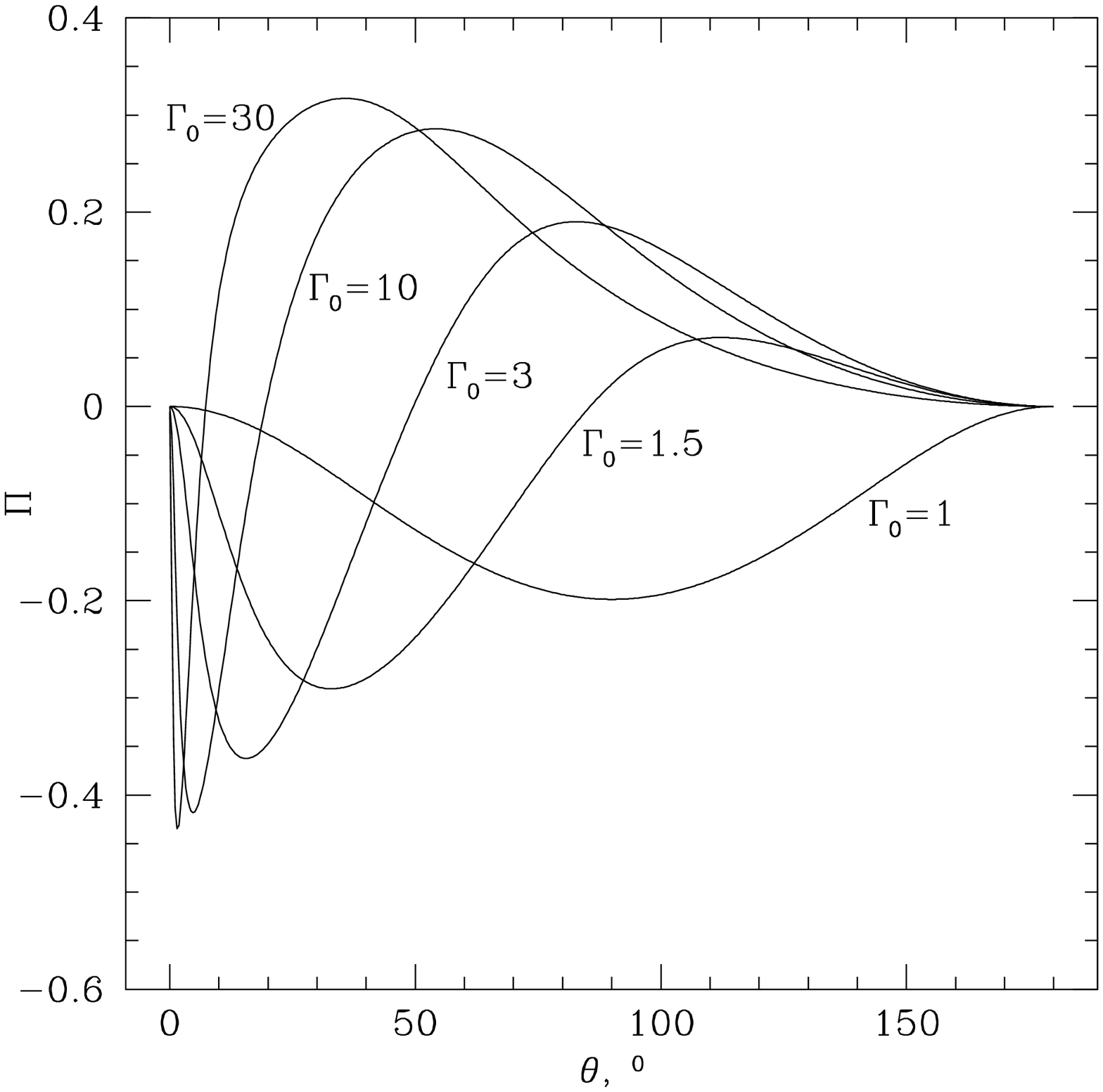}
\vskip .01 truein
\hskip 0.01 truein (a) \hskip 2.5 truein (b) \\
\caption{Polarization for the sheared diffuse pinch (for discussion see
Section \ref{sheared}) for $p=2.4$ integrated over the whole jet. 
(a) The density of relativistic particles  $K_{\rm e}\propto j'^2$,
(b) The density of relativistic particles  $K_{\rm e}\propto B'^2$.
}
\label{jetshear}
\end{figure}

As in Section~\ref{filled}, we investigate two example cases for the 
energy density of the relativistic emitting particles:
$K_{\rm e} \propto {\bf j}^{\prime 2}$ and
$K_{\rm e} \propto B^{\prime 2}$ (Fig.~\ref{jetshear}). 
Since there is no global reference frame in which the whole jet is
at rest, the current density in the local comoving frame 
${\bf j}^{\prime}$ is now, in general, not equal to the curl 
of ${\bf B}^{nr}$. To calculate ${\bf j}^{\prime}$, we first calculate
the current density ${\bf j}=\nabla\times{\bf B}/(4\pi)$ and charge density
$\rho_{\rm e}=\nabla\cdot{\bf E}/(4\pi)$ in the laboratory frame.
Then, we make a Lorentz transformation of the current 4-vector to the 
local comoving frame of the plasma element to obtain ${\bf j}^{\prime}$.
The result is that still ${\bf j}^{\prime}=\nabla\times{\bf B}^{nr}/(4\pi)$,
i.e., terms containing $d\Gamma/dr$ cancel out in the expression
for ${\bf j}^{\prime}$. The reason for this cancellation is that the 
radial force balance is not broken by the introduction of the shear in 
the axial flow in the jet (see Appendix~\ref{FF}).

At small viewing angles $\theta \leq 1/\Gamma_0$, 
the emission is dominated
by the central regions, where the magnetic field in the rest frame of
the moving plasma
is mostly poloidal, so the polarization is perpendicular to the jet axis. 
At larger viewing angles $\theta \geq 1/\Gamma_0$, the emission from 
the  central parts of the jet is beamed away, so that the average 
polarization may be along the jet. 

\section{Electromagnetically dominated jets}
\label{EMjet}

In this paper, we present polarization calculations that indicate that
large-scale magnetic fields may be responsible for the
polarization observed in parsec-scale jets in AGNs. In order to produce the 
substantial degrees of polarization that are observed in these jets, the
total energy density of the ordered component of the magnetic field must
be at least comparable to the random component.
This raises the question of how strong the magnetic fields in AGN jets can be.

Below we discuss the 
 possibility, that the energy is transported along the jets
mainly in the form of the Poynting flux. 
It may be possible to 
overcome the difficulties of pure electron--positron beam models 
as well as ion-dominated models.
 Qualitatively, electromagnetic jets
(or, rather, magnetic helices) will be toroidally dominated
in the observer frame on 
parsec scales due to the conservation of the poloidal magnetic 
flux and the poloidal current. As we have shown in this paper, this 
is consistent with observations. If a jet has a net poloidal current,
the self-pinching of this current will tend to collimate it.
In turn, the poloidal currents can become unstable 
to  the excitation of anomalous resistivity and the
subsequent development of resistive instabilities. The
result will be the dissipation of poloidal  currents or, equivalently, of
the toroidal magnetic field. The dissipated energy can go primarily into the
acceleration of electrons, similar to the case of solar flares.  Thus, {\it 
a current that contributes to the jet collimation provides a natural source
of energy for the re-energization of the jet via magnetic 
dissipation} (\eg reconnection).

If the jets are magnetically dominated, the energy  ultimately comes from
the rotational energy of the central source (disk or black hole).
It is first converted into
magnetic energy by the dynamo  action of the  unipolar inductor 
(e.g., via the Blandford--Znajek mechanism),
propagated in the form of a Poynting-flux-dominated flow
and then dissipated by current instabilities at large 
distances from the sources.

Let us summarize the advantages of the magnetically dominated
model for the energy transport in AGN jets  \citep[see also][]{bla02,Bolog}.
\begin{itemize}

\item {\bf ``High quality''.} 
The energy stored in the low entropy electromagnetic
outflow has a  ``high quality''. It can be efficiently converted into 
high-frequency electromagnetic
radiation {\it far from the source}. For example,
recent RHESSI and TRACE observations of the Sun indicate that the primary 
energy output in reconnection
is in the form of {\it nonthermal electrons}, and only a small fraction 
goes into heating and bulk motion \citep{benz03}.

\item {\bf Variability}  
Recent observations of high-energy  emission from blazars with very short
variability time scales (as short as 20 minutes in Mrk 421, \cite{gaid96,cui04})
have stressed again the need for {\it in situ}
acceleration, since the synchrotron cooling times for the X-ray and especially
$\gamma$-ray emitting electrons are an order of magnitude shorter than
the light-travel time from the core (and often shorter than the light-crossing
time of the emitting region itself).

Magnetic fields are strongly nonlinear systems: slow evolution
during which magnetic stresses build up can lead to the accumulation
of a large amount of free energy, which is 
released in explosive events on a short time
scale (of the order of the \Alfven time scale) as the system crosses the
stability threshold. Such events happening in the central engine, close 
to the black hole and accretion disk, could be the source of  
flaring events that result in the emergence of new bright knots.
Small-scale reconnection events could produce quasi-steady high-energy 
emission, analogous to the micro-flare paradigm
for the quiescent emission of the solar corona (e.g., \citealt{benz98, 
krucker00}). In addition, variations in the surrounding 
medium may lead to restructuring of the jet,
accompanied by dissipation \citep{ck86}. 
The magnetic field dissipation can be internal, and does not necessarily lead
to the global disruption of the system, as happens, 
for example, in sawtooth oscillations
in TOKAMAKs (e.g., \citealt{kad75}).   Current-driven 
instabilities in jets may proceed in a similar way \citep{Appl00}.

\item {\bf Particle acceleration}
Dissipation of the magnetic field leads to particle acceleration that produces
a power-law distribution of accelerated particles, as is exemplified by
solar flares, TOKAMAK measurements, and laboratory Z-pinches.
However, the physics of magnetic dissipation (with reconnection
being the prime example) is highly uncertain: it depends critically on the
kinetic and geometric properties of the plasma, which are very hard to
measure observationally. This situation can be contrasted with shock
acceleration schemes, where a qualitatively correct result for the spectrum
of accelerated particles (a kinetic property!) can be derived from simple
{\it macroscopic } considerations.

In spite of these difficulties, acceleration 
(or pre-acceleration) of electrons in some type of reconnection
layers may provide an alternative possibility 
(or the first stage) for shock acceleration.
First, we know that it works on the Sun. Second, magnetic dissipation
 can accelerate particles from a thermal
bath, without a need for pre-acceleration. 
(Recall that only electrons with $\gamma_e \geq \Gamma_s m_p/m_e$ 
can be 
accelerated at a relativistic shock; here $\Gamma_s$ is the shock Lorentz
factor, and $m_p$ and $m_e$ are the proton and electron masses). 
Third, acceleration in reconnection layers can 
produce very hard power-law spectra, $dn/d\gamma \sim \gamma^{-1}$, 
\citep{hos02,llm02}.
The fact that reconnection models can produce spectra 
which are prohibitively hard for shock acceleration can serve as 
a distinctive property of electromagnetic models.

\item {\bf Large scale stability and small scale dissipation}.
Relativistic electromagnetic jets are expected to be less susceptible to 
instabilities due to interaction with the external medium, 
as well as internal instabilities.
In the case of interactions with an external medium, the magnetic field 
has a stabilizing effect on the surface modes, especially 
if there is longitude shear and rotation of the jet \citep{istpar94,istpar96}.
On the other hand, the presence of velocity shear would lead  
to the formation of critical layers
where dissipative effects become important \citep{pariev03}. 
Thus, rotation and shear can lead to large-scale
stabilization and small-scale de-stabilization of the jet. The
energy released due to small-scale 
dissipation could be used to power the jet luminosity.

\item {\bf Knots}.
AGN jets have distinct structures along the
jets in the form of knots, fainter bridges of emission, etc. These structures
could be regions of enhanced dissipation, in which case they are likely to
break cylindrical and/or translational symmetry along the jet; 
alternatively, they could be the jet structures that emit preferentially in our
direction. Quasi-periodic bright knots are observed in the jets of many
AGNs, such as M87 \citep{marshall02}, 3C273 \citep{marshall01}, and
4C~19.44/1354+195 \citep{sambruna02}.  An intriguing possibility 
is that the quasi-periodic bright knots are associated with
large-amplitude waves propagating in the jet \citep{istpar96, pariev03}.
These waves can be excited due to explosive reconnection events 
at the base of the jet. Alternatively, accretion disk instabilities
and the evacuation and reformation of parts of an accretion disk 
could be the source 
of these large amplitude waves \citep{lovelace94, lnr97}.
The propagation of such waves in a radially inhomogeneous jet 
can be weakly dissipative, providing a means to energize particles 
in the jet \citep{beresnyak03}. 
\end{itemize}

At present, hydrodynamical models of AGN jets are much better developed 
\citep[\eg][]{hardee03}. It remains to be seen if 
electromagnetic jets are able to reproduce equally well 
the large-scale dynamical behavior. 
In addition, radiation modeling of AGN spectra allowing for 
a two-zone structure of the magnetic field
(low inside the acceleration region, high in the bulk) needs to be done.

\section{Discussion and Conclusions}
\label{discussion}

We have analyzed the polarization properties of the synchrotron emission
of relativistic cylindrical jets with helical magnetic fields, taking 
into account the relativistic motion of the jet. We considered uniform
radial profiles of the velocity of the jet as well as sheared velocity 
profiles in which the velocity is higher at the center of the jet and 
decreases to zero at the outer boundary of the jet.
We have explored a number 
of fairly general magnetic-field structures and emission profiles,
on which we base the following conclusions.
\begin{enumerate}
\item We stress again that, in case of a relativistically moving
optically thin jets, the observed direction of the polarization {\it is not
generally orthogonal} to the projected direction of the magnetic field. We
accordingly encourage observers to always plot the direction of the electric
field of the wave, not the ``inferred direction of the magnetic field.''
It is important to employ a correct Lorentz transformation of the polarization,
regardless of the model considered: calculations of the polarization from
internal shocks must take this effect into account as well.
\item To produce polarization along the jet axis, the emissivity weighted
pitch angle of the magnetic field in the jet frame cannot be small. 
This can be achieved
either (i) if a jet has a pinch configuration with no, one or many reversals
{\it and} the jet emissivity is confined to a narrow region of radii where
the toroidal magnetic field is of order of the poloidal field in the jet
frame, or (ii) if the velocity of the jet changes with radius such that 
the relativistic beaming is maximum in a narrow range of radii closer
to the outer boundary of the jet, where the toroidal magnetic field is of 
order of the poloidal field in the jet frame. 
\item The polarization electric vector position angle (EVPA) 
integrated over the jet cross section shows a
bimodal distribution, being directed either along or orthogonal to the jet, 
in accordance with observations \citep{caw93,gab00}.
\item Strongly beamed jets (viewed at $\theta \leq 1/\Gamma$) can  have
polarization directed along the jet axis only if the toroidal magnetic field
strongly dominates over the poloidal field in the observer's frame,
$B_\phi / B_z \geq \Gamma \gg 1$. This will be true if the toroidal field
is at least comparable to the poloidal field in the jet frame. In this case,
we have in the jet frame $B_\phi ' \sim B_z '$, so that the jet is expected
to be dynamically stable to pinches and kinks. When an orthogonal jump
of the polarization position angle is seen along such jets, we can estimate
 $B_\phi / B_z \sim \Gamma \gg 1$. 
The occurrence of orthogonal jumps of polarization
should be more prominent in sources with flatter spectra. 
\item In resolved jets, 
the asymmetry of the polarization structure  
(\eg with respect to the intensity profile)
can be used to determine whether the spin axis of the central object 
is aligned or anti-aligned with the jet axis.
\item In a resolved cylindrical shell with  $B_\phi ' \sim B_z '$, 
the polarization in the central parts is along the jet, while the 
polarization at the jet edges is orthogonal to the jet axis, in accordance 
with observations \citep{att99, push02, push04}. This provides further 
evidence
that the jet emissivity is confined to a narrow region of radii at the 
periphery of the jet, where the toroidal magnetic field is of the order of the 
poloidal field in the jet frame.
\end{enumerate}

\begin{figure}[ht]
\includegraphics[width=0.9\linewidth,angle=270]{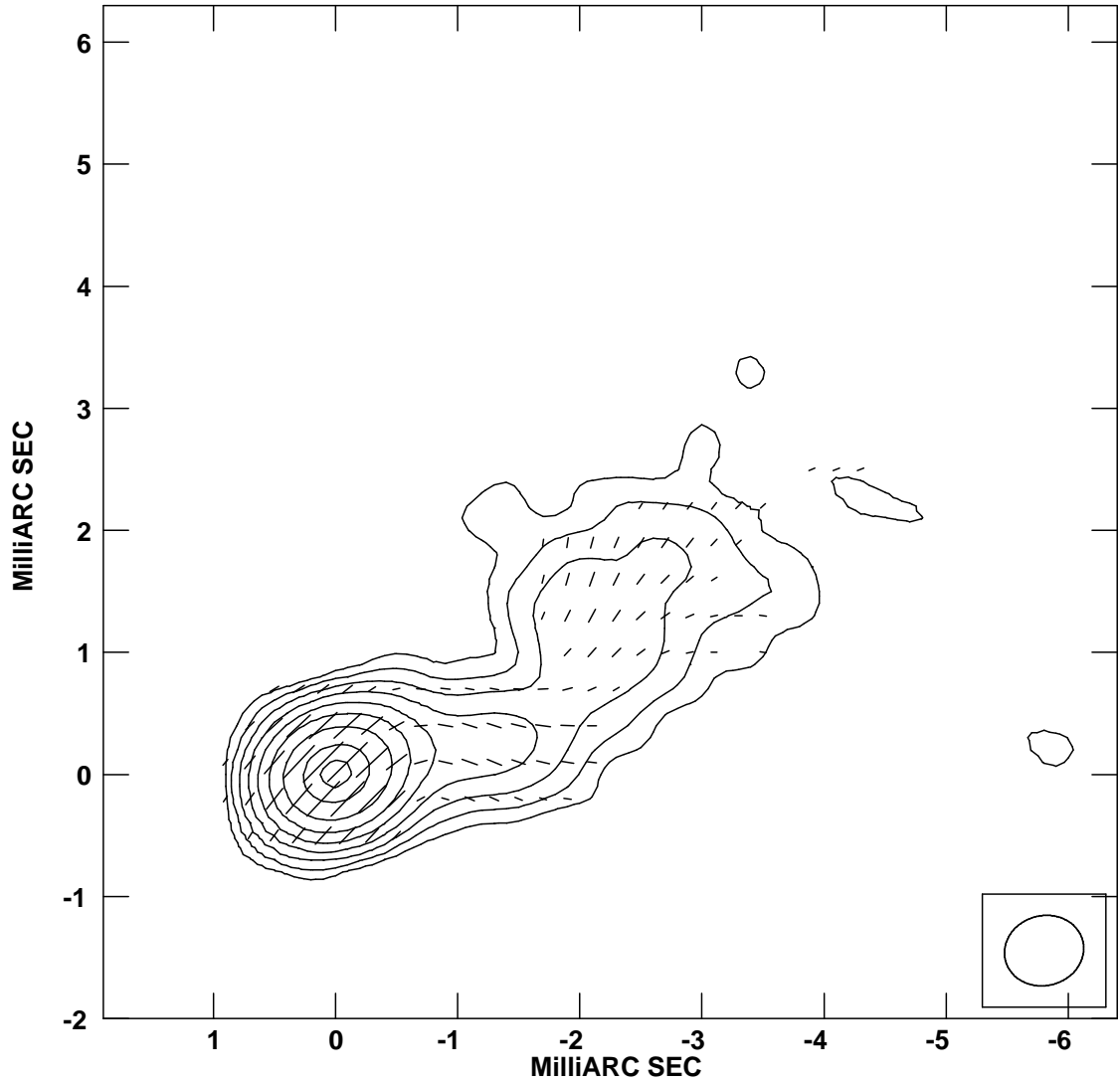}
\caption{Observed 4~cm total-intensity contours and superposed
linear-polarization vectors for the BL~Lac object 1749+701: the polarization
remains aligned with the jet direction as the jet bends (reproduced from
\protect\citealt{gab03a}; data processing is described in 
\protect\citealt{gpg01}).
}
\label{follow_1749}
\end{figure}

These results support the
possibility that the characteristic behavior of the polarization of
AGN jets arises precisely because these jets carry helical magnetic
fields. Additional support for this scenario comes from the observations of
differential Faraday rotation across the jets of BL~Lac objects, which can
be understood as a natural consequence of a toroidal or helical magnetic-field
configuration \citep{gab04}, and the evidence that the circular
polarization of AGN cores display a constant sign over long intervals
\citep{haw01}. Both of these last two effects argue in favor of the dominant
role of large-scale magnetic fields in the jets.  It is likewise not difficult
to account for possible exceptions to the general tendencies stated above.
For example a non-axisymmetric structure of the jet can give rise to a
relative orientation of the EVPA relative to the jet that differs from 
0 or $90^{\circ}$ (there are theoretical grounds for non-axisymmetric 
jets, \citet{ck86}).
The curved jet in Fig.~\ref{follow_1749} may be viewed close to its axis,
so that it may be represented by two quasi-cylindrical pieces smoothly 
changing direction at a merger point. Then, the resolved and unresolved 
EVPA will be approximately aligned with the axis of the cylindrical pieces 
because of the bimodality described above. The largest deviation of 
the EVPA from such alignment is expected to happen at the point of the 
jet bending. This is indeed the case for the EVPA plotted on 
Fig.~\ref{follow_1749}: at the location of the bend, the EVPA makes an 
angle of almost $45^{\circ}$ with the local direction of the middle 
axis of the brightness distribution there.

The conditions for having appreciable longitudinal polarization in a 
relativistic jet are more restrictive than those for orthogonal polarization. 
This is a purely geometrical effect. For a given jet structure, the highest 
degree of longitudinal polarization will be observed when the jet is 
viewed at the angle $\theta \sim 1/\Gamma$. In this case, the line of sight is
orthogonal to the jet axis in the jet rest frame. It is thought that,
statistically, $\theta \sim 1/\Gamma$ for the pc-scale jets of
core-dominated AGN, while most kpc-scale jets
are viewed at angles $\theta \gg  1/\Gamma$ in the observer's rest frame,
so that the line of sight makes a small angle with the jet axis in the 
comoving frame. Therefore, the observed kpc-scale jets should have
orthogonal polarization more frequently than pc-scale jets.
This stresses again
that, {\it due to  relativistic kinematic effects, it is not possible to make
a firm conclusion about whether a jet is dominated by toroidal or poloidal
fields based purely on the observed direction of polarization. }

Another implication of our calculations is that the polarization properties
of an intrinsically similar jet and counter jet can be very 
different (if the counter jet is detected), due
to the different angles that the line of sight makes with the jet magnetic
field in the two cases. It follows from Fig.~\ref{polariztot} 
that, in the counter jet,
high degrees of polarization, $\geq 10\%$, can be directed only orthogonal
to the jet axis, while low degrees of polarization can be either longitudinal
or orthogonal.
In the strongly relativistic limit $\Gamma \gg 1$, jets viewed at
angles $\theta = C/\Gamma$ and $\theta_{ob} = 1/(\Gamma C)$, where 
$C$ is some constant, $C \ll \Gamma$, should have the same 
linear polarization properties (in this case, the lines of sight make 
the same angles with the jet axis in the jet rest frame).

The typical degrees of polarization observed in compact AGN are fairly low,
of the order of a few to ten or fifteen percent. 
Modest polarization is observed
from any jet that has comparable toroidal and poloidal  fields in the jet rest
frame, $B'_\phi \sim B'_z$. (Such relatively weakly polarized relativistic
jets are still strongly toroidally dominated in the observer's frame.) 
The condition  $B'_\phi \sim B'_z$ can be reached self-consistently: 
over-expansion of a jet would lead to an increase of the ratio 
$B'_\phi / B'_z$, making the jet more likely to be unstable. 
The development of instabilities would
lead to the dissipation of poloidal current and a decrease in 
$B'_\phi /B'_z$ \citep{colgate98}. Calculations of relativistic
force-free expanding helices produced by the twisting of the magnetic field
lines by the accretion disk also indicate that $B'_\phi \sim B'_z$
in the inner collimated part of the expanding magnetic helix
\citep{lovelace02}.
Thus, if the modest observed degrees of polarization are indeed due to
the presence of comparable toroidal and poloidal  fields in the jet frame,
this would imply that {\it all relativistic jets are dominated
by the toroidal magnetic field component in the observer's frame}.

Toroidally dominated jets are also expected on theoretical grounds.
Suppose that AGN
jets are launched from a black-hole--disk system with large-scale poloidal
and toroidal fields at the surface of the disk.  
During the  acceleration of the flow by magnetic
stresses, the flow expands so that the ratio of $B_\phi$  
to $B_z$ will increase approximately linearly with cylindrical 
radius. If the jets are toroidally dominated on a cylindrical scale
$\varpi  \sim 0.3$ pc with $B_\phi/B_z \sim \Gamma \sim 10 $, this implies 
that the \Alfven point (where $ B_\phi/B_z \sim 1$) is 
located at $ \varpi \sim 10^{17} $~cm, so that the ratio of the
fields is $ B_\phi/B_z \sim 0.1$ at the outer edge of the disk, 
located at  $\varpi \sim 500  R_G \sim  10^{16}$~cm 
($R_G \sim 3 \times 10^{13}$~cm  is the Schwarzschild radius of 
the supermassive black hole). Further in, the pinch angle depends on the 
details of the EMF distribution in the disk (\eg Blandford 1976). 
This helps to reconcile the strong toroidal dominance in the jets 
expected on theoretical grounds with the observed EVPA orthogonal 
to the jet axis (as in Fig.~\ref{resolved_1652}).

In this paper, we have considered jets that have achieved asymptotic
cylindrical collimation. Close to the central object, the jets are often
observed to be conically divergent. Though a separate set of calculations
for conical jets are necessary, qualitatively, we expect that the ratio
of the toroidal and poloidal fields increases with distance, so that the
jets are likely to have polarization along the jet axis at larger distances.
This type of behavior (predominance of longitudinal polarization at larger
distances) is seen in some cases \citep{dmm00,gab01a}.

\acknowledgements

VP acknowledges support from DOE grant DE-FG02-00ER54600.

\appendix

\section{Comparison with \citet{blandford79}}
\label{BK}

For comparison with previous calculations  \citep{blandford79},
we adopt a frame aligned with the direction of motion.
In this frame  (quantities measured in this frame are 
labeled with the subscript ${BK}$),
\ba &&
\B_{BK} = \{ \cos \eta_{BK} \sin \psi'_{BK}, - \sin \eta_{BK},
\cos \eta_{BK} \cos \psi'_{BK}\}
\nn &&
{\bf n} =  \{ \cos \theta_{BK} , 0, \sin \theta_{BK} \}
\nn &&
{\bf v} = \beta \{1,0,0\}
\nn &&
{\bf e}_{BK} = \{- \cos \xi_{BK}  \sin \theta_{BK}, \sin  \xi_{BK},
\cos \xi_{BK} \cos \theta_{BK}  \} \mbox{.}
\ea
We find  from (\ref{eee})
\be
\tan \xi_{BK} = \cot \eta_{BK} { \cos( \theta_{BK}+ \psi'_{BK}) - \beta  \cos \psi'_{BK}
\over
1- \beta \cos \theta_{BK}}  \mbox{,}
\ee
reproducing Eq.~(16) in \citet{blandford79}.
The angle between ${\bf \hat{B}}$ and ${\bf \hat{e}}$ is
\ba &&
\cos \zeta = ({\bf \hat{B}} \cdot {\bf \hat{e}}) =
\cos \eta_{BK} \cos \xi_{BK} \cos(\theta_{BK} +\psi'_{BK}) - \sin \eta_{BK} \sin \xi_{BK}=
\nn &&
{ \beta \cos \eta_{BK} \sin \theta_{BK} \sin ( \theta_{BK}+ \psi'_{BK})
\over \sqrt{(1- \beta \cos \theta_{BK} )^2
+ \cot^2 \eta_{BK} ( \cos (\theta_{BK} +\psi'_{BK}) - \beta \cos \psi'_{BK})^2 }}
\ea

\section{Relativistic force-free jets}
\label{FF}

Conventionally, the structure of relativistic force-free jets has been
considered in terms of the relativistic Grad--Shafranov equation, 
which for cylindrical geometry becomes  a second-order ordinary
differential equation  
\citep[\eg][]{applcam93}. Only few exact solutions have been
found \citep{applcam93,istpar94}.
The structure of relativistic  jets (pinches) is more complicated than
that of non-relativistic jets. In addition to the 
poloidal and toroidal magnetic fields,
a relativistic jet has a radial electric field, which can be of the same order
of magnitude
as the total magnetic field, and a charge density, which can approach
$\rho_{\rm e} \sim j/nec$ (so that  electric forces are comparable to 
the magnetic forces, $\rho_{\rm e}  E \sim j B$).

In a force-free approximation \citep{michel73,uch97,kom02,lb03}, 
it is assumed that inertia and pressure forces
are not important, so that the force balance is determined only by
electromagnetic forces
$
\E {\nabla \cdot {\bf E}}/{4\pi} + {\bf j } \times \B=0\mbox{.}
\label{rff1}
$
This, combined  with Maxwell equations
and the ideal condition
${\bf E} \cdot {\bf B} =0 $,
allows one to express the current in terms of the fields
\be
\label{relohm}
{\bf j}={({\bf E}\times{\bf B})\nabla\cdot{\bf E}+
({\bf B}\cdot\nabla\times{\bf B}-{\bf E}\cdot
\nabla\times{\bf E}){\bf B}\over 4\pi B^2}\mbox{.}
\ee
In addition, it  is assumed that $B^2-E^2 >0$ (this ensures that 
there is a frame in which the electric field is zero).
It is possible to  define an electromagnetic velocity, which is equal to the 
plasma velocity across the magnetic field
\be
{\bf \beta}_{EM} = { \E \times \B \over B^2}\mbox{.}
\ee
The velocity along the magnetic field is not defined and can be chosen 
arbitrarily, subject to the physical constraint that the total velocity 
is lower that the speed of light.

For stationary  cylindrical jets, the above relations
imply that the non-zero components of the electromagnetic fields are
$B_\phi$, $B_z$ and $E_r$, which are functions of $r$ only.
There is a charge density
$\displaystyle \rho = \frac{\nabla\cdot{\bf E}}{4\pi} = 
\frac{\partial_r (r E_r)}{4\pi r}$ and  current
$\displaystyle {\bf j} = \frac{1}{4\pi} \nabla\times {\bf B}$.
 We find then  \citep{istpar94}
\be
B_z^2 = { 1 \over 2 r} \partial_r r^2
\left( B_z^2 + B_\phi^2 -E_r^2 \right) \label{eqn_cylrel}\mbox{.}
\ee
This is an equation describing the equilibrium of a {\it relativistic} 
force-free jet.
It reduces to the non-relativistic force-free equilibrium if $E_r=0$.
Next, we note that  both sides of this equation
are unchanged under a Lorentz boost along $z$:
($B_z^2 + B_\phi^2 -E_r^2={\bf B}^2 - {\bf E}^2$ is a relativistic
invariant and $B_z$ does not change under a boost along
$z$). If $|E_r(r)| < |B_\phi (r)|$, it is possible to choose the 
velocity of a radius-dependent Lorentz boost $\beta(r)$ such that
the electric field will vanish. Then, the magnetic fields $B_\phi^{nr}(r)$
and $B_z^{nr}(r)$ in the frame where $E_r=0$
will satisfy the non-relativistic force-free jet equilibrium.
This implies that any solution for a {\it relativistic} 
force-free jet with $|E_r(r)| < |B_\phi (r)|$
can be obtained from a known solution for a non-relativistic
pinch, $B_z^{nr} (r) $ and $B^{nr} _\phi(r)$, 
by making a {\it formal radius-dependent
boost} along the $z$ axis:
\be
E_r = \beta \Gamma B_\phi^{nr}, \, B_\phi = B_\phi^{nr}  \Gamma
,\, B_z=B_z^{nr}\mbox{,}
\label{b}
\ee
where $\beta \equiv \beta (r)$ and $\Gamma =1/\sqrt{1 - \beta^2}$.
We stress that this transformation is not a physical  Lorentz transformation --
it can have a radius-dependent $z$ velocity, so that $\beta$ and
$\Gamma$ are functions of $r$. 

This method does not allow us to find solutions 
when $|E_r(r)| \geq |B_\phi (r)|$ (but $|E_r(r)| < |B(r)|$). 
In the latter case, there is no
local frame where $E_r$ would be zero. The equation~(\ref{eqn_cylrel}) 
can be integrated to give
\be
\int_0^r r'^2 \frac{d B_z^2(r')}{dr'} \, dr' = r^2 (E_r^2(r) - B_{\phi}^2(r))
\label{eqn_intrel}\mbox{.}
\ee 
One can see that, if the poloidal magnetic field $B_z$ is a monotonic decreasing
function of radius, then $|E_r(r)| < |B_\phi (r)|$ for any $r$ and 
such solutions can be obtained using the method described here. 
If $dB^2_z/dr \geq 0$ for some $r$, then our method may not find all solutions 
having such $B_z(r)$. On the contrary, any solution that cannot be 
obtained as a result of a formal Lorentz boosting of a non-relativistic
force-free solution must have $dB^2_z/dr \geq 0$ in at least some interval 
of radii.  
An additional  relativistic-pinch solution can be obtained by adding
a line charge
$E_r \sim 1/r$. Although this provides intuitive physical insight into 
possible relativistic force-free configurations by relating them 
to the non-relativistic solutions, the  radius-dependent Lorentz 
boost method allows us to obtain only a subclass of all
possible relativistic solutions. Nevertheless, in this work, we limit 
our analysis to solutions that can be obtained by this method
and have $|E_r(r)| < |B_\phi (r)|$.

For cylindrical jets,
the axial and azimuthal electromagnetic velocities are
$\beta_{EM,z}= E_r B_\phi/B^2$ and
$\beta_{EM,\phi} = - B_z E_r /B^2 $.
The plasma velocity along the magnetic field is not constrained by the 
force-free equations.
In the previous sections, when calculating the emissivity and Doppler boosts,
we assumed that the total plasma velocity is along
the $z$ axis and that there is no rotation. 
This implies that the emitting plasma
indeed slides along the magnetic field in such a way as to make 
the azimuthal velocity equal to zero. 
This may not be true in real  jets; then one
needs to modify the relations allowing for radial-dependent axial and azimuthal
velocities. An alternative possible choice of the velocity of the plasma 
could be one that minimizes the kinetic energy of the flow. In this case,
the velocity of the plasma will be the electromagnetic velocity $\beta_{EM}$,  
since the absolute value of the velocity cannot be smaller than the 
electromagnetic velocity. This choice for the velocity field of the plasma 
was used in the  previous relativistic calculations of polarization from 
cylindrical relativistic force-free jets by \citet{pariev03}. 
Since \citet{pariev03} also assumed that $B_z$ 
is constant with radius, their field configuration cannot be obtained by 
the method described in this Appendix.

\newpage

\newpage

\newpage

\newpage

\newpage

\newpage

\newpage

\newpage

\newpage

\newpage
\newpage

\newpage

\newpage

\newpage

\newpage

\end{document}